\documentclass[twocolumn,times,tighten]{aastex62}
\newcommand\ksmpc{{~km~s$^{-1}$~Mpc$^{-1}$}}
\newcommand\lmass{{\log(M_\ast/M_\odot)}}
\newcommand\uv{{$U\!-\!V$}}
\newcommand\vj{{$V\!-\!J$}}
\newcommand\highz{{high-\emph{z}}}
\newcommand\fdm{{$f_\mathrm{DM}[<r_\mathrm{e}]$}}
\newcommand\mldm{{$(M/L)_\mathrm{\ast,NFW}$}}
\newcommand\mlmfl{{$(M/L)_\mathrm{MFL}$}}
\newcommand\mldyn{{$M_\mathrm{dyn}/L$}}

\newcommand\sigmaeff{{$\sigma_\mathrm{e}$}}
\newcommand\qint{{$q_\mathrm{int}$}}
\newcommand\qobs{{$q_\mathrm{obs}$}}
\newcommand\rsma{{$r_\mathrm{e}^\mathrm{sma}$}}

\newcommand\spark{{\texttt{SPARK}}}

\usepackage{etoolbox}
\usepackage{amsmath}
\usepackage{graphicx}
\hypersetup{colorlinks,breaklinks, citecolor=blue,linkcolor=blue,urlcolor=blue}

\makeatletter
\patchcmd{\NAT@citex}
    {\@citea\NAT@hyper@{\NAT@nmfmt{\NAT@nm}\NAT@date}}
    {\@citea\NAT@nmfmt{\NAT@nm}\NAT@hyper@{\NAT@date}}
    {}
    {}
\patchcmd{\NAT@citex}
    {\@citea\NAT@hyper@{%
         \NAT@nmfmt{\NAT@nm}%
         \hyper@natlinkbreak{\NAT@aysep\NAT@spacechar}{\@citeb\@extra@b@citeb}%
         \NAT@date}}
    {\@citea\NAT@nmfmt{\NAT@nm}%
     \NAT@aysep\NAT@spacechar%
     \NAT@hyper@{\NAT@date}}
    {}
    {}
\patchcmd{\NAT@citex}
    {\@citea\NAT@hyper@{%
         \NAT@nmfmt{\NAT@nm}%
         \hyper@natlinkbreak
         {\NAT@spacechar\NAT@@open\if*#1*\else#1\NAT@spacechar\fi}%
         {\@citeb\@extra@b@citeb}%
         \NAT@date}}
    {\@citea\NAT@nmfmt{\NAT@nm}%
        \NAT@spacechar\NAT@@open\if*#1*\else#1\NAT@spacechar\fi%
        \NAT@hyper@{\NAT@date}}
    {}
    {}
\makeatother

\submitjournal{The Astrophysical Journal}
\accepted{June 23, 2020}

\shorttitle{Stellar kinematics at $1.4 < \MakeLowercase{z} < 2.1$}
\shortauthors{Mendel et al.}

\begin{document}

\title{The kinematics of massive quiescent galaxies at $1.4 < \MakeLowercase{z} < 2.1$: dark matter fractions, IMF variation, and the relation to local early-type galaxies\footnote{Based on observations obtained at the Very Large Telescope (VLT) of the European Southern Observatory (ESO), Paranal, Chile (ESO program IDs 092.A-0091, 093.A-0079, 093.A-0187, and 094.A-0287).}}

\correspondingauthor{J.\ Trevor Mendel}
\email{trevor.mendel@anu.edu.au}

\author[0000-0002-6327-9147]{J.\ Trevor Mendel}
\affiliation{Universit\"{a}ts-Sternwarte M\"{u}nchen, Scheinerstr. 1, 81679 M\"{u}nchen, Germany}
\affiliation{Max-Planck-Institut f\"{u}r Extraterrestrische Physik, Giessenbachstr. 1, 85748 Garching, Germany}
\affiliation{Research School of Astronomy and Astrophysics, Australian National University, Canberra, ACT 2611, Australia}
\affiliation{ARC Centre of Excellence for All Sky Astrophysics in 3 Dimensions (ASTRO 3D)}

\author[0000-0001-8017-6097]{Alessandra Beifiori}
\affiliation{Universit\"{a}ts-Sternwarte M\"{u}nchen, Scheinerstr. 1, 81679 M\"{u}nchen, Germany}
\affiliation{Max-Planck-Institut f\"{u}r Extraterrestrische Physik, Giessenbachstr. 1, 85748 Garching, Germany}

\author[0000-0003-0378-7032]{Roberto P. Saglia}
\affiliation{Universit\"{a}ts-Sternwarte M\"{u}nchen, Scheinerstr. 1, 81679 M\"{u}nchen, Germany}
\affiliation{Max-Planck-Institut f\"{u}r Extraterrestrische Physik, Giessenbachstr. 1, 85748 Garching, Germany}

\author[0000-0001-7179-0626]{Ralf Bender}
\affiliation{Universit\"{a}ts-Sternwarte M\"{u}nchen, Scheinerstr. 1, 81679 M\"{u}nchen, Germany}
\affiliation{Max-Planck-Institut f\"{u}r Extraterrestrische Physik, Giessenbachstr. 1, 85748 Garching, Germany}

\author[0000-0003-2680-005X]{Gabriel B. Brammer}
\affiliation{Cosmic Dawn Center, Niels Bohr Institute, University of Copenhagen, Juliane Maries Vej 30, DK-2100 Copenhagen, Denmark}

\author[0000-0001-6251-3125]{Jeffrey Chan}
\affiliation{Department of Physics and Astronomy, University of California, Riverside, CA 92521, USA}

\author[0000-0003-4264-3381]{Natascha M. F\"orster Schreiber}
\affiliation{Max-Planck-Institut f\"{u}r Extraterrestrische Physik, Giessenbachstr. 1, 85748 Garching, Germany}

\author[0000-0002-9043-8764]{Matteo Fossati}
\affiliation{Universit\"{a}ts-Sternwarte M\"{u}nchen, Scheinerstr. 1, 81679 M\"{u}nchen, Germany}
\affiliation{Max-Planck-Institut f\"{u}r Extraterrestrische Physik, Giessenbachstr. 1, 85748 Garching, Germany}
\affiliation{Dipartimento di Fisica G. Occhialini, Universit\`a degli Studi di Milano-Bicocca, Piazza della Scienza 3, 20126 Milano, Italy}

\author{Audrey Galametz}
\affiliation{Department of Astronomy, University of Geneva, 1205, Versoix, Switzerland}

\author[0000-0003-1665-2073]{Ivelina G. Momcheva}
\affiliation{Space Telescope Science Institute, Baltimore, MD 21218, USA}

\author[0000-0002-7524-374X]{Erica J. Nelson}
\affiliation{Max-Planck-Institut f\"{u}r Extraterrestrische Physik, Giessenbachstr. 1, 85748 Garching, Germany}
\affiliation{Harvard-Smithsonian Center for Astrophysics, Cambridge, USA}

\author[0000-0002-1822-4462]{David J. Wilman}
\affiliation{Universit\"{a}ts-Sternwarte M\"{u}nchen, Scheinerstr. 1, 81679 M\"{u}nchen, Germany}
\affiliation{Max-Planck-Institut f\"{u}r Extraterrestrische Physik, Giessenbachstr. 1, 85748 Garching, Germany}

\author[0000-0003-3735-1931]{Stijn Wuyts}
\affiliation{Department of Physics, University of Bath, Claverton Down,Bath, BA2 7AY, UK}

\begin{abstract}

We study the dynamical properties of massive quiescent galaxies at $1.4 < z < 2.1$ using deep \emph{Hubble Space Telescope} WFC3/F160W imaging and a combination of literature stellar velocity dispersion measurements and new near-infrared spectra obtained using KMOS on the ESO VLT. We use these data to show that the typical dynamical-to-stellar mass ratio has increased by $\sim$0.2 dex from $z = 2$ to the present day, and investigate this evolution in the context of possible changes in the stellar initial mass function (IMF) and/or fraction of dark matter contained within the galaxy effective radius, \fdm{}. Comparing our high-redshift sample to their likely descendants at low-redshift, we find that \fdm{} has increased by a factor of more than 4 since $z \approx 1.8$, from \fdm{} = $6.6\pm1.0$\% to $\sim$24\%.  The observed increase appears robust to changes in the methods used to estimate dynamical masses or match progenitors and descendants. We quantify possible variation of the stellar IMF through the offset parameter $\alpha$, defined as the ratio of dynamical mass in stars to the stellar mass estimated using a Chabrier IMF. We demonstrate that the correlation between stellar velocity dispersion and $\alpha$ reported among quiescent galaxies at low-redshift is already in place at $z = 2$, and argue that subsequent evolution through (mostly minor) merging should act to preserve this relation while contributing significantly to galaxies overall growth in size and stellar mass.

\end{abstract}


\section{Introduction}
\label{section.intro}

Spectroscopic surveys of the high-redshift Universe have shown that well-known scaling relations such as the fundamental and mass planes were already in place by at least $z = 2$ \citep[e.g.][]{toft2012,bezanson2013a,van-de-sande2014,beifiori2017,prichard2017}, despite the fact that individual galaxies appear to evolve significantly from the time they join the passive population to the present day.  The most conspicuous signature of this evolution is seen in galaxy sizes, where massive quiescent galaxies at $z>1$ are significantly smaller than their local counterparts at fixed stellar mass \citep[e.g.][but see also \citealp{carollo2013}]{daddi2005,trujillo2006,van-dokkum2008,cimatti2012,van-der-wel2014,chan2016,chan2018}, but it is also apparent in measurements of galaxy stellar velocity dispersions and surface brightness profiles \citep[e.g.][]{kriek2009,cenarro2009,van-der-wel2011,van-de-sande2013,chang2013}. However, the exact degree to which individual galaxies change as they evolve is still unclear: although some amount of inferred evolution can be explained by a bias in the matching of progenitor and descendent populations \citep[progenitor bias, e.g.][]{van-dokkum1996,saglia2010,valentinuzzi2010,keating2015}, some evolution is still required to reproduce properties of the full population \citep[e.g.][]{belli2015}.

Guided by the intrinsically hierarchical assembly of structure in $\Lambda$CDM models, the most attractive explanation for the continued structural evolution of quiescent galaxies is by gas-poor merging after the cessation of star formation.  Both major (mass ratio $\mu_\ast > 0.25$) and minor ($\mu_\ast < 0.25$) mergers can significantly alter galaxy light profiles, leading to a disproportionate increase in (half-light/-mass) size relative to stellar mass \citep[e.g.][]{oser2010,hilz2012,hilz2013}, which seems all but demanded in the most compact, massive high-$z$ galaxies \citep[e.g.][]{damjanov2011}.  Detailed photometric and kinematic analyses of nearby passive galaxies appear to support the idea of a ``two-phase'' formation scenario characterized by early, rapid formation and subsequent assembly through repeated mergers \citep[e.g.][]{arnold2011,arnold2014,de-la-rosa2016,foster2016}.  But while it appears that mergers with $0.1 < \mu_\ast < 1$ can account for the evolution of galaxy sizes and velocity dispersions since $z \sim 1$, they have more difficulty explaining the dramatic increase in average sizes at earlier epochs \citep[e.g.][]{newman2012}, suggesting that other mechanisms such as stellar mass loss or feedback from active galactic nuclei (AGN) may also play some role \citep[e.g.][]{fan2008,damjanov2009,fan2010}.

While different evolutionary scenarios predict different physical characteristics for the resulting galaxy population, the persistence of the fundamental plane, mass plane, and other scaling relations over time limits the parameter space available to models describing the evolution of galaxy properties.  The existence of a fundamental plane for quiescent galaxies can be understood as a manifestation of the virial relation, where for relaxed systems the dynamical mass $M_\mathrm{dyn} \propto \sigma_\ast^2  r_\mathrm{e}$, with $\sigma_\ast$ and $r_\mathrm{e}$ the stellar velocity dispersion and half-light size respectively.  Given measurements of $\sigma_\ast$ and $r_\mathrm{e}$, the remaining unknown is the dynamical mass-to-light ratio, \mldyn{}.  Following \citet{graves2009a}, \mldyn{} can be rewritten in terms of its underlying physical dependencies as

\begin{equation}
\frac{M_\mathrm{dyn}}{L} = \frac{M_\mathrm{dyn}}{M_\mathrm{tot}} \times \frac{M_\mathrm{tot}}{M_\mathrm{\ast}} \times \frac{M_\mathrm{\ast}}{M_\mathrm{\ast,IMF}} \times \frac{M_\mathrm{\ast,IMF}}{L}.
\label{eqn.mdyn}
\end{equation}

\noindent  The first and last terms, $M_\mathrm{dyn}/M_\mathrm{tot}$ and $M_\mathrm{\ast,IMF}/L$, depend on our ability to model certain galaxy properties: the former encapsulates offsets between the derived dynamical mass $M_\mathrm{dyn}$ and the true total mass of the system $M_\mathrm{tot}$, while the latter is the stellar mass-to-light ratio for some fiducial stellar initial mass function (IMF), usually obtained by modelling multi-band photometric data.  The fact that dynamical studies of nearby early-type galaxies can recover the virial relation suggests that, given appropriate assumptions, $M_\mathrm{dyn}/M_\mathrm{tot} \approx 1$ \citep{[e.g.][]hyde2009,cappellari2013a}.  Uncertainties in the derivation of $M_\mathrm{\ast,IMF}/L$ from multi-band photometry, on the other hand, can be significant (of order 0.1--0.2 dex) depending on the treatment of star-formation history, metallicity, and dust \citep[e.g.][]{leja2019}.  The short formation timescales and low attenuation generally inferred for passive galaxies helps to reduce these uncertainties considerably \citep[e.g.][]{pforr2012}, but the extent to which these assumptions remain valid at higher redshift remains to be seen.

The remaining terms of Equation \ref{eqn.mdyn}, $M_\mathrm{tot}/M_\mathrm{\ast}$ and $M_\mathrm{\ast}/M_\mathrm{\ast,IMF}$, encapsulate the relationship between different physical components of the galaxy and are the most likely to be affected by evolutionary processes.  $M_\mathrm{tot}/M_\mathrm{\ast}$ is the ratio of total to stellar mass, and is related to the balance of baryonic and dark matter (DM) within a given aperture---typically the effective radius, $r_\mathrm{e}$---while $M_\mathrm{\ast}/M_\mathrm{\ast,IMF}$ accounts for differences between the assumed and true stellar IMF.  Variation of the IMF might be expected due to the evolution of interstellar medium (ISM) properties with redshift and stellar mass, but there is no clear theoretical consensus as to how these changes might manifest in the observed galaxy population \citep[see, e.g.,][and references therein]{chabrier2014,krumholz2014}.

In nearby galaxies, deep photometric and spectroscopic data can be used to study the relationship between galaxies, their stellar populations, and the properties of their dark matter halos in great detail.  \citet{van-dokkum2010a} used stellar population models to show that massive early-type galaxies host a large population of low-mass stars in their cores ($\lesssim r_\mathrm{e}/8$), suggesting a very bottom heavy IMF compared to the Milky Way (MW) and other nearby star-forming galaxies. These results were consistent with a complementary analysis of strong lensing systems by \citet{treu2010}, who additionally found evidence for \emph{systematic} variation of the IMF from MW-like at low stellar velocity dispersions to \citet{salpeter1955} or heavier in the most massive galaxies.  \citet{cappellari2012} obtained similar results based on modelling the spatially-resolved stellar kinematics of galaxies in the ATLAS$^\mathrm{3D}$ survey.  Stellar population results from studies like \citet{van-dokkum2010a} are uniquely sensitive to a galaxy's stellar content, but dynamical IMF constraints cannot necessarily distinguish between IMF variation and changes in the central DM fraction.  \citet{cappellari2013a} showed that the typical dark matter fraction within $r_\mathrm{e}$, \fdm{}, is relatively low (9--17\%) and, while \fdm{} tends to increase with increasing galaxy mass, this variation cannot account for the observed trends in \emph{total} $M/L$, supporting their conclusion of a systematically varying IMF \citep{cappellari2013}; unfortunately the picture becomes complicated if there is no clear distinction between the baryonic and dark matter distributions \citep[e.g.][]{thomas2011b}.  Even though there is no consensus on the exact correlations between IMF normalization (or shape) and observed galaxy properties, variability of the IMF is now supported by a number of different studies using a wide range of stellar population, lensing, and dynamical techniques \citep[e.g.][but see also \citealp{smith2015}]{thomas2011b,conroy2012,dutton2012,cappellari2013,conroy2013,ferreras2013,spiniello2014,martin-navarro2015,parikh2018}.

At intermediate redshift, \citet{tortora2018} used data from the Kilo Degree Survey (KiDS) and Sloan Digital Sky Survey (SDSS) to show that the locally-observed correlations between stellar mass, dynamical mass, stellar velocity dispersion, and structural parameters are already in place by $z\sim0.65$, but that high redshift quiescent galaxies likely have lower \fdm{} at fixed stellar velocity dispersion than nearby galaxies \citep[see also][]{beifiori2014,tortora2014}.  \citet{shetty2015} found a similar decrease in the central dark matter fraction for massive galaxies at $z\approx0.8$, while at the same time reporting a Salpeter-like IMF consistent with massive galaxies at $z=0$ \citep[see also][]{shetty2014,sonnenfeld2015,martin-navarro2015a}.  Extending such studies of kinematic scaling relations beyond $z > 1$ remains challenging. While an abundance of massive, compact red galaxies have been identified using deep Hubble Space Telescope (\emph{HST}) and ground-based imaging \citep[e.g.][]{cimatti2004,daddi2005,whitaker2011,whitaker2013}, kinematic data for individual galaxies have been notoriously difficult to obtain \citep[e.g.][]{kriek2009}.  The development of efficient, highly multiplexed near-infrared spectrographs such as MOSFIRE at Keck \citep{mclean2012} and KMOS at the ESO VLT \citep{sharples2012,sharples2013} has led to rapid growth in the number of kinematic measurements at $z > 1.4$, but individual samples remain relatively small and have been analysed using a wide variety of methods that makes combining the results from different surveys difficult.

In this paper we undertake a homogeneous re-analysis of currently-available kinematic data at high redshift in order to study the key parameters governing the behaviour of Equation \ref{eqn.mdyn} over cosmic time, namely the central dark matter fraction and normalization of the stellar IMF.  Our sample comprises 58 quiescent galaxies at $1.4 \leq z \leq 2.1$ with stellar velocity dispersion measurements and high resolution HST/WFC3 imaging available.  These data include 17 new stellar velocity dispersion measurements obtained as part of the VLT IR IFU Absorption Line Survey (VIRIAL; Mendel et al. in prep), in addition to measurements from a variety of samples in the literature.  We derive dynamical properties based on both a straightforward application of the virial theorem as well as more complex dynamical models, allowing us to test the influence of different assumptions about galaxy structure on the study of high-redshift stellar kinematics.  

The outline of this paper is as follows: in Section \ref{section.sample} we describe the compilation of high-redshift galaxies, along with a comparison sample at $z=0$.  In Section \ref{section.modelling} we discuss our modeling of galaxy surface brightness profiles and the calculation of dynamical masses.  The main results of this work---the relationship between dynamical and stellar masses, central dark matter fraction, and dynamical constraints on the normalization of the stellar IMF---are presented in Section \ref{section.results}.  In Section \ref{section.evolution} we discuss our results in the context of the high- and low-redshift galaxy populations.  We summarize our conclusions in Section \ref{section.conclusions}.

Throughout this paper we use AB magnitudes \citep{oke1983} and adopt a flat $\Lambda$CDM cosmology with $\Omega_{\Lambda} = 0.7$, $\Omega_{\mathrm{M}} = 0.3$ and $H_0 = 70$\ksmpc.

\section{Samples and Data}
\label{section.sample}

\subsection{KMOS observations at $1.5 < z < 2.0$}
\label{section.virial}

\begin{figure}
\centering
\includegraphics[scale=1]{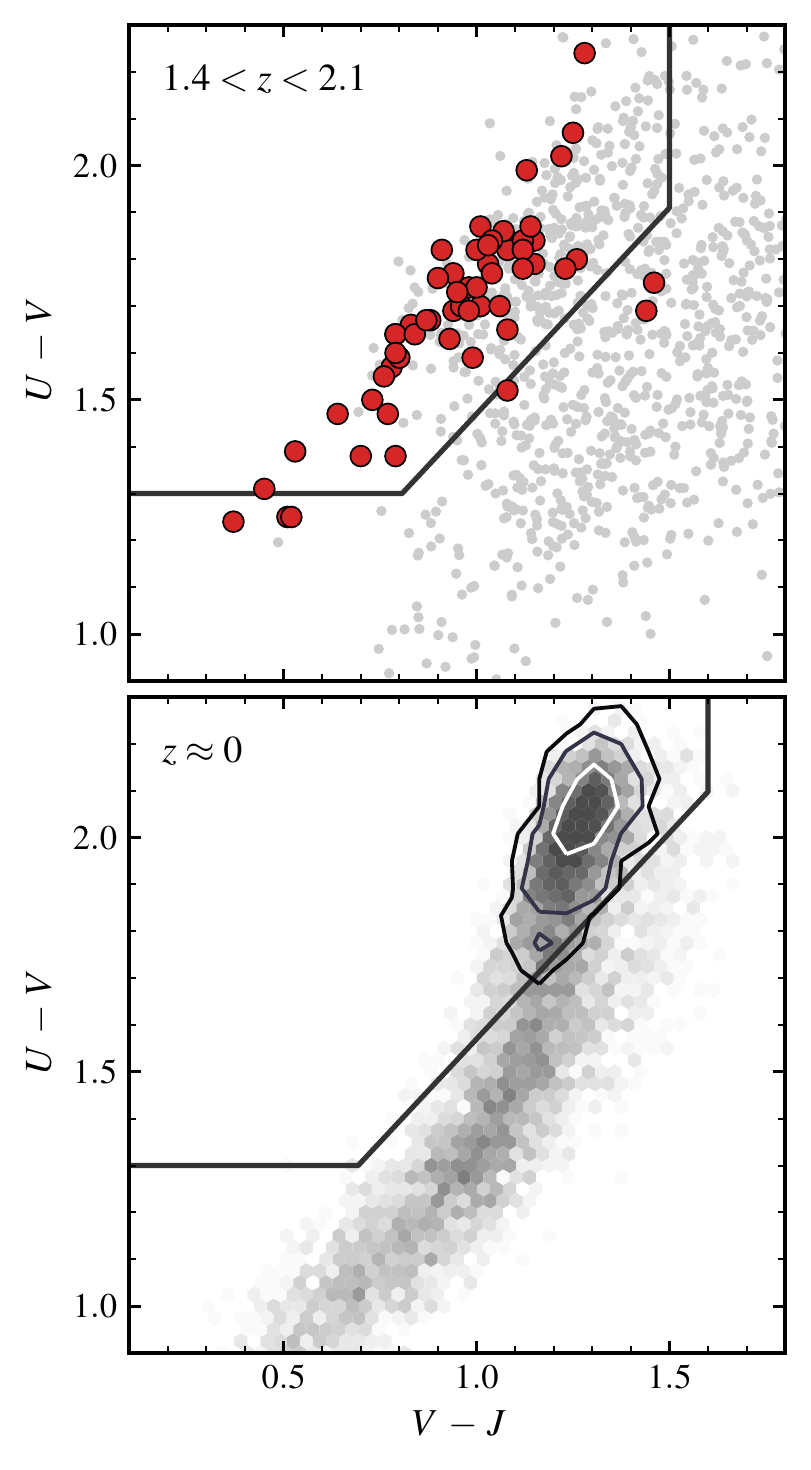}
\caption{Color-color selection used to identify quiescent galaxies for the high redshift sample (top panel) and the low-redshift GAMA/SDSS data (bottom panel).  The \emph{UVJ} selection window is taken from \citet{whitaker2011}.  In the top panel, small (grey) points show the underlying distribution of galaxies with $1.4 < z < 2.1$ from 3D-HST, while large filled (red) circles indicate the distribution of high-redshift data discussed in Sections \ref{section.virial} and \ref{section.literature}.  Note that although some galaxies in the high redshift sample fall outside of the \emph{UVJ} selection window we nevertheless include them in our analysis based on the presence of strong absorption features and the relative lack of emission lines in their spectra.  Contours in the bottom panel show the distribution of $U-V$ and $V-J$ colors for galaxies with mass-weighted stellar ages older than 9 Gyr.}
\label{fig.uvj}
\end{figure}

Our analysis includes new spectroscopic data for 17 galaxies in the redshift range $1.5 < z < 2.0$ observed as part of the VIRIAL GTO survey  \citep[Mendel et al.~in prep.]{mendel2015} using KMOS \citep{sharples2012,sharples2013}.  These galaxies were selected from 3D-HST \citep{brammer2012,skelton2014,momcheva2016} in the COSMOS, GOODS-S, and UDS fields to have $m_\mathrm{F140W} \leq 22.5$ mag and be classified as quiescent according to their rest-frame \uv{} and \vj{} colors using the criteria described by \citet[][see also \citealp{williams2009}]{whitaker2011}, shown in the top panel of Figure \ref{fig.uvj}. Their general properties are given in Table \ref{tab.virial_targets}.

\begin{deluxetable*}{lrrrcccr}
\tabletypesize{\footnotesize}
\tablewidth{0pt}
\tablecaption{General properties of 17 VIRIAL targets\label{tab.virial_targets}}
\tablehead{
\colhead{Field} & \colhead{ID} & \colhead{R.A.} & \colhead{Decl.} & \colhead{$m_\mathrm{F160W}$} & \colhead{$(U-V)_\mathrm{rf}$} & \colhead{$(V-J)_\mathrm{rf}$} & \colhead{exposure} \\
 & & \colhead{(J2000)} & \colhead{(J2000)} & \colhead{(mag)} & \colhead{} & \colhead{} & \colhead{(min.)}}
\startdata
UDS 		& 22480  	& 34.3353 	& -5.2017 		& 20.85 & 1.83 & 1.03 & 670 \\
UDS 		& 24891  	& 34.4458 	& -5.1940 		& 21.37 & 1.65 & 1.08 & 735 \\
GOODS-S	& 39364 	& 53.0628 	& -27.7265 	& 20.99 & 1.70 & 1.06 & 475 \\
GOODS-S	& 42113 	& 53.1279 	& -27.7189 	& 20.95 & 1.99 & 1.13 & 495 \\
GOODS-S	& 43548  	& 53.1294 	& -27.7073 	& 21.82 & 1.47 & 0.77 & 505 \\
COSMOS		& 6977  	& 150.0695 	& 2.2500 		& 21.62 & 1.73 & 0.95 & 650 \\
UDS			& 22802 	& 34.4469 	& -5.2007 		& 21.05 & 1.70 & 0.96 & 635 \\
UDS			& 29352 	& 34.4696 	& -5.1786 		& 21.44 & 1.69 & 0.94 & 740 \\
UDS			& 10237 	& 34.3148 	& -5.2433 		& 20.75 & 1.78 & 1.12 & 440 \\
COSMOS		& 7411 	& 150.1770 	& 2.2552 		& 21.37 & 1.82 & 1.00 & 630 \\
UDS			& 35111 	& 34.4536 	& -5.1589 		& 21.63 & 1.67 & 0.87 & 740 \\
UDS			& 32892 	& 34.3896 	& -5.1681		& 21.17 & 1.55 & 0.76 & 660 \\
UDS			& 38073 	& 34.3365 	& -5.1490 		& 21.30 & 1.38 & 0.79 & 635 \\
COSMOS		& 6396 	& 150.1728 	& 2.2441 		& 21.89 & 1.69 & 0.98 & 615\\
COSMOS		& 9227 	& 150.0618 	& 2.2737 		& 21.47 & 1.60 & 0.79 & 620 \\ 
COSMOS		& 7391 	& 150.0773 	& 2.2548 		& 22.01 & 1.39 & 0.53 & 650 \\
COSMOS		& 2816 	& 150.1411 	& 2.2085 		& 21.43 & 1.84 & 1.04 & 650 \\
\enddata
\end{deluxetable*}

\subsubsection{Observations and data reduction}
\label{section.obs}

Observations of VIRIAL galaxies were carried out between 2014 and 2016 using the KMOS $YJ$ band (1--1.36$\mu$m).  Data were taken using a standard object-sky-object pattern with individual exposure times of 300s.  Each science exposure was offset by between 0\farcs{1} and 0\farcs{6} in order to avoid bad pixels in the final extracted spectra.  Along with our science targets, we assigned one IFU from each of the three KMOS spectrographs to a reference star which we used to monitor the ambient conditions (seeing, atmospheric transmission, etc.), pointing accuracy, and point spread function (PSF) shape.  Due to the relatively small angular size of the KMOS IFUs (2\farcs{8}$\times$2\farcs{8}), sky exposures were taken nodding completely off source. Total on-source integration times range from 440 to 740 minutes (see Table \ref{tab.virial_targets}).

Data were reduced using a combination of the Software Package for Astronomical Reductions with KMOS pipeline tools \citep[\spark{};][]{davies2013} and custom \texttt{Python} scripts.  In the following we briefly outline the steps used to produce calibrated one-dimensional spectra.  Details of the VIRIAL reduction will be described in a future paper (Mendel et al.~in prep.).  Calibration exposures (dark, arc, and flat) were reduced using standard \spark{} routines to produce flat field, wavelength, and spatial calibration frames.  When processing science frames we first corrected each raw image for a readout channel dependent bias term estimated from reference pixels around the perimeter of each detector.  We then adjusted the wavelength and spatial illumination calibrations for each exposure based on the positions and relative flux of bright sky lines before subtracting the object and sky images.  The brightness of atmospheric OH lines can vary significantly between object and sky exposures \citep[$\sim$10\% on 5--10 minute timescales;][]{ramsay1992,davies2007}, often leading to significant systematic residuals in the initial sky-subtracted frames.  In order to limit the impact of these systematics on our final spectra we performed a second-order correction to the sky for each IFU using residuals measured in other IFUs in the same detector, excluding the IFU of interest.  

One-dimensional spectra were extracted directly from the flat fielded, illumination corrected, and sky subtracted detector frames for each exposure separately.  Since VIRIAL targets are typically undetected in individual 300s exposures we used the available 3D-HST/CANDELS F125W imaging \citep{grogin2011,koekemoer2011,skelton2014} to model the source flux distribution and mask neighboring objects in the optimal extraction.  The HST images were convolved to match the KMOS PSF measured from the reference stars in each exposure, which were also used to adjust for changes in transmission between exposures.  Individual optimally-extracted spectra were then corrected for telluric absorption using synthetic atmospheric models computed with {\tt MOLECFIT} \citep{kausch2014}, and combined using inverse variance weights. Uncertainties on the output spectra were estimated using bootstrap combines of the individual 1D spectra for each object. The typical spectral resolution in the extracted 1D spectra (as measured from sky lines) ranges from \emph{R} = 3000 to 3500 ($\sigma_\mathrm{inst} \approx 36-42$ km s$^{-1}$) depending on arm and detector \citep[see also][]{wisnioski2019}.

\subsubsection{Stellar masses and velocity dispersions}
\label{section.spec_fitting}

We estimated stellar velocity dispersions for VIRIAL galaxies using a simultaneous fit to the observed KMOS spectrum and multi-band photometry from 3D-HST \citep{skelton2014}.  We generated model spectral energy distributions (SEDs) using FSPS v2.4 \citep{conroy2009,conroy2010b} assuming a lognormal star formation history (SFH) with

\begin{equation}
\mathrm{SFR}(t) = \begin{cases}
\frac{1}{t \sqrt{2\pi\tau^2}} e^{-\slantfrac{(\ln t - \ln t_0)^2}{2 \tau^2}} & \text{if $t\leq t_\mathrm{trunc}$}\\
0  & \text{if $t > t_\mathrm{trunc}$},
\end{cases}
\label{eqn.ssfr}
\end{equation}

\noindent where $t$ is the age of the universe, $t_0$ is the delay time, and $\tau$ controls the width of the distribution \citep[see also][]{gladders2013}.  The additional parameter $t_\mathrm{trunc}$ allows for star formation to be abruptly truncated, and provides added flexibility when modeling the star-formation histories of quiescent galaxies.  We stress that our adoption of a lognormal SFH is motivated by its flexibility compared to more commonly used $\tau$ or delayed-$\tau$ models, rather than an assumption that galaxies star-formation histories are intrinsically lognormal \citep[e.g.][]{gladders2013,abramson2016,diemer2017}.  In Appendix \ref{appendix.modeling} we show that our derived velocity dispersions are not biased by the use of a parametric SFH.  We modeled the effects of dust using a two-component extinction law which includes a foreground screen and additional attenuation towards young stellar populations ($<10^7$ yr), which are assumed to remain embedded within their birth clouds \citep[see, e.g.,][]{charlot2000}.  We used the reddening curve of \citet{calzetti2000} and, following \citet{wuyts2013}, adopted a relationship between the total $V$-band extinction $A_V$ and the additional extinction towards young stellar population $A_{extra}$ such that $A_{extra} = 0.9 A_V - 0.15A_V ^2$.  For simplicity we assume a fixed solar metallicity; in Appendix \ref{appendix.modeling} we show that changing the metallicity by $\pm0.2$ dex leads to systematic shifts in the derived velocity dispersion of $\lesssim 2$\%.

\begin{figure*}[t]
\centering
\includegraphics[scale=1.0]{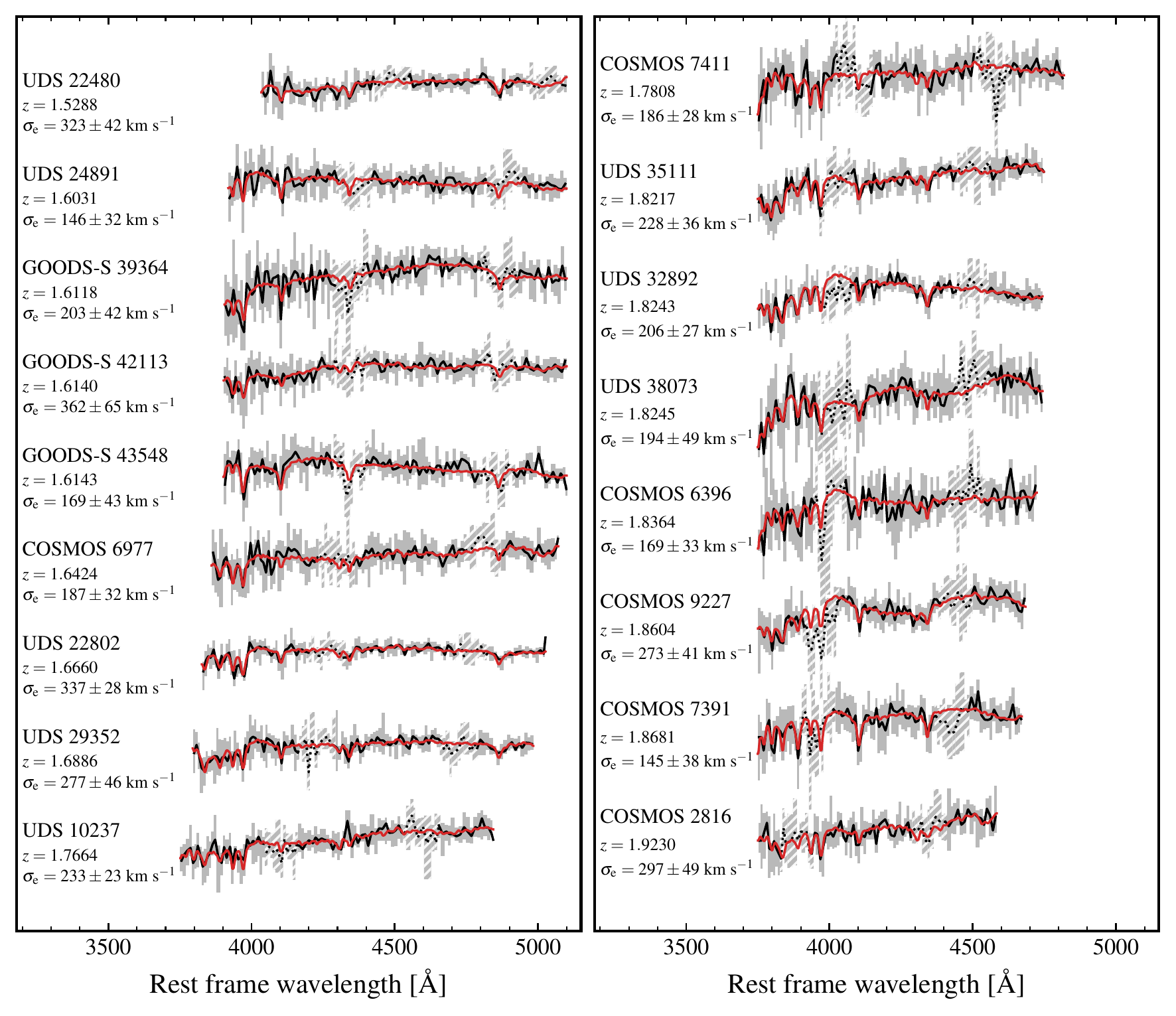}
\caption{KMOS spectra for the 17 galaxies described in Section \ref{section.virial}.  The extracted one-dimensional spectra (black) and uncertainties (grey) have been median rebinned in a 15 pixel ($\sim$10 \AA{} rest-frame) moving window for display purposes.  The best-fit model is overplotted in red.   Dotted lines and hatching indicate regions of the spectra which are significantly contaminated by sky emission and absorption features.}
\label{fig.spectra}
\end{figure*}

Before fitting, templates were smoothed to match the wavelength-dependent KMOS resolution measured from sky lines in extracted 1D spectra.  The final matched templates include an additional (constant) offset of $\sigma_\mathrm{offset}=65$ km s$^{-1}$ to account for the resolution difference between KMOS ($\sigma_\mathrm{inst}\approx 40$ km s$^{-1}$) and the adopted MILES spectral library \citep[$\sigma_\mathrm{MILES}\approx75$ km s$^{-1}$;][]{beifiori2011}. This effectively sets a floor for our velocity dispersion measurements of 65 km s$^{-1}$. We limited our fits to the wavelength range from 3750 to 5300 \AA, and included a 9$^\mathrm{th}$ order additive polynomial---corresponding to $\sim$1 order per 10,000 km s$^{-1}$---to minimize the effects of template mismatch on our final velocity dispersion measurements. We verified that our results are not sensitive to the adoption of an additive, as opposed to multiplicative, polynomial.  In the end our model has a total of 7 free parameters: redshift, $z$; stellar mass, $M_\mathrm{SPS}$\footnote{For clarity we will refer to stellar masses derived via SED fitting as $M_\mathrm{SPS}$ in order to distinguish them from those derived using dynamical methods.  In the context of Equation \ref{eqn.mdyn} these represent $M_\mathrm{\ast,IMF}$, the stellar mass derived using a fiducial, in this case \citet{chabrier2003}, IMF.}; three parameters which describe the star-formation history, $\tau$, $t_0$, and $t_\mathrm{trunc}$; absolute $V$-band extinction, $A_V$; and stellar velocity dispersion, $\sigma_\ast$.  Samples from the posterior distribution were generated using \texttt{emcee} \citep{foreman-mackey2013}, and our final estimates of velocity dispersion and stellar mass were taken as the medians of their respective marginal posterior distributions, with 1$\sigma$ uncertainties estimated from the 16$^\mathrm{th}$ and 84$^\mathrm{th}$ percentiles.  We have confirmed that the derived stellar masses do not change significantly if we re-fit objects using only the available photometric data (i.e. excluding spectra).  The final redshifts, stellar masses, and velocity dispersions are provided in Table \ref{tab.virial_derived}\footnote{The dispersions quoted in Table \ref{tab.virial_derived} have been corrected for the effects of seeing and scaled to the luminosity-weighted mean within the half-light radius following the procedure outlined by \citet{van-de-sande2013}.  The derived corrections range between 1.02 and 1.1, and are consistent with similar corrections derived directly from the dynamical modelling discussed in Section \ref{section.jam}}.  One-dimensional spectra and the corresponding best-fit models are shown in Figure \ref{fig.spectra}.

\begin{figure}
\centering
\includegraphics[scale=1.0]{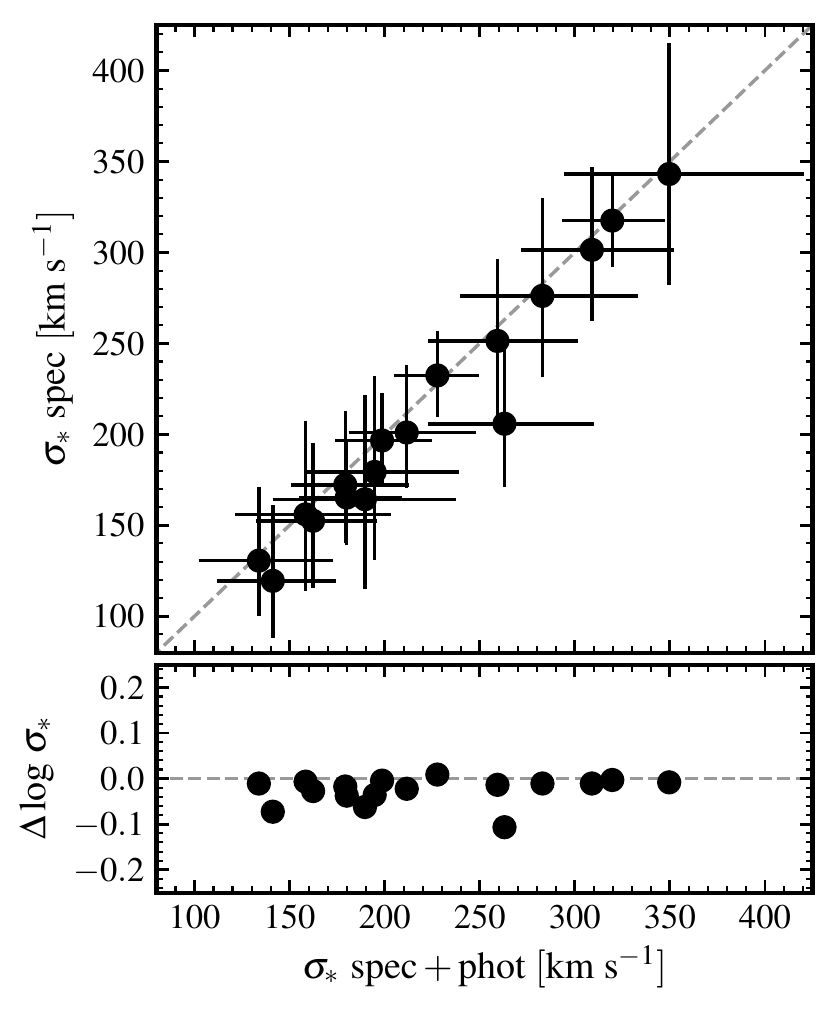}
\caption{Comparison of velocity dispersions derived from fits to the KMOS spectra alone to those which include multi-band photometry in the fits.  Uncertainties are $\pm1\sigma$ estimated from the marginalized posterior distribution described in Section \ref{section.spec_fitting}.  The dashed line marks a one-to-one correlation between the two measurements.  There is a clear systematic offset between the two measurements such that the spectra-plus-photometry fits predict slightly higher velocity dispersions on average.}
\label{fig.spec_phot}
\end{figure}

In Figure \ref{fig.spec_phot} we show a comparison of stellar velocity dispersions obtained with and without the inclusion of photometric data in the fit.  The two estimates are generally consistent within their quoted uncertainties, though there is a clear systematic offset in the sense that spectra-only fits return velocity dispersions which are $\sim$5\% lower on average than those which also incorporate photometric data.  This stems from the fact that the photometric data generally down-weight the youngest spectral templates---as one might expect from our \textit{a priori} selection of galaxies based on their \uv{} and \vj{} colors---preferring instead solutions with smaller contributions from (rapidly-rotating) early stellar types.

\begin{deluxetable}{lrccc}
\tabletypesize{\footnotesize}
\tablewidth{0pt}
\tablecaption{Redshift, velocity dispersion and stellar masses of KMOS galaxies \label{tab.virial_derived}}
\tablehead{
\colhead{Field} & \colhead{ID} & \colhead{$z$} & \colhead{$\log\,(M_\mathrm{SPS}/M_\odot)$} & \colhead{$\sigma_\mathrm{e}$\tablenotemark{a}} \\
 & & & & \colhead{(km s$^{-1}$)} }
\startdata
UDS 				& 22480  	& 1.5288 & 11.08 	& $323\pm 42$\\
UDS\tablenotemark{b} 	& 24891  	& 1.6031 & 10.99 	& $146\pm 32$\\
GOODS-S 			& 39364  	& 1.6118 & 11.10 	& $203\pm 42$\\
GOODS-S 			& 42113 	& 1.6140 & 11.20 	& $362\pm 65$\\
GOODS-S 			& 43548  	& 1.6143 & 10.64 	& $169\pm 43$\\
COSMOS 			&  6977  	& 1.6424 & 10.86 	& $187\pm 32$\\
UDS\tablenotemark{b} 	& 22802  	& 1.6660 & 11.13 	& $337\pm 28$\\
UDS\tablenotemark{b} 	& 29352  	& 1.6886 & 10.91 	& $277\pm 46$\\
UDS 				& 10237  	& 1.7664 & 11.38 	& $233\pm 23$\\
COSMOS 			& 7411  	& 1.7808 & 11.09 	& $186\pm 28$\\
UDS 				& 35111  	& 1.8217 & 10.95 	& $228\pm 36$\\
UDS 				& 32892  	& 1.8243 & 11.02 	& $206\pm 27$\\
UDS 				& 38073  	& 1.8245 & 10.94 	& $194\pm 49$\\
COSMOS 			& 6396  	& 1.8364 & 10.90 	& $169\pm 33$\\
COSMOS 			& 9227  	& 1.8604 & 10.98 	& $273\pm 41$\\
COSMOS 			& 7391  	& 1.8681 & 10.54 	& $145\pm 38$\\
COSMOS 			& 2816  	& 1.9230 & 11.26 	& $297\pm 49$
\enddata
\tablecomments{The formal statistical uncertainties on stellar masses derived from our SED fitting is of order 0.02 dex.  Where relevant we include an additional 0.15 dex uncertainty on $\log\,(M_\ast/M_\odot)$ in quadrature to account for systematic uncertainties in the determination of stellar masses \citep[see, e.g.][]{conroy2009,mendel2014}.}
\tablenotetext{a}{Velocity dispersion corrected to $r_\mathrm{e}$ following the procedure described by \citet{van-de-sande2013}.}
\tablenotetext{b}{These galaxies are in common with the \citet{belli2017} sample.  See discussion in Section \ref{section.literature}}
\end{deluxetable}

\subsection{Literature data at $1.4 < z < 2.1$}
\label{section.literature}

\begin{figure}
\centering
\includegraphics[scale=1.0]{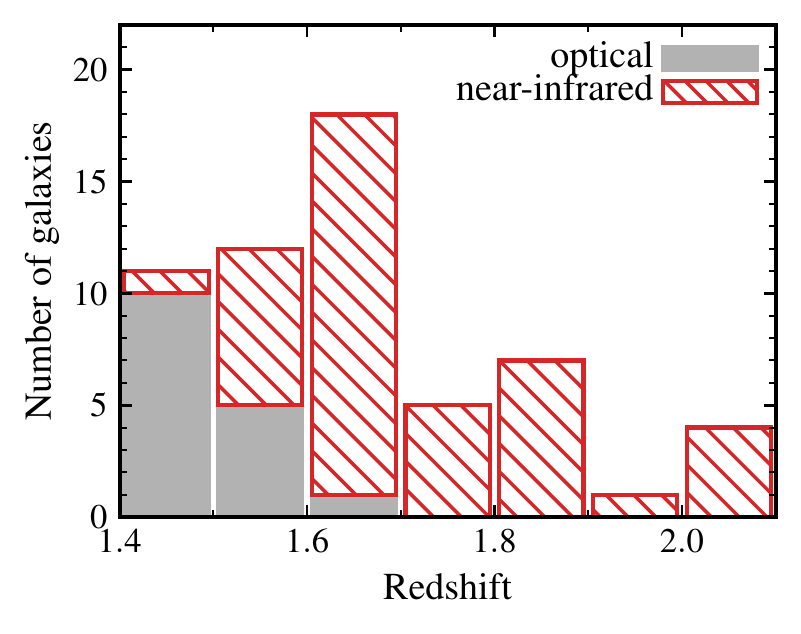}
\caption{Redshift distribution of our passive galaxy sample.  The solid (grey) histograms shows the distribution of data observed using red-sensitive optical detectors, while the hatched (red) distribution shows the contribution of near-infrared observations.}
\label{fig.redshift}
\end{figure}

In addition to our KMOS data, we have compiled a sample of quiescent galaxies with $1.4 < z < 2.1$ from the literature where HST/WFC3 F160W imaging, multi-wavelength photometric catalogs, and stellar velocity dispersion measurements were available.  A full accounting of the literature data is given in Table \ref{tab.lit_params}, along with a few general galaxy properties.  In Figure \ref{fig.redshift} we show the redshift distribution of our full high-redshift galaxy sample.

This literature sample includes 15 galaxies from \citet{newman2010}, \citet{bezanson2013}, and \citet{belli2014a} observed with Keck LRIS at $1.4 < z < 1.6$, as well as 2 galaxies from the GMASS spectroscopic sample \citep{cimatti2008} with velocity dispersions published by \citet{cappellari2009}.  \citet{belli2014a} incorporate the LRIS data from \citet{newman2010} in their analysis, and there is one galaxy in common between \citet{belli2014a} and \citet{cappellari2009}.  Velocity dispersion measurements derived using NIR spectroscopy are available for 29 additional galaxies with $1.4 < z < 2.1$ from \citet{toft2012} and \citet{van-de-sande2013}, obtained using VLT XShooter, and from \citet{belli2014}, \citet{barro2016}, and \citet{belli2017} using Keck MOSFIRE.  The sample of \citet{barro2016} includes one galaxy in common with the LRIS sample of \citet{newman2010} and \citet{belli2014a}, and there are several galaxies in common between \citet{toft2012}, \citet{van-de-sande2013}, \citet{belli2014}, and \citet{belli2017}.  See Table \ref{tab.lit_params} for details.

There are three galaxies in common between \citet{belli2017} and the KMOS sample described in Section \ref{section.virial}---UDS 24891, UDS 29352 and UDS 22802---which are highlighted in Tables \ref{tab.virial_derived} and \ref{tab.lit_params}. For UDS 22802, the two independent velocity dispersion measurements are in relatively good agreement ($337\pm28$ vs. $291\pm31$ km s$^{-1}$), however for the other two galaxies the discrepancy is larger: $277\pm46$ vs. $146\pm31$ km s$^{-1}$ for UDS 29352 (2.4-$\sigma$ offset) and $146\pm32$ vs. $391\pm71$ km s$^{-1}$ for UDS 24891 (3.1-$\sigma$ offset). Although we are not in a position to assess which of these measurements are ``correct'', we note that adopting \sigmaeff{} as measured by \citet{belli2017} for these galaxies results in large offsets between their dynamical and stellar masses (see Section \ref{results.mdyn_mstar}), such that UDS 29352 (UDS 24891) would have the highest (lowest) dynamical-to-stellar mass ratio in the sample. Nevertheless, in the absence of additional data we adopt a final \sigmaeff{} for these objects based on an error-weighted average of the quoted measurements, with an increased uncertainty to reflect the large discrepancy between quoted values; in Section \ref{section.vdisp} we describe in more detail how we combine data for galaxies with multiple velocity dispersion measurements.

In order to ensure that our high-redshift sample is as homogeneous as possible, we re-measured stellar masses for all galaxies using the SED fitting procedure described in Section \ref{section.spec_fitting}.  In most cases, multi-wavelength photometric catalogs were available from either the Newfirm Medium Band Survey \citep[NMBS;][]{whitaker2011} or 3D-HST \citep{skelton2014}.  Several galaxies in the UDS field---UDS 55531 and UDS 53937 from \citet{bezanson2013}, as well as UDS 19627 from \citet{toft2012} and \citet{van-de-sande2013}---fall outside of the 3D-HST footprint, and for these objects we adopted the combined Subaru/XMM-Newton Deep Survey \citep[SXDS;][]{furusawa2008} and UKIRT Infrared Deep Sky Survey \citep{lawrence2007} catalogs described by \citet{simpson2012}.  We supplemented these data with deep \emph{Spitzer}/IRAC 3.6 and 4.5$\mu m$ flux measurements from \citet{ashby2013}, which were corrected to match the 3$^{\prime\prime}$ apertures used by \citet{simpson2012} using the UKIDSS \emph{K}-band mosaics.  Stellar masses derived for the literature sample are provided in Table \ref{tab.lit_params}.

\setlength\tabcolsep{0.3cm}
\begin{deluxetable*}{lrccccrlr}
\tabletypesize{\footnotesize}
\tablewidth{0pc}
\tablecaption{Properties of the high-$z$ literature sample\label{tab.lit_params}}
\tablehead{
\colhead{Field} 	& \colhead{3D-HST ID}& \colhead{$z$}  & \colhead{$\log\,(M_\mathrm{SPS}/M_\odot)$\tablenotemark{a}} & \colhead{$(U-V)$} & \colhead{$(V-J)$} 	& \colhead{Ref. ID} & \colhead{$\sigma_\mathrm{e}$} & \colhead{Reference} \\
& & & & & & & \colhead{(km s$^{-1}$)} &
}
\startdata
COSMOS 			& 30145 	& 1.4010	& 10.90	& 1.84 	& 1.09 	& 19498 		& $250\pm39$					& \citet{belli2014a} \\
AEGIS 				& 5087	& 1.4060	& 11.00	& 1.84 	& 1.15 	& 42109		& $369\pm48$					& \citet{belli2014a} \\
					&		&		&		&		&		& E9	 		& $295\pm69$\tablenotemark{b}	& \citet{newman2010} \\
GOODS-S			& 40623	& 1.4149	& 10.89	& 2.07	& 1.25	& 2239		& $116\pm36$\tablenotemark{b}	& \citet{cappellari2009} \\ 
GOODS-S			& 42466	& 1.4150	& 11.07	& 1.82	& 1.08	& 5020     		& $181\pm54$					& \citet{belli2014a} \\
					&		&		&		&		&		& 2470 		& $147\pm27$\tablenotemark{b}	& \citet{cappellari2009} \\ 
GOODS-S			& 43042 	& 1.4190	& 11.32	& 2.24	& 1.28	& 4906		& $298\pm26$ 					& \citet{belli2014a} \\
AEGIS				& \ldots	& 1.4235	& 11.26	& 1.57	& 0.78	& A17300 		& $276\pm7$\tablenotemark{b}		& \citet{bezanson2013} \\
COSMOS				& 21628	& 1.4320	& 10.82	& 1.79	& 1.15	& 13880		& $169\pm70$					& \citet{belli2014a} \\
COSMOS\tablenotemark{d}	& 31780	& 1.4390	& 10.78	& 1.75	& 1.46	& 20841		& $267\pm52$					& \citet{belli2014a} \\
COSMOS				& 31136	& 1.4420	& 10.93	& 1.86	& 1.07	& 20275		& $221\pm70$					& \citet{belli2014a} \\
UDS					& 1854	& 1.4560	& 11.49	& 1.74	& 0.98	& 29410		& $355\pm98$					& \citet{van-de-sande2013} \\ 
UDS\tablenotemark{d}	& \ldots	& 1.4848	& 11.53	& 1.52	& 1.08	& U55531 	& $260\pm24$\tablenotemark{b}	& \citet{bezanson2013} \\
COSMOS				& \ldots	& 1.5222	& 11.34	& 1.77	& 0.94	& C20866		& $284\pm24$\tablenotemark{b}	& \citet{bezanson2013} \\
COSMOS				& \ldots	& 1.5223	& 11.26	& 1.66	& 0.83	& C21434		& $229\pm17$\tablenotemark{b}	& \citet{bezanson2013} \\
COSMOS				& 17364 	& 1.5260	& 11.02	& 1.84	& 1.12	& 17364 		& $168\pm84$					& \citet{belli2017} \\
COSMOS				& 17361 	& 1.5270	& 10.86	& 1.63	& 0.93	& 17361 		& $169\pm43$					& \citet{belli2017} \\
COSMOS				& 17641 	& 1.5280	& 10.79	& 1.70	& 1.01	& 17641 		& $142\pm54$					& \citet{belli2017} \\
COSMOS				& 17089 	& 1.5280	& 11.37	& 2.02	& 1.23	& 17089 		& $348\pm57$					& \citet{belli2017} \\
AEGIS				& 17926 	& 1.5730 	& 11.14	& 1.79	& 1.03	& 17926 		& $231\pm39$					& \citet{belli2017} \\
AEGIS           			& 22719 	& 1.5790 	& 11.13	& 1.87	& 1.14	& 22719 		& $262\pm51$					& \citet{belli2017} \\
COSMOS				& 28523	& 1.5825	& 11.38	& 1.82	& 0.91	& 34265 		& $377\pm54$					& \citet{belli2014a} \\
					&		& 		&		&		&		&18265		& $400\pm72$					& \citet{van-de-sande2013} \\
AEGIS				& \ldots	& 1.5839	& 11.24	& 1.47	& 0.64	& A21129		& $275\pm10$\tablenotemark{b}	& \citet{bezanson2013} \\
GOODS-N			& 17678	& 1.5980	& 11.00	& 1.59	& 0.80	& 2653 		& $174\pm27$					& \citet{belli2014a} \\
					&		&		&		&		&		& GN5 		& $245\pm37$\tablenotemark{b}	& \citet{newman2010} \\ 
					&		& 		&		&		&		& 12632		& $187\pm36$\tablenotemark{b}	& \citet{barro2016} \\ 
UDS\tablenotemark{c}      & 24891 	& 1.6035	& 10.99	& 1.65	& 1.08	& 24891 		& $391\pm71$					& \citet{belli2017} \\
UDS             			& 35616 	& 1.6090	& 11.19	& 1.64	& 0.79	& 35616 		& $198\pm49$					& \citet{belli2017} \\
UDS             			& 30737 	& 1.6200	& 11.37	& 1.77	& 1.04	& 30737 		& $307\pm82$					& \citet{belli2017} \\
UDS					& \ldots	& 1.6210	& 10.93	& 1.31	& 0.47	& U53937 	& $251\pm21$\tablenotemark{b}	& \citet{bezanson2013} \\
UDS             			& 43367 	& 1.6240	& 11.26	& 1.80	& 1.26	& 43367 		& $299\pm74$					& \citet{belli2017} \\
UDS             			& 30475 	& 1.6330	& 10.83	& 1.38	& 0.70	& 30475 		& $296\pm109$				& \citet{belli2017} \\
UDS             			& 32707 	& 1.6470	& 11.25	& 1.82	& 1.12	& 32707 		& $174\pm30$					& \citet{belli2017} \\
COSMOS         			& 16629 	& 1.6570	& 10.67	& 1.64	& 0.79	& 16629 		& $358\pm76$					& \citet{belli2017} \\
UDS             			& 37529 	& 1.6650 	& 11.13	& 1.78	& 1.23	& 37529 		& $232\pm60$					& \citet{belli2017} \\
UDS\tablenotemark{c}      & 22802 	& 1.6665	& 11.13	& 1.70	& 0.96	& 22802 		& $291\pm31$					& \citet{belli2017} \\
GOODS-N\tablenotemark{d}	& 11470	& 1.6740	& 10.77	& 1.25	& 0.51	& 8231 		& $221\pm36$\tablenotemark{b}	& \citet{barro2016} \\ 
GOODS-N			& 24033	& 1.6740	& 10.80	& 1.64	& 0.84	& 17360		& $155\pm31$\tablenotemark{b}	& \citet{barro2016} \\ 
GOODS-N			& 3604	& 1.6750	& 10.69	& 1.67	& 0.88	& 2617		& $317\pm118$	\tablenotemark{b}	& \citet{barro2016} \\ 
UDS\tablenotemark{c}      & 29352 	& 1.6895	& 10.91	& 1.69	& 0.94	& 29352 		& $146\pm31$					& \citet{belli2017} \\
COSMOS         			& 19958 	& 1.7220	& 10.75	& 1.50	& 0.73	& 19958 		& $169\pm87$					& \citet{belli2017} \\
COSMOS         			& 17255 	& 1.7390	& 10.97	& 1.74	& 1.00	& 17255 		& $147\pm40$					& \citet{belli2017} \\
AEGIS           			& 25526 	& 1.7520	& 10.84	& 1.59	& 0.99	& 25526 		& $134\pm36$					& \citet{belli2017} \\
COSMOS\tablenotemark{d}	& \ldots	& 1.8000	& 11.31	& 1.24	& 0.37	& 7447		& $287\pm53$					& \citet{van-de-sande2013} \\
UDS\tablenotemark{d}	& \ldots	& 2.0360	& 11.20	& 1.25	& 0.52	& 19627		& $304\pm41$					& \citet{van-de-sande2013} \\
					& 		& 		&		&		&		& 19627		& $335\pm56$\tablenotemark{b}	& \citet{toft2012} \\ 
COSMOS         			& 13083 	& 2.0880	& 11.10	& 1.76	& 0.90	& 13083 		& $197\pm52$					& \citet{belli2017} \\
COSMOS				& 11494 	& 2.0920	& 11.58	& 1.87	& 1.01	& 7865		& $446\pm57$					& \citet{van-de-sande2013} \\
					& 		& 		& 		&		&		& 31719	 	& $358\pm30$					& \citet{belli2014} \\
					& 		&		&		&		&		& 11494		& $319\pm26$					& \citet{belli2017} \\
COSMOS\tablenotemark{d}	& 12020	& 2.0960	& 11.34	& 1.69	& 1.44	& 31769		& $312\pm65$					& \citet{belli2014} \\
\enddata
\tablenotetext{a}{Stellar masses are re-derived in this work following the method described in Section \ref{section.spec_fitting}.}
\tablenotetext{b}{Dispersions corrected to $R_\mathrm{e}$ following the Appendix B of \citet{van-de-sande2013}.}
\tablenotetext{c}{These galaxies are in common with the KMOS sample.  See discussion in Section \ref{section.literature}}
\tablenotetext{d}{These galaxies fall outside of the \emph{UVJ} quiescent selection defined by \citet{whitaker2011}; however, their spectra show strong absorption features characteristic of post-starburst galaxies as well as weak or absent [\ion{O}{2}] emission, so we include them in our analysis.}
\end{deluxetable*}

\subsection{Comparison sample at $z \approx 0$}
\label{section.lowz}

We identified a comparison sample of quiescent galaxies at low redshift from the SDSS Legacy Survey \citep[Sloan Digital Sky Survey;][]{abazajian2009} using the same color-based selection criteria as at high redshift. We select galaxies with $0.02 \leq z \leq 0.2$ and that also have stellar velocity dispersions measured by the Portsmouth group \citep[see][]{thomas2013} using pPXF \citep{cappellari2004}.  In order to avoid potential biases in the SED fitting between our high- and low-redshift data we limit our selection to galaxies in the GAMA DR2 survey area \citep[Galaxy and Mass Assembly;][]{driver2011,liske2015}, where \citet{wright2016} provide aperture-matched photometric catalogs covering from the ultraviolet to infrared. 

We computed rest-frame colors for these galaxies using \texttt{EAZY} \citep{brammer2008} and the resulting distribution is shown in the bottom panel of Figure \ref{fig.uvj}, where we again adopt the color criteria of \citet{whitaker2011} to select quiescent galaxies.  Stellar masses for the 4546 galaxies satisfying this selection were estimated from fits to their far-UV to \emph{K}-band photometry using the procedures described in Sections \ref{section.spec_fitting} and \ref{section.literature}. Based on these data we derive a redshift-dependent stellar mass limit following the approach of \citet{sohn2017} and \citet{zahid2019}, such that our final sample of 3108 galaxies is mass complete at the 97.5\% level.

The passive galaxy population appears to grow significantly from $z=2$ to the present day, suggesting our low-redshift data contains galaxies which are too young to be descendants of the galaxies in our \highz{} sample. While there is no consensus on the magnitude of such progenitor bias effects \citep[e.g.][]{carollo2013,belli2015,fagioli2016}, it is nevertheless important for account for them in our analysis.  We use here the mass-weighted stellar ages derived by \citet{comparat2017} for SDSS galaxies using \texttt{FIREFLY} \citep{wilkinson2017}, and select those galaxies with ages older than 9~Gyrs as the most likely descendants of our \highz{} sample.  This identifies a sub-sample of 792 galaxies, or $\sim$27\% of the full quiescent sample.  In the following we will discuss results for both the full and age-selected samples.

\section{Dynamical modeling}
\label{section.modelling}

The main focus of this work is a discussion of the dynamical constraints afforded by current high-redshift quiescent galaxy samples and a comparison with low-redshift data.  In this section we describe the key quantities required for this analysis---stellar velocity dispersions, structural parameters (sizes, Sersic indices, etc.)---as well as our estimates of dynamical masses and their related quantities.

\subsection{Stellar velocity dispersions}
\label{section.vdisp}

Galaxies in our high-redshift sample have stellar velocity dispersions derived within a range of apertures, and are based on data obtained with a variety of instruments and extraction methods. These measurements therefore require some degree of homogenisation in order to be meaningfully combined.  In many cases authors quote velocity dispersions corrected such that they represent the luminosity-weighted mean with one effective radius, $\sigma_e$, and we adopt these values when available.  Where velocity dispersions are quoted within a different aperture---as is the case for \citet{cappellari2009}, \citet{newman2010}, \citet{toft2012}, \citet{bezanson2013}, and \citet{barro2016}---we correct the quoted velocity dispersion to one effective radius following the procedure outlined in \citet{van-de-sande2013}.  

In cases where multiple velocity dispersion measurements were available we used an inverse variance weighted average of the published dispersions, after correcting them to a common $r_\mathrm{e}$ aperture.  As well as their propagated uncertainties, we included an additional term (in quadrature) to account for large offsets between quoted dispersions, taken as half of the range of dispersion measurements.  The one exception to this procedure is GOODS-N 17678, where the velocity dispersion measured by \citet{newman2010} differs significantly from the measurements of \citet{belli2014} and \citet{barro2016}; for this object we used an average of only the \citet{belli2014} and \citet{barro2016} dispersions.

In our low redshift sample all dispersions were measured from spectra within a common 3$^{\prime\prime}$ aperture, corresponding to the SDSS fibre diameter.  Where we quote individual stellar velocity dispersions, these aperture values have been corrected to one effective radius, again following the procedure described by \citet{van-de-sande2013} and using structural parameters described below. However, in our dynamical modelling (see Section \ref{section.masses}) we fit directly to model dispersions computed within the 3$^{\prime\prime}$ (fibre) aperture, accounting for seeing effects.  Although the physical scale subtended by the 3$^{\prime\prime}$ SDSS fibers increases dramatically over the redshift range of our low-$z$ sample, the physical quantities derived from our dynamical models are \emph{independent} of redshift at fixed stellar mass, suggesting that the use of aperture measurements does not bias our results.

\subsection{Structural parameters}
\label{seciton.structural_parameters}

We adopted two different approaches to measuring structural properties for our galaxy samples: first using \texttt{galfit} \citep{peng2002} to model their two-dimensional surface brightness distributions using a single S\'ersic profile \citep{sersic1963}, and second using the Multi-Gaussian Expansion (MGE) approach described by \citet[][see also \citealp{cappellari2002}]{emsellem1994}.

\subsubsection{S\'ersic profile fits}
\label{section.sersic_fits}

In our high-redshift sample 50/58 galaxies fall within the \emph{HST} WFC3/F160W imaging footprint of the CANDELS survey \citep{grogin2011,koekemoer2011}, and for these objects we used the mosaics and composite point spread functions (PSFs) described by \citet{skelton2014}\footnote{\url{http://3dhst.research.yale.edu/Data.php}}.  The remaining 8 galaxies were observed separately using  \emph{HST} WFC3/F160W as part of HST-GO-12167 (PI: Franx; AEGIS 17300, AEGIS 21129, COSMOS 21434, COSMOS 20866, COSMOS 07447, UDS 53937, and UDS 55531) and HST-GO-13002 (PI: Williams; UDS 19627) for a single orbit each with total exposure times of 2611 or 2411 s, respectively.  Level 2 data products were retrieved from the Hubble Legacy Archive (HLA)\footnote{\url{http://hla.stsci.edu/}} and we constructed empirical PSFs for these objects by stacking the images of bright unsaturated stars in each combined frame.  We generated segmentation maps for the HLA images using \texttt{SExtractor} \citep{bertin1996} with parameters similar to those given by \citet{skelton2014} for 3D-HST.

We used \texttt{galfit} to model the surface brightness distribution of the primary galaxy, while also including in the fit neighboring galaxies with $m_\mathrm{F160W} < 25$ and projected separations $r_\mathrm{p} < 5 (r_\mathrm{primary}+r_\mathrm{neighbor})$.  Initial estimates of the galaxy sizes, i.e. $r_\mathrm{primary}$ and $r_\mathrm{neighbor}$, were taken from the \texttt{SExtractor} output.  The local sky background for each object was estimated using the full image by first masking all pixels within 3 Kron radii of nearby sources using the ellipse parameters produced by \texttt{SExtractor}.  We then identified the nearest 10,000 un-masked pixels as sky.  An initial estimate of the background was taken as the mode of these sky pixels, which was then iteratively refined to obtain our final estimates of the local sky background. Postage stamps for individual objects were then extracted and the local sky background removed; the background level was subsequently held fixed during fitting.  The structural parameters derived in this way are consistent with those available in the literature; a direct comparison with literature values is given in Appendix \ref{appendix.highz}.  An example of our photometric modeling for COSMOS 30145 is shown in Figure \ref{fig.sb_example}, with figures for the remaining galaxies included in Appendix \ref{appendix.phot_figs}.

Although there are numerous existing catalogs of structural parameter measurements for the SDSS and GAMA \citep[e.g.][]{simard2002,kelvin2012,meert2015}, for consistency with our \highz{} data we chose to re-derive these quantities using the methodology described above.  We retrieved ``corrected'' \emph{r}-band images from the SDSS Data Archive Server (DAS), along with their associated mask and PSF files.  We then used \texttt{SExtractor} to generate segmentation images following the procedures described by \citet{simard2011}, and the local sky background for each source was estimated using the method described above. Individual postage stamps and PSFs\footnote{Source-specific PSFs were extracted from the SDSS drField files using the {\tt read\textunderscore PSF} routine described at \url{http://classic.sdss.org/dr7/products/images/read_psf.html}} for each galaxy were then extracted and the background removed.  \texttt{galfit} was used to simultaneously fit the primary galaxy and any neighboring sources with $m_r < 22$ and $r_\mathrm{p} < 5 (r_\mathrm{primary} + r_\mathrm{neighbor})$.  All other sources were masked during the fit.  A comparison of our measurements with several different literature catalogs can be found in Appendix \ref{appendix.lowz}.

At both high and low redshift we derive sizes in fixed photometric bands (\emph{HST} WFC3/F160W and SDSS $r$-band, respectively), which probe different \emph{rest-frame} wavelengths at different redshifts.  In the presence of strong color gradients this shift in rest-frame wavelength can systematically bias our size measurements and must be taken into account. Following \citet{van-der-wel2014} we define the corrected, in this case $r$-band, semi-major axis size $r_\mathrm{e}^\mathrm{sma}$ as

\begin{equation}
r_\mathrm{e}^\mathrm{sma} = r_\mathrm{e, obs}^\mathrm{sma} \left ( \frac{1+z}{1+z_p} \right ) ^{\frac{\Delta \log r_\mathrm{e}}{\Delta \log \lambda}},
\label{eqn.size_correction}
\end{equation} 

\noindent where $r_\mathrm{e, obs}^\mathrm{sma}$ is the measured half-light size in either the WFC3/F160W or SDSS $r$-band filter and $z_p$ is the ``pivot redshift''. $z_p = 0$ by definition for the GAMA/SDSS sample as we are correcting to the rest-frame $r$-band size, while for F160W imaging $z_p = 1.49$.  \citet{kelvin2012} used GAMA data to show that $\Delta \log r_\mathrm{e} / \Delta \log \lambda = -0.3$ for early-type galaxies on average.  \citet{chan2016} and \citet{van-der-wel2014} derive similar values based on their analyses of quiescent galaxies high redshift, and we therefore adopt $\Delta \log r_\mathrm{e} / \Delta \log \lambda = -0.3$ for all galaxies in our sample.  The typical correction derived in this way is of order 2-3\%, and we adopt these corrected $r$-band sizes for the remainder of this work.

\subsubsection{Multi-Gauss Expansion fits}
\label{section.mge_fits}

While the single-component S\'ersic fits described in Section \ref{section.sersic_fits} provide a straightforward summary of the overall surface brightness profile, S\'ersic models have several drawbacks which complicate their use in constructing dynamical models.  As well as providing a poor description of multi-component profiles (e.g. bulge + disk),  the coupling between inner and outer profile shapes makes the S\'ersic models extremely sensitive to sky background: over-/under-subtraction of the sky level can significantly affect the inferred inner profile shape. In addition, with the exception of a few special cases, S\'ersic profiles cannot be de-projected analytically, making their use for constructing dynamical models computationally expensive compared to simpler functional forms. In this context, modeling galaxies as a sum of individual Gaussian components---so-called multi-Gaussian expansion \citep[MGE;][]{emsellem1994,cappellari2002}---provides a flexible description of surface brightness profiles which does not require any extrapolation of the profile to large radii, can accommodate multiple photometric components, and can be easily de-projected to obtain an estimate of the three-dimensional luminosity density (see Section \ref{section.jam}).

\begin{figure*}
\centering
\includegraphics[scale=1]{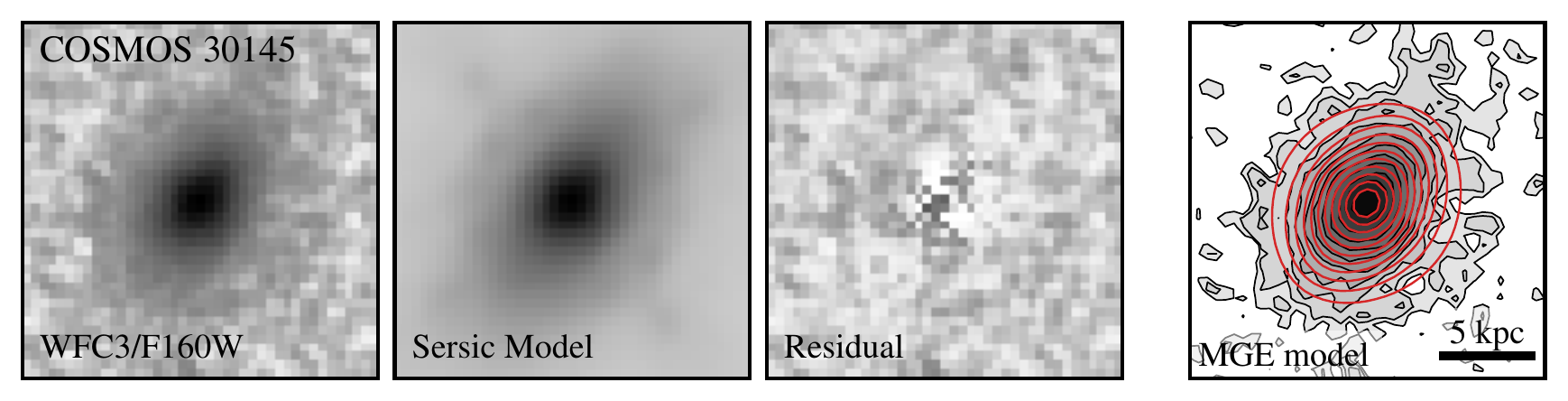}
\caption{An example of the photometric models adopted for galaxies in our high-redshift sample. From left to right panels show the observed \emph{HST} WFC3/F160W image from CANDELS/3D-HST \citep{skelton2014}, the best-fit \texttt{galfit} model, and the resulting image residual.  The right-most panel shows contours from the best-fit MGE model (red) overlayed on the WFC3/F160W image. In all cases the images are plotted in surface brightness units, and contours are evenly spaced in steps of 0.5 mag arcsec$^{-2}$.  Models for all galaxies in our high-redshift sample are show in Appendix \ref{appendix.phot_figs}. }
\label{fig.sb_example}
\end{figure*}

The starting points for our MGE models were the background-subtracted postage stamps produced as described in Section \ref{section.sersic_fits}.  We used the results of our S\'ersic model fits to subtract neighboring sources before identifying the primary object and producing binned two-dimensional surface brightness measurements using the {\tt find\textunderscore galaxy} and {\tt sectors\textunderscore photometry} routines described by \citet{cappellari2002}\footnote{Available at \url{http://purl.org/cappellari/software}.}.  A model of the surface photometry in terms of nested Gaussians was then derived using the {\tt mge\textunderscore fit\textunderscore sectors} method of \citet{cappellari2002}.  For high-redshift galaxies we constructed MGE-based PSF models per field using either composite PSFs provided by \citet{skelton2014} for galaxies within the CANDELS/3D-HST footprint, or else the stacked images of bright stars within the same field for stand-alone observations.   In the right-hand panel of Figure \ref{fig.sb_example} we show a comparison of the observed and MGE derived surface brightness contours for one object, COSMOS 30145.

MGE PSF models for the low-redshift SDSS/GAMA data were constructed on a galaxy-by-galaxy basis using the PSF extracted from the SDSS drField files.  In all cases---that is, both high and low redshift---we tied the ellipticity of the fitted Gaussian components together to avoid large variations in the derived axis ratios for low-surface-brightness components; however, we confirmed that our results are not qualitatively sensitive to this assumption.

\subsection{Dynamical masses}
\label{section.masses}

The final piece of information we require is an estimate of total galaxy mass, including both stellar and dark matter components.  For this work we investigate two broad approaches to estimating dynamical masses in order to test their sensitivity to underlying assumptions: the first is based on a simple application of the virial theorem and scaling relations derived for nearby galaxies, while the second relies on more detailed dynamical modelling of the stellar density profile and velocity dispersion.

\subsubsection{Virial mass estimates}

As outlined in Section \ref{section.intro}, the tight relationship between size, stellar velocity dispersion, and mass for nearby early-type galaxies can be understood as a consequence of virial equilibrium, where for a pressure-supported system the total mass is given by 

\begin{equation}
M_\mathrm{vir} = \kappa(n) \frac{\sigma_\mathrm{e}^2 r_\mathrm{e}^\mathrm{sma}}{G}.
\label{eqn.virial_mass}
\end{equation}

\noindent Here, $\kappa(n)$ is the so-called virial coefficient, and in this case is taken as an analytic function of S\'ersic index that encapsulates the effects of structural and orbital non-homology \citep[e.g.][]{bertin2002, cappellari2006}.  We adopt the relation derived by \citet{cappellari2006} based on spherical, isotropic models,

\begin{equation}
\kappa(n) = 8.87 - 0.831n + 0.0241n^2,
\label{eqn.virial_coeff}
\end{equation}

\noindent which has been shown to provide a reliable estimate of the total mass for nearby early-type galaxies in the SAURON and ATLAS$^\mathrm{3D}$ samples \citep[e.g.][]{cappellari2006,cappellari2013a}.  Note that in Equation \ref{eqn.virial_mass} we used the semi-major axis size, \rsma{}, following the discussion of \citet[their figure 14]{cappellari2013a}.  The semi-major axis size is expected to be more robust to systematic changes in galaxy shapes than the harmonic mean size (e.g.~$\sqrt{ab}$, where $a$ and $b$ are the semi-major and semi-minor axis sizes), especially for (thin) disk galaxies where the observed $b/a$ is an indicator of inclination rather than intrinsic shape.

\subsubsection{Jeans models}
\label{section.jam}

The assumptions of spherical symmetry and isotropy discussed above appear to be reasonable at low redshift, however \highz{} quiescent galaxies are known to be flatter on average---that is, have intrinsically lower $b/a$---than their low-redshift counterparts \citep[e.g.][]{van-der-wel2011,chang2013}, leading to a possible bias in their derived masses when using Equation \ref{eqn.virial_coeff}.  We therefore consider an alternative approach to computing dynamical masses based on the Jeans Anisotropic MGE (JAM) method discussed by \citet{cappellari2008}, which allows us to relax these assumptions.  The modelling requires as input the MGE-derived surface brightness profile describe in Section \ref{section.mge_fits} and a measurement of the stellar velocity dispersion (see Section \ref{section.spec_fitting} and \ref{section.literature}).

Following \citet{cappellari2002} the deprojected luminosity density can be computed from the best-fit MGE decomposition given assumptions about the inclination, which is related to the \emph{intrinsic} axis ratio of an oblate ellipsoid \qint{} by

\begin{equation}
\cos i = \sqrt{\frac{q_\mathrm{obs}^2 - q_\mathrm{int}^2}{1-q_\mathrm{int}^2}},
\label{eqn.inclination}
\end{equation}

\noindent where $i$ is the inclination and \qobs{} is the observed axis ratio.  Since in this work we are concerned with the sensitivity of our dynamical mass estimates to possible changes of the intrinsic axis ratio, we computed JAM models over a grid of \qint{} from $0.05 \leq q_\mathrm{int} \leq \mathrm{min}(0.95, q_\mathrm{obs})$ in steps of 0.05; unless otherwise stated our results are based on marginalizing over \qint{}.  In our default modeling we assumed that the velocity ellipsoid is marginally anisotropic with an anisotropy parameter $\beta \equiv 1 - \sigma_z^2/\sigma_R^2 = 0.2$ (where $z$ and $R$ define directions parallel and perpendicular to the symmetry axis for an axisymmetric system) based on local early-type galaxies \citep[e.g.][]{cappellari2007,thomas2009}.  We explored possible systematic effects over a range of anisotropies from $0 \leq \beta \leq 0.8$ and found they resulted in variations of the derived dynamical masses of at most a few per cent, consistent with previous results \citep[e.g.][]{wolf2010,dutton2013}; all results are therefore quoted adopting our fiducial value of $\beta = 0.2$.

We adopt two different implementations of the JAM modelling procedure distinguished by their treatment of baryonic and dark matter components.  In the first instance we assume that the total mass is proportional to the light at all radii, i.e. mass-follows-light (MFL).  This provides a self-consistent estimate of the dynamical mass-to-light ratio \mlmfl{}.  MFL models have been shown to reliably recover the total mass within relatively small apertures ($\lesssim r_\mathrm{e}$) even in the presence of multiple mass components \citep[e.g.][]{cappellari2006,williams2010}, and provide a baseline comparison for dynamical masses computed following Equation \ref{eqn.virial_coeff}.  A similar approach was used by \citet{shetty2014} to study quiescent galaxies at $z\sim0.8$ in the DEEP2 survey.  In the MFL case the best-fitting value of \mlmfl{} for a given combination of $q_\mathrm{int}$ and $\beta$ is simply given by \mlmfl{} $ = (\sigma_\mathrm{e, obs} / \sigma_\mathrm{e,model})^2$, where $\sigma_\mathrm{e,obs}$ is the observed aperture velocity dispersion and $\sigma_\mathrm{e,model}$ is the model prediction assuming \mlmfl{} $ = 1$.  Instead, our second implementation includes an explicit dark matter component described by a spherical NFW halo profile \citep{navarro1996}. With sufficient sampling of the velocity field it is possible to independently constrain the stellar mass-to-light ratio, \mldm{}, and properties of the dark matter halo \citep[e.g.][]{cappellari2013a,ubler2018}.  However, the aperture velocity dispersions used here cannot be used to break the degeneracy between stellar and dark matter components, leading us to impose additional constraints on the properties of the dark matter halo.  Starting from our photometric estimates of galaxy stellar mass, we assigned dark matter halo masses based on the evolving stellar-to-halo mass relation derived by \citet{moster2013}.  We then used the calculations of \citet{diemer2015} to assign a halo concentrations.  This halo profile was then fed back into the JAM modelling procedure along with the MGE-based stellar density profile, and a grid search was used to determine the mass-to-light ratio of the stellar component, \mldm{}, as a function of \qint{} and $\beta$.  

In the explicit DM halo case we obtain an estimate of the dark matter fraction within $r_e$, \fdm{}, defined as

\begin{equation}
f_\mathrm{DM}[< r_\mathrm{e}] = \frac{M_\mathrm{DM}}{M_\mathrm{\ast,NFW} + M_\mathrm{DM}}.
\label{eqn.fdm_nfw}
\end{equation}

\noindent We compute \fdm{} within a volume defined by $4 \pi r_\mathrm{e}^3 / 3$, where for consistency with the literature $r_\mathrm{e}$ is the circularized half-light radius ($\equiv r_\mathrm{e}^\mathrm{sma} \times \sqrt{q_\mathrm{obs}}$), and the relevant masses are computed using the derived $M/L$ values and deprojected MGE luminosity densities.  For all galaxies we use the rest-frame \emph{r}-band sizes computed following Equation \ref{eqn.size_correction}.

\begin{longrotatetable}
\renewcommand{\arraystretch}{1.1}
\begin{deluxetable*}{lrcccclccccc}
\tablewidth{0pt}
\tablecolumns{10}
\tablecaption{Derived properties of high-redshift galaxies\label{table.derived}}
\tablehead{
 \colhead{Field} & \colhead{ID\tablenotemark{a}} & \colhead{$r_\mathrm{e}^\mathrm{sma}$\tablenotemark{$\dagger$}} & \colhead{$n$} & \colhead{$q_\mathrm{obs}$} & \colhead{$\log(L_r)$} & \colhead{$\sigma_\ast$} & \colhead{$\log(M/L_r)_\mathrm{VIR}$} & \colhead{$\log(M/L_r)_\mathrm{MFL}$} & \colhead{$\log(M/L_r)_{\ast, \mathrm{NFW}}$} & \colhead{$\log(f_\mathrm{DM})$} & \colhead{Ref.} \\
 \colhead{} & \colhead{} & \colhead{$[\mathrm{kpc}]$} & \colhead{} & \colhead{} & \colhead{$[L_\odot]$} & \colhead{$[\mathrm{km~s^{-1}}]$} & \colhead{$[M_\odot/L_\odot]$} & \colhead{$[M_\odot/L_\odot]$} & \colhead{$[M_\odot/L_\odot]$} & \colhead{} & \colhead{}
 }
\startdata
COSMOS & 30145  & $ 1.25\pm 0.26$ & $4.31$ & $0.42$ & $11.28\pm0.01$ & $250\pm 39$ & $-0.30\pm0.17$ & $-0.32\pm0.14$ & $-0.33\substack{+0.11 \\[-1pt] -0.18}$ & $-1.77\substack{+0.18 \\[-1pt] -0.11}$ & 2\\
AEGIS & 05087  & $ 1.18\pm 0.26$ & $2.53$ & $0.41$ & $11.36\pm0.01$ & $345\pm 54$ & $-0.02\pm0.17$ & $-0.12\pm0.14$ & $-0.12\substack{+0.11 \\[-1pt] -0.18}$ & $-2.02\substack{+0.18 \\[-1pt] -0.11}$ & 2, 3\\
GOODS-S & 40623  & $ 2.29\pm 0.27$ & $2.98$ & $0.88$ & $11.18\pm0.01$ & $116\pm 36$ & $-0.52\pm0.27$ & $-0.50\pm0.27$ & $-0.61\substack{+0.12 \\[-1pt] -1.51}$ & $-0.63\substack{+0.34 \\[-1pt] -0.18}$ & 4\\
GOODS-S & 42466  & $ 2.53\pm 0.30$ & $6.42$ & $0.91$ & $11.45\pm0.01$ & $154\pm 30$ & $-0.75\pm0.18$ & $-0.51\pm0.17$ & $-0.61\substack{+0.13 \\[-1pt] -0.45}$ & $-0.62\substack{+0.21 \\[-1pt] -0.12}$ & 2, 4\\
GOODS-S & 43042  & $ 3.48\pm 0.30$ & $5.62$ & $0.63$ & $11.68\pm0.01$ & $298\pm 26$ & $-0.20\pm0.09$ & $-0.11\pm0.08$ & $-0.19\substack{+0.07 \\[-1pt] -0.11}$ & $-0.72\substack{+0.08 \\[-1pt] -0.07}$ & 2\\
AEGIS & A17300\tablenotemark{b} & $ 2.93\pm 0.29$ & $5.48$ & $0.68$ & $11.89\pm0.01$ & $276\pm  7$ & $-0.54\pm0.06$ & $-0.47\pm0.02$ & $-0.52\substack{+0.02 \\[-1pt] -0.03}$ & $-0.92\substack{+0.02 \\[-1pt] -0.02}$ & 5\\
COSMOS & 21628  & $ 1.09\pm 0.26$ & $3.07$ & $0.70$ & $11.18\pm0.01$ & $169\pm 70$ & $-0.52\pm0.37$ & $-0.48\pm0.36$ & $-0.50\substack{+0.16 \\[-1pt] -1.35}$ & $-1.42\substack{+0.73 \\[-1pt] -0.20}$ & 2\\
COSMOS & 31780  & $ 2.45\pm 0.27$ & $1.28$ & $0.32$ & $11.11\pm0.01$ & $267\pm 52$ & $\hphantom{-}0.39\pm0.18$ & $\hphantom{-}0.30\pm0.17$ & $\hphantom{-}0.29\substack{+0.13 \\[-1pt] -0.25}$ & $-1.74\substack{+0.24 \\[-1pt] -0.13}$ & 2\\
COSMOS & 31136  & $ 2.01\pm 0.28$ & $4.24$ & $0.45$ & $11.33\pm0.01$ & $221\pm 70$ & $-0.24\pm0.28$ & $-0.29\pm0.28$ & $-0.31\substack{+0.16 \\[-1pt] -0.78}$ & $-1.36\substack{+0.50 \\[-1pt] -0.17}$ & 2\\
UDS & 01854  & $ 2.40\pm 0.27$ & $2.69$ & $0.49$ & $11.99\pm0.01$ & $355\pm 98$ & $-0.32\pm0.24$ & $-0.39\pm0.24$ & $-0.43\substack{+0.15 \\[-1pt] -0.69}$ & $-1.12\substack{+0.40 \\[-1pt] -0.16}$ & 6\\
UDS & U55531\tablenotemark{b} & $ 8.01\pm 0.40$ & $3.90$ & $0.75$ & $11.94\pm0.01$ & $260\pm 24$ & $-0.09\pm0.08$ & $-0.05\pm0.08$ & $-0.43\substack{+0.10 \\[-1pt] -0.56}$ & $-0.22\substack{+0.08 \\[-1pt] -0.07}$ & 5\\
COSMOS & C20866\tablenotemark{b} & $ 2.71\pm 0.27$ & $3.45$ & $0.66$ & $11.84\pm0.01$ & $284\pm 24$ & $-0.35\pm0.09$ & $-0.38\pm0.07$ & $-0.43\substack{+0.07 \\[-1pt] -0.10}$ & $-0.90\substack{+0.08 \\[-1pt] -0.06}$ & 5\\
COSMOS & C21434\tablenotemark{b} & $ 1.99\pm 0.26$ & $3.43$ & $0.72$ & $11.87\pm0.01$ & $229\pm 17$ & $-0.70\pm0.09$ & $-0.69\pm0.06$ & $-0.74\substack{+0.06 \\[-1pt] -0.08}$ & $-0.97\substack{+0.07 \\[-1pt] -0.06}$ & 5\\
COSMOS & 17364  & $ 2.86\pm 0.28$ & $2.77$ & $0.48$ & $11.37\pm0.01$ & $168\pm 84$ & $-0.28\pm0.44$ & $-0.35\pm0.43$ & $-0.42\substack{+0.13 \\[-1pt] -1.88}$ & $-0.90\substack{+0.57 \\[-1pt] -0.26}$ & 7\\
COSMOS & 17361  & $ 1.91\pm 0.26$ & $2.13$ & $0.70$ & $11.45\pm0.01$ & $169\pm 43$ & $-0.49\pm0.23$ & $-0.58\pm0.22$ & $-0.61\substack{+0.15 \\[-1pt] -0.51}$ & $-1.07\substack{+0.34 \\[-1pt] -0.15}$ & 7\\
COSMOS & 17089  & $ 5.65\pm 0.40$ & $4.58$ & $0.86$ & $11.85\pm0.01$ & $348\pm 57$ & $\hphantom{-}0.06\pm0.15$ & $\hphantom{-}0.12\pm0.14$ & $\hphantom{-}0.01\substack{+0.12 \\[-1pt] -0.34}$ & $-0.54\substack{+0.16 \\[-1pt] -0.11}$ & 7\\
COSMOS & 17641  & $ 1.19\pm 0.27$ & $5.32$ & $0.92$ & $11.26\pm0.02$ & $142\pm 54$ & $-0.86\pm0.35$ & $-0.68\pm0.33$ & $-0.70\substack{+0.16 \\[-1pt] -1.20}$ & $-1.14\substack{+0.57 \\[-1pt] -0.19}$ & 7\\
UDS & 22480  & $ 1.84\pm 0.26$ & $4.54$ & $0.55$ & $11.62\pm0.01$ & $323\pm 42$ & $-0.26\pm0.13$ & $-0.24\pm0.11$ & $-0.25\substack{+0.09 \\[-1pt] -0.15}$ & $-1.54\substack{+0.14 \\[-1pt] -0.09}$ & 1\\
AEGIS & 17926  & $ 5.45\pm 0.43$ & $4.19$ & $0.72$ & $11.61\pm0.02$ & $231\pm 39$ & $-0.04\pm0.15$ & $-0.05\pm0.15$ & $-0.20\substack{+0.12 \\[-1pt] -0.51}$ & $-0.47\substack{+0.17 \\[-1pt] -0.11}$ & 7\\
AEGIS & 22719  & $ 2.25\pm 0.30$ & $6.08$ & $0.94$ & $11.54\pm0.01$ & $262\pm 51$ & $-0.39\pm0.18$ & $-0.26\pm0.17$ & $-0.29\substack{+0.13 \\[-1pt] -0.30}$ & $-0.88\substack{+0.22 \\[-1pt] -0.12}$ & 7\\
COSMOS & 28523  & $ 1.88\pm 0.26$ & $2.93$ & $0.25$ & $11.95\pm0.01$ & $385\pm 45$ & $-0.32\pm0.12$ & $-0.43\pm0.10$ & $-0.46\substack{+0.09 \\[-1pt] -0.13}$ & $-1.49\substack{+0.12 \\[-1pt] -0.09}$ & 2, 6\\
AEGIS & A21129\tablenotemark{b} & $ 1.96\pm 0.26$ & $7.15$ & $0.49$ & $11.97\pm0.01$ & $275\pm 10$ & $-0.94\pm0.08$ & $-0.64\pm0.03$ & $-0.66\substack{+0.03 \\[-1pt] -0.04}$ & $-1.40\substack{+0.03 \\[-1pt] -0.03}$ & 5\\
GOODS-N & 17678  & $ 1.29\pm 0.26$ & $8.00$ & $0.70$ & $11.61\pm0.01$ & $179\pm 23$ & $-1.25\pm0.16$ & $-0.91\pm0.11$ & $-0.94\substack{+0.10 \\[-1pt] -0.15}$ & $-1.13\substack{+0.13 \\[-1pt] -0.09}$ & 2, 3, 8\\
UDS & 24891  & $ 1.88\pm 0.26$ & $2.45$ & $0.88$ & $11.48\pm0.01$ & $187\pm126$ & $-0.46\pm0.59$ & $-0.43\pm0.58$ & $-0.45\substack{+0.16 \\[-1pt] -1.92}$ & $-1.10\substack{+0.74 \\[-1pt] -0.31}$ & 1, 7\\
UDS & 35616  & $ 4.47\pm 0.48$ & $6.07$ & $0.65$ & $11.82\pm0.02$ & $198\pm 49$ & $-0.61\pm0.23$ & $-0.50\pm0.21$ & $-0.67\substack{+0.12 \\[-1pt] -1.31}$ & $-0.50\substack{+0.25 \\[-1pt] -0.16}$ & 7\\
GOODS-S & 39364  & $ 1.58\pm 0.26$ & $2.97$ & $0.95$ & $11.64\pm0.01$ & $203\pm 42$ & $-0.64\pm0.19$ & $-0.56\pm0.18$ & $-0.61\substack{+0.13 \\[-1pt] -0.34}$ & $-1.07\substack{+0.26 \\[-1pt] -0.13}$ & 1\\
GOODS-S & 42113  & $ 1.96\pm 0.27$ & $6.53$ & $0.78$ & $11.66\pm0.01$ & $362\pm 65$ & $-0.33\pm0.17$ & $-0.16\pm0.16$ & $-0.19\substack{+0.12 \\[-1pt] -0.23}$ & $-1.17\substack{+0.20 \\[-1pt] -0.12}$ & 1\\
GOODS-S & 43548  & $ 0.94\pm 0.25$ & $3.91$ & $0.62$ & $11.30\pm0.01$ & $169\pm 43$ & $-0.74\pm0.25$ & $-0.66\pm0.22$ & $-0.68\substack{+0.15 \\[-1pt] -0.42}$ & $-1.57\substack{+0.37 \\[-1pt] -0.15}$ & 1\\
UDS & 30737  & $ 3.32\pm 0.27$ & $2.71$ & $0.51$ & $11.82\pm0.01$ & $307\pm 82$ & $-0.13\pm0.24$ & $-0.21\pm0.23$ & $-0.30\substack{+0.13 \\[-1pt] -1.21}$ & $-0.77\substack{+0.35 \\[-1pt] -0.16}$ & 7\\
UDS & U53937\tablenotemark{b} & $ 0.63\pm 0.25$ & $3.78$ & $0.78$ & $11.74\pm0.01$ & $251\pm 21$ & $-1.01\pm0.19$ & $-0.86\pm0.07$ & $-0.87\substack{+0.07 \\[-1pt] -0.08}$ & $-1.91\substack{+0.08 \\[-1pt] -0.06}$ & 5\\
UDS & 43367  & $ 2.69\pm 0.28$ & $5.18$ & $0.53$ & $11.60\pm0.01$ & $299\pm 74$ & $-0.18\pm0.22$ & $-0.16\pm0.21$ & $-0.23\substack{+0.14 \\[-1pt] -0.73}$ & $-0.83\substack{+0.32 \\[-1pt] -0.15}$ & 7\\
UDS & 30475  & $ 0.98\pm 0.25$ & $3.04$ & $0.75$ & $11.55\pm0.01$ & $296\pm109$ & $-0.44\pm0.34$ & $-0.39\pm0.32$ & $-0.39\substack{+0.17 \\[-1pt] -0.94}$ & $-1.91\substack{+0.72 \\[-1pt] -0.18}$ & 7\\
COSMOS & 06977  & $ 1.49\pm 0.25$ & $1.43$ & $0.79$ & $11.41\pm0.01$ & $187\pm 32$ & $-0.43\pm0.17$ & $-0.48\pm0.15$ & $-0.51\substack{+0.12 \\[-1pt] -0.23}$ & $-1.23\substack{+0.20 \\[-1pt] -0.12}$ & 1\\
UDS & 32707  & $ 1.75\pm 0.26$ & $3.62$ & $0.25$ & $11.70\pm0.01$ & $174\pm 30$ & $-0.83\pm0.16$ & $-0.92\pm0.15$ & $-0.99\substack{+0.12 \\[-1pt] -0.29}$ & $-0.96\substack{+0.22 \\[-1pt] -0.12}$ & 7\\
COSMOS & 16629  & $ 0.74\pm 0.25$ & $2.40$ & $0.72$ & $11.32\pm0.01$ & $358\pm 76$ & $-0.13\pm0.24$ & $-0.07\pm0.18$ & $-0.08\substack{+0.13 \\[-1pt] -0.28}$ & $-2.29\substack{+0.27 \\[-1pt] -0.13}$ & 7\\
UDS & 37529  & $ 2.32\pm 0.28$ & $3.83$ & $0.64$ & $11.52\pm0.01$ & $232\pm 60$ & $-0.29\pm0.23$ & $-0.28\pm0.23$ & $-0.33\substack{+0.14 \\[-1pt] -0.68}$ & $-0.98\substack{+0.35 \\[-1pt] -0.15}$ & 7\\
UDS & 22802  & $ 1.50\pm 0.25$ & $2.33$ & $0.36$ & $11.66\pm0.01$ & $316\pm 31$ & $-0.27\pm0.11$ & $-0.23\pm0.09$ & $-0.24\substack{+0.07 \\[-1pt] -0.10}$ & $-1.85\substack{+0.10 \\[-1pt] -0.07}$ & 1, 7\\
GOODS-N & 11470  & $ 2.85\pm 0.29$ & $4.00$ & $0.71$ & $11.60\pm0.01$ & $221\pm 36$ & $-0.33\pm0.15$ & $-0.31\pm0.14$ & $-0.35\substack{+0.12 \\[-1pt] -0.21}$ & $-1.15\substack{+0.19 \\[-1pt] -0.11}$ & 8\\
GOODS-N & 24033  & $ 1.09\pm 0.25$ & $3.20$ & $0.72$ & $11.43\pm0.01$ & $155\pm 31$ & $-0.84\pm0.20$ & $-0.85\pm0.17$ & $-0.88\substack{+0.13 \\[-1pt] -0.28}$ & $-1.25\substack{+0.24 \\[-1pt] -0.13}$ & 8\\
GOODS-N & 03604  & $ 0.82\pm 0.25$ & $2.55$ & $0.25$ & $11.31\pm0.01$ & $317\pm118$ & $-0.19\pm0.35$ & $-0.25\pm0.32$ & $-0.25\substack{+0.17 \\[-1pt] -0.92}$ & $-2.47\substack{+0.82 \\[-1pt] -0.17}$ & 8\\
UDS & 29352  & $ 1.08\pm 0.25$ & $4.79$ & $0.77$ & $11.52\pm0.01$ & $187\pm 70$ & $-0.87\pm0.34$ & $-0.73\pm0.33$ & $-0.75\substack{+0.16 \\[-1pt] -1.05}$ & $-1.41\substack{+0.65 \\[-1pt] -0.18}$ & 1, 7\\
COSMOS & 19958  & $ 2.62\pm 0.29$ & $2.94$ & $0.84$ & $11.45\pm0.01$ & $169\pm 87$ & $-0.39\pm0.45$ & $-0.39\pm0.45$ & $-0.46\substack{+0.13 \\[-1pt] -1.86}$ & $-0.91\substack{+0.59 \\[-1pt] -0.26}$ & 7\\
COSMOS & 17255  & $ 1.54\pm 0.26$ & $2.93$ & $0.60$ & $11.46\pm0.01$ & $147\pm 40$ & $-0.75\pm0.25$ & $-0.76\pm0.24$ & $-0.81\substack{+0.15 \\[-1pt] -0.69}$ & $-1.05\substack{+0.40 \\[-1pt] -0.16}$ & 7\\
AEGIS & 25526  & $ 0.83\pm 0.25$ & $2.48$ & $0.51$ & $11.37\pm0.01$ & $134\pm 36$ & $-0.98\pm0.27$ & $-0.99\pm0.23$ & $-1.02\substack{+0.15 \\[-1pt] -0.52}$ & $-1.39\substack{+0.41 \\[-1pt] -0.15}$ & 7\\
UDS & 10237  & $ 3.07\pm 0.29$ & $4.07$ & $0.67$ & $11.88\pm0.01$ & $233\pm 23$ & $-0.54\pm0.10$ & $-0.58\pm0.09$ & $-0.79\substack{+0.10 \\[-1pt] -0.20}$ & $-0.39\substack{+0.09 \\[-1pt] -0.07}$ & 1\\
COSMOS & 07411  & $ 2.04\pm 0.27$ & $3.95$ & $0.85$ & $11.64\pm0.01$ & $186\pm 28$ & $-0.66\pm0.14$ & $-0.69\pm0.13$ & $-0.77\substack{+0.11 \\[-1pt] -0.23}$ & $-0.73\substack{+0.16 \\[-1pt] -0.10}$ & 1\\
COSMOS & C07447\tablenotemark{b} & $ 1.67\pm 0.25$ & $5.58$ & $0.68$ & $12.19\pm0.01$ & $287\pm 53$ & $-1.03\pm0.18$ & $-0.87\pm0.16$ & $-0.92\substack{+0.13 \\[-1pt] -0.29}$ & $-0.95\substack{+0.22 \\[-1pt] -0.12}$ & 6\\
UDS & 35111  & $ 0.78\pm 0.25$ & $2.56$ & $0.31$ & $11.56\pm0.01$ & $228\pm 36$ & $-0.74\pm0.19$ & $-0.75\pm0.14$ & $-0.76\substack{+0.11 \\[-1pt] -0.18}$ & $-2.02\substack{+0.18 \\[-1pt] -0.11}$ & 1\\
UDS & 32892  & $ 1.63\pm 0.26$ & $4.05$ & $0.87$ & $11.73\pm0.01$ & $206\pm 27$ & $-0.76\pm0.14$ & $-0.68\pm0.11$ & $-0.72\substack{+0.10 \\[-1pt] -0.16}$ & $-1.12\substack{+0.14 \\[-1pt] -0.09}$ & 1\\
UDS & 38073  & $ 2.78\pm 0.29$ & $7.06$ & $0.84$ & $11.68\pm0.02$ & $194\pm 49$ & $-0.78\pm0.23$ & $-0.54\pm0.22$ & $-0.59\substack{+0.14 \\[-1pt] -0.62}$ & $-0.84\substack{+0.31 \\[-1pt] -0.15}$ & 1\\
COSMOS & 06396  & $ 1.44\pm 0.25$ & $1.38$ & $0.86$ & $11.47\pm0.01$ & $169\pm 33$ & $-0.58\pm0.19$ & $-0.61\pm0.17$ & $-0.65\substack{+0.13 \\[-1pt] -0.31}$ & $-1.09\substack{+0.25 \\[-1pt] -0.13}$ & 1\\
COSMOS & 09227  & $ 1.16\pm 0.25$ & $2.75$ & $0.68$ & $11.65\pm0.01$ & $273\pm 41$ & $-0.51\pm0.16$ & $-0.55\pm0.13$ & $-0.56\substack{+0.11 \\[-1pt] -0.17}$ & $-1.60\substack{+0.17 \\[-1pt] -0.10}$ & 1\\
COSMOS & 07391  & $ 0.69\pm 0.25$ & $8.00$ & $0.80$ & $11.40\pm0.01$ & $145\pm 38$ & $-1.48\pm0.29$ & $-0.94\pm0.23$ & $-0.96\substack{+0.15 \\[-1pt] -0.45}$ & $-1.56\substack{+0.39 \\[-1pt] -0.15}$ & 1\\
COSMOS & 02816  & $ 1.93\pm 0.25$ & $1.82$ & $0.24$ & $11.74\pm0.01$ & $297\pm 49$ & $-0.25\pm0.15$ & $-0.30\pm0.14$ & $-0.33\substack{+0.12 \\[-1pt] -0.22}$ & $-1.40\substack{+0.20 \\[-1pt] -0.11}$ & 1\\
UDS & U19627\tablenotemark{b} & $ 2.08\pm 0.25$ & $3.48$ & $0.48$ & $12.06\pm0.01$ & $315\pm 37$ & $-0.57\pm0.12$ & $-0.65\pm0.10$ & $-0.69\substack{+0.09 \\[-1pt] -0.14}$ & $-1.15\substack{+0.12 \\[-1pt] -0.09}$ & 6, 9\\
COSMOS & 13083  & $ 1.38\pm 0.24$ & $3.43$ & $0.88$ & $11.75\pm0.01$ & $197\pm 52$ & $-0.84\pm0.24$ & $-0.73\pm0.23$ & $-0.78\substack{+0.14 \\[-1pt] -0.73}$ & $-1.02\substack{+0.40 \\[-1pt] -0.16}$ & 7\\
COSMOS & 11494  & $ 2.70\pm 0.28$ & $4.68$ & $0.80$ & $12.19\pm0.01$ & $348\pm 66$ & $-0.58\pm0.17$ & $-0.53\pm0.17$ & $-0.66\substack{+0.13 \\[-1pt] -0.53}$ & $-0.62\substack{+0.22 \\[-1pt] -0.13}$ & 6, 7, 10\\
COSMOS & 12020  & $ 1.98\pm 0.26$ & $3.87$ & $0.59$ & $11.70\pm0.01$ & $312\pm 65$ & $-0.26\pm0.19$ & $-0.38\pm0.18$ & $-0.46\substack{+0.13 \\[-1pt] -0.46}$ & $-0.70\substack{+0.23 \\[-1pt] -0.13}$ & 10\\
\enddata

\tablecomments{References: (1) this work, (2) \citet{belli2014a}, (3) \citet{newman2010}, (4) \citet{cappellari2009}, (5) \citet{bezanson2013}, (6) \citet{van-de-sande2013}, (7) \citet{belli2017}, (8) \citet{barro2016}, (9) \citet{toft2012}, (10) \citet{belli2014}}
\tablenotetext{a}{Unless otherwise noted, IDs correspond to those provided by \citet{skelton2014} for galaxies in the 3D-HST fields.}
\tablenotetext{b}{These galaxies fall outside the 3D-HST footprint; IDs therefore correspond to those given in their originating publications.}
\tablenotetext{\dagger}{Sizes have been corrected to the rest frame $r$ band according to Equation \ref{eqn.size_correction}.}

\end{deluxetable*}
\end{longrotatetable}

\section{Results}
\label{section.results}

In this Section we present the main results of this work, which are focused in two areas: the relationship between dynamical and stellar masses, and the interplay between dark matter content and the IMF at high-redshift.  Derived quantities for our high-redshift sample are provided in Table \ref{table.derived}, and they are described in more detail in Sections \ref{section.sample} and \ref{section.modelling}.

\subsection{The relationship between dynamical and stellar mass}
\label{results.mdyn_mstar}

\begin{figure}
\centering
\includegraphics[scale=1.0]{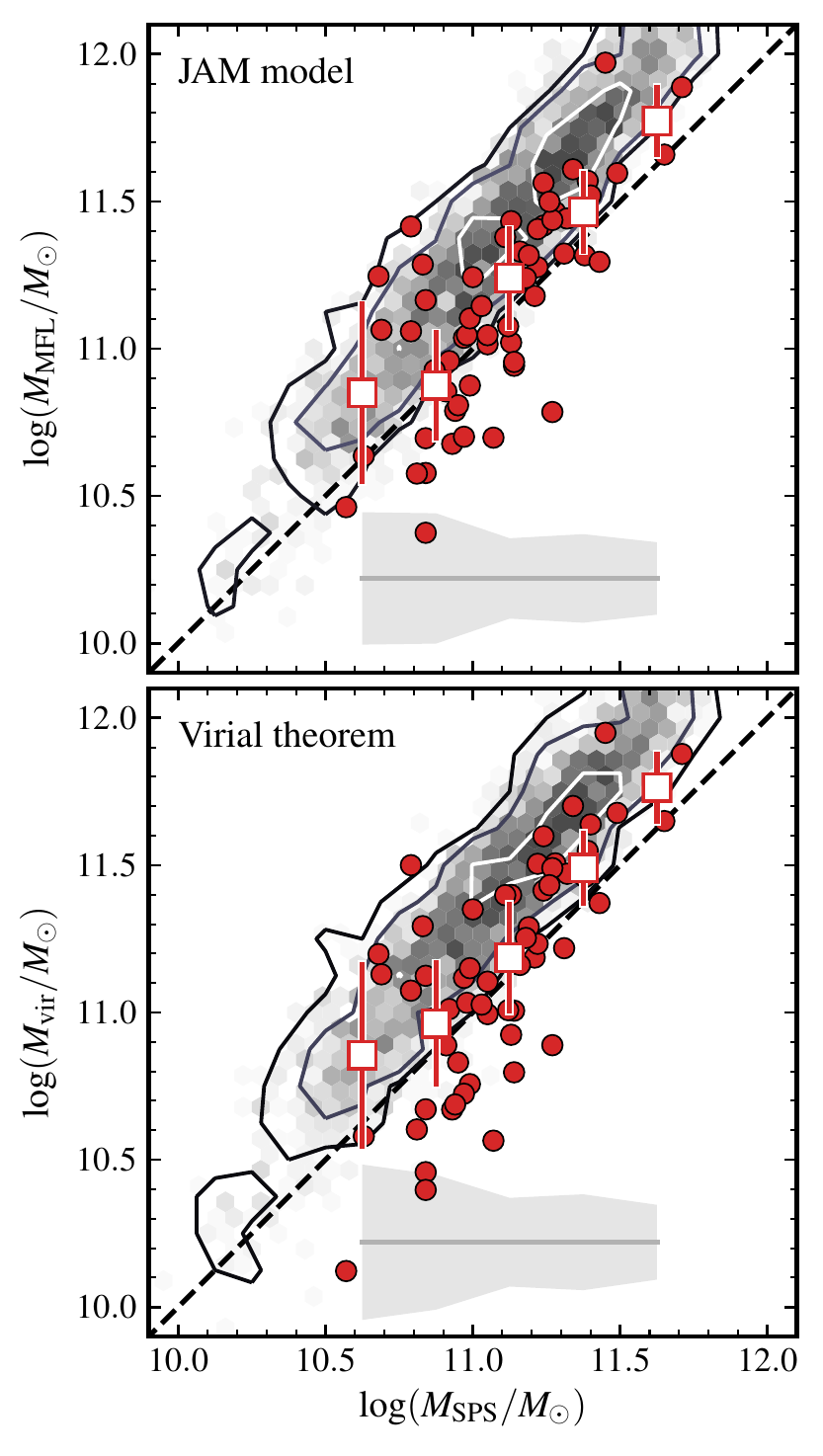}
\caption{Dynamical vs. stellar masses for two different dynamical mass estimates based on either JAM models (top panel; see Section \ref{section.jam}) or a simple \emph{n}-dependent virial coefficient (bottom panel; Equations \ref{eqn.virial_mass} and \ref{eqn.virial_coeff}).  Quiescent galaxies at $1.4 < z < 2.1$ are shown as red circles, with large red squares indicating their (binned) median and scatter.  The grey shading in the bottom of each panel shows the average uncertainty of individual dynamical mass estimates. The distribution of galaxies in our low-redshift SDSS/GAMA sample is indicated by the background shading.  Contours show the 30th, 60th, and 90th percentile distribution of dynamical and stellar masses for local galaxies with ages $>9$ Gyr, i.e. old enough to be the descendants of galaxies in our high-redshift sample.}
\label{fig.mstar_mdyn}
\end{figure}

\begin{figure*}
\centering
\includegraphics[scale=1.0]{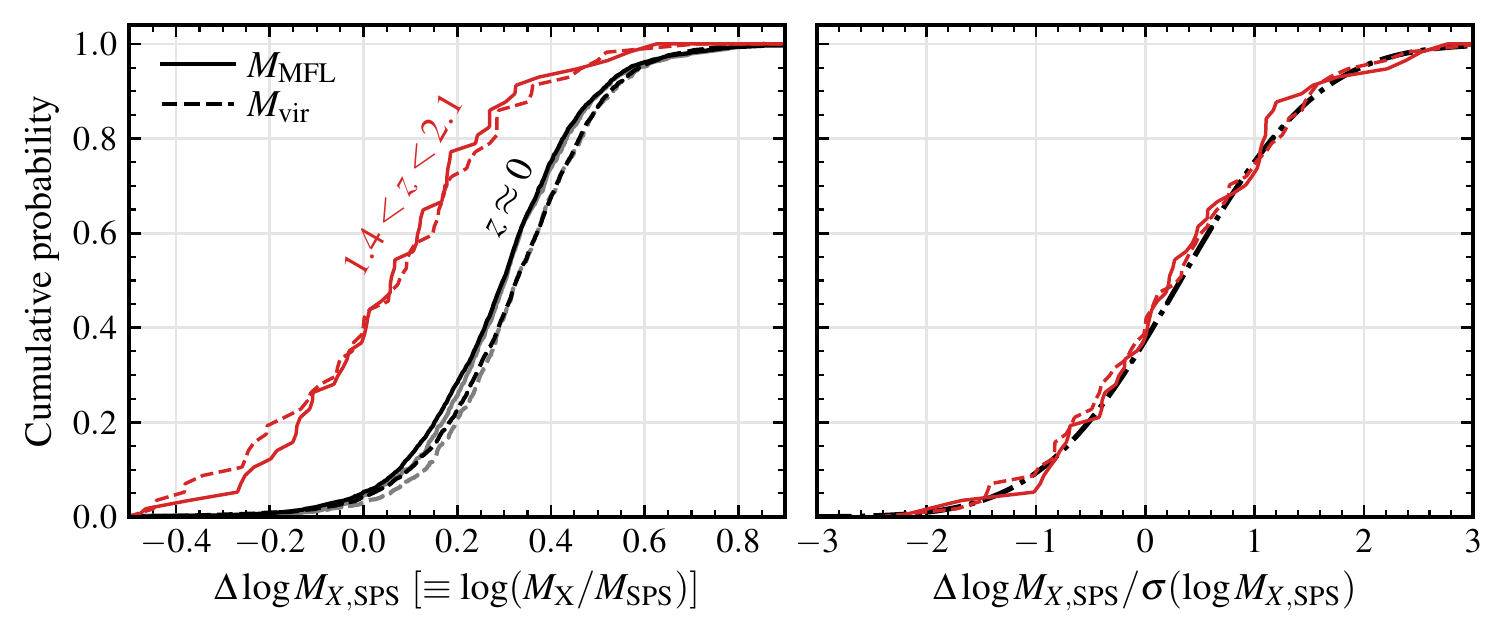}
\caption{Cumulative distribution of the dynamical-to-stellar mass ratio, comparing different redshifts and dynamical mass estimators.  In the Left panel, red curves show high-redshift data, while the $z=0$ GAMA/SDSS data are shown in black.  The light gray curves show the results for galaxies in the GAMA/SDSS sample with mass-weighted stellar ages $>9$ Gyr (i.e. accounting for possible progenitor bias effects).  Solid and dashed curves indicate dynamical masses derived using JAM models assuming mass-follow-light (MFL) and a simple virial estimator, respectively.  The median dynamical-to-stellar mass ratio is $\sim$0.2 dex lower at high redshift compared to $z=0$ regardless of the adopted mass estimator.  In the right panel we show the error normalized distribution of dynamical-to-stellar mass ratios.  Red curves again show high-redshift data, while the dot-dashed curve shows the expectation based on a standard normal distribution with the same mean as the observed data.  A number of galaxies have stellar masses apparently larger than their associated dynamical mass estimates, however the overall population is consistent (within uncertainties) with a positive dynamical-to-stellar mass ratio on average.} 
\label{fig.mstar_mdyn_cumul}
\end{figure*}

In Figure \ref{fig.mstar_mdyn} we show a comparison of dynamical and stellar masses for the different dynamical mass estimates described in Section \ref{section.masses}.  Two features are apparent.  First, fixed stellar mass high-redshift galaxies appear to have dynamical masses which are $\sim$0.20 dex lower on average than their low-redshift counterparts.  This offset appears regardless of the dynamical mass estimate used (i.e. $M_\mathrm{MFL}$ vs. $M_\mathrm{vir}$).  Second, the correlation between dynamical and stellar mass is super-linear regardless of redshift, in the sense that the ratio of dynamical-to-stellar mass increases with increasing stellar mass.  Such a ``tilt'' in the relationship between dynamical and stellar mass has been studied extensively at low-redshift, and has commonly been interpreted as variation of the central dark matter fraction and/or stellar IMF \citep[e.g.][]{renzini1993,dutton2013,cappellari2013}; in the following sections we will consider evidence for changes in the dark matter fraction and stellar IMF among high-redshift galaxies in more detail.

Finally, there are a number of galaxies in Figure \ref{fig.mstar_mdyn} with stellar mass estimates formally larger than their derived dynamical masses.  While this cannot physically be the case, there a number of factors that influence the apparent trend, particularly at low stellar masses. Observational uncertainties at $\log(M_\mathrm{SPS} / M_\odot) < 11$ increase dramatically, driven primarily by the increased uncertainty on galaxy size as one pushes down the size-mass relation.  While these increased uncertainties cannot in and of themselves explain the apparent shift towards low dynamical masses, when combined with the tilt of the relation described above they can nevertheless increase the fraction of galaxies with low dynamical-to-stellar mass ratios. In addition, we will show in Section \ref{sect.rotation} that our dynamical modelling likely underestimates the dynamical mass for galaxies that are intrinsically flat. Enforcing a flat structure for face-on galaxies can increase dynamical mass estimates by as much as $\sim$0.2 dex.  Indeed, nearly 59\% of galaxies with \qobs{} $\geq 0.7$ have dynamical masses smaller than their derived stellar mass, compared to only 24\% for galaxies with \qobs{} $<0.7$.

\subsubsection{Central dark matter fractions}
\label{section.fdm}

The tendency for high-redshift quiescent galaxies to have lower dynamical-to-stellar mass ratios compared to low redshift has been reported in a number of previous studies \citep[e.g.][]{toft2012,van-de-sande2013,belli2017}, and is generally interpreted as reflecting a systematic decrease in the central dark matter fraction, \fdm{}. This decline in dynamical-to-stellar mass ratio appears to occur relatively smoothly with increasing redshift, as shown by a number of studies based on large spectroscopic surveys at $z < 1$ \citep[e.g.][]{beifiori2014,tortora2014,tortora2018}.  Figure \ref{fig.mstar_mdyn_cumul} shows the cumulative distribution of dynamical-to-stellar mass ratio in the samples considered here.  We find an offset in the mean dynamical-to-stellar mass ratio of $-0.20$ dex when moving to high redshift---$\log\,(M_\mathrm{MFL}/M_\ast) = 0.29\pm0.01$ at $z=0$ compared to $0.09\pm0.05$ at $1.4 \leq z \leq 2.1$---which is consistent with the results of previous studies.  The magnitude of this offset is independent of the dynamical mass estimator used (e.g. $M_\mathrm{MFL}$ vs. $M_\mathrm{vir}$), and does not change when considering only the oldest galaxies at $z = 0$ (shown as contours in Figure \ref{fig.mstar_mdyn} and light gray lines in Figure \ref{fig.mstar_mdyn_cumul}). The right panel of Figure \ref{fig.mstar_mdyn_cumul} shows the same distribution of dynamical-to-stellar mass ratios as the left, but with individual measurements normalized by their uncertainties. These can be compared to the dot-dashed (black) line, which shows the prediction for a standard normal distribution.  Although nearly 40\% of galaxies in the high-redshift sample have photometrically-derived stellar masses that exceed their dynamical masses, given measurement uncertainties the overall distribution is consistent with a positive (albeit small) dynamical-to-stellar mass ratio on average.

\begin{figure}
\centering
\includegraphics[scale=1.0]{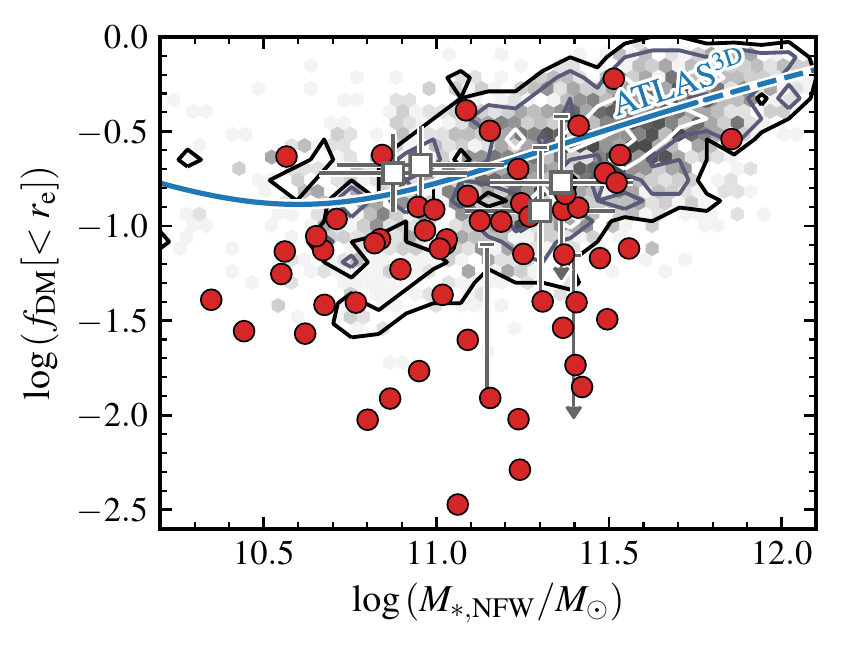}
\caption{Estimated dark matter fraction within the half-light radius, \fdm{}, as a function of the dynamical mass in the stellar component $M_\mathrm{\ast,NFW}$. Circles (red) show the measured dark matter fractions for our sample of quiescent galaxies at $1.4 < z < 2.1$.  The blue curve shows the results from \citet{cappellari2013a} derived using IFU data from the local ATLAS$^{\mathrm{3D}}$ sample, while background shading shows the distribution of \fdm{} for galaxies in our GAMA/SDSS at $z=0$. Squares and upper limits represent dark matter fractions measured by \citet{genzel2017} for a sample of massive disk galaxies at $z > 1$, which are consitent with our quiescent galaxy data at high-redshift.  Overall, the dark matter fractions in high-redshift quiescent galaxies appear lower by a factor of $>$2 on average compared to galaxies of the same stellar mass at $z = 0$.}
\label{fig.mstar_fdm}
\end{figure}

We can examine the evolution of \fdm{} more directly using our dynamical models that include an explicit dark matter component, where dark matter fractions are computed following Eqn. \ref{eqn.fdm_nfw}.  In Figure \ref{fig.mstar_fdm} we show \fdm{} as a function of $M_\mathrm{\ast,NFW}$, the dynamical mass of the stellar component.  While there is significant uncertainty in the individual measurements of \fdm{} at high redshift, the overall trends support our interpretation of Figures \ref{fig.mstar_mdyn} and \ref{fig.mstar_mdyn_cumul} in terms of an evolution in the central dark matter fraction: galaxies at $1.4 < z < 2.1$ have a mean \fdm{} = $6.6\pm1.0$\%, a factor of $>$2 lower than galaxies of a similar mass in our SDSS/GAMA sample at $z=0$ (\fdm{} $\approx 16.3\pm0.3$\%; c.f. 17\% from \citealp{cappellari2013a}). Furthermore, our low redshift \fdm{} measurements are consistent with the values derived by \citet{thomas2011b} and \citet{cappellari2013a} based on more detailed dynamical modelling of low-redshift galaxies, suggesting that the observed offset in \fdm{} between different redshifts is unlikely to be due to differences in the modelling approach.  The above comparison between high- and low-redshift galaxies at fixed mass must nevertheless be made with some caution, as individual galaxies are expected to evolve from $z=2$ to 0; we will revisit the evolution of \fdm{} using more carefully matched progenitor and descendant samples in Section \ref{section.evolution}.

Using data from the SINS survey, \citet{forster-schreiber2009} found that star-forming galaxies at $z\sim2$ are strongly baryon dominated, even for a \citet{chabrier2003} IMF, suggesting little room for either a bottom-heavy Salpeter IMF or significant dark matter.  These results have recently been supported by kinematic data for hundreds of early star-forming disks in the KMOS$^{\mathrm{3D}}$ \citep{wisnioski2015,wisnioski2019} and MOSDEF \citep{kriek2015} surveys \citep[e.g.][]{price2016,wuyts2016,lang2017,ubler2017}, as well as the detailed analysis of outer rotation curves for individual high-redshift disks \citep[see also \citealp{genzel2020}]{genzel2017}.  For comparison, in Figure \ref{fig.mstar_fdm} we show the dark matter fractions derived by \citet[shown as squares and upper limits]{genzel2017}, which are in good agreement with the \fdm{} measurements derived here.

\subsubsection{The effects of unresolved rotation}
\label{sect.rotation}

\begin{figure}
\centering
\includegraphics[scale=1.0]{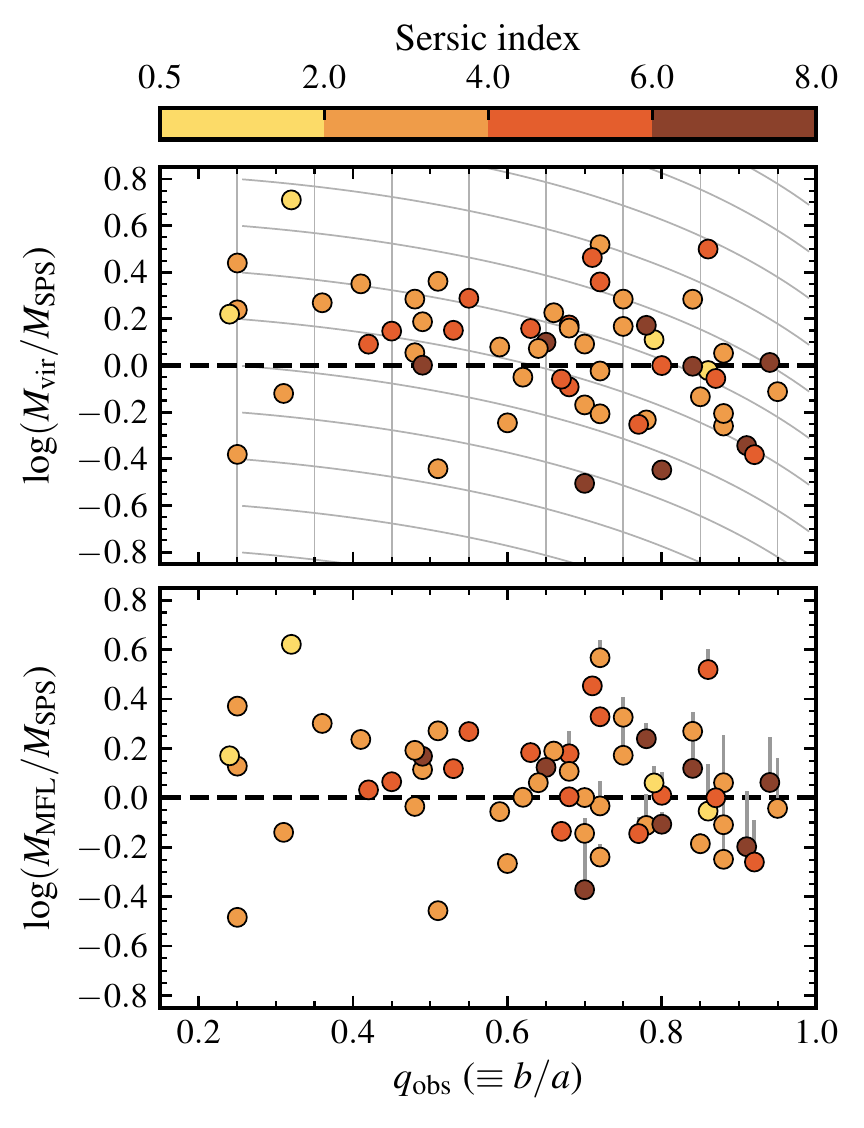}
\caption{Dynamical-to-stellar mass ratio as a function of \qobs{}, the observed axis ratio. The top and bottom panels show results for two different estimates of the dynamical mass based on the virial theorem (i.e. Equations \ref{eqn.virial_mass} and \ref{eqn.virial_coeff}; top panel) or JAM models (bottom panel).  Individual points are color-coded according to their S\'ersic indices as indicated by the color bar at the top of the Figure.  In the top panel, thin grey lines show the expected dependence of the dynamical-to-stellar mass ratio on \qobs{} for an anisotropic model with \qint{}$=0.25$ and $\beta = 0.7(1-q_\mathrm{int})$ following Equation \ref{eqn.mass_ratio}.  In the bottom panel, grey vertical lines show show the effect of adopting a Gaussian prior on the intrinsic axis ratio such that \qint{} $= N(0.25,0.05)$.  In both cases the apparent anti-correlation between dynamical-to-stellar mass ratio and \qobs{} is consistent with some portion of the population harboring significant rotational support.}
\label{fig.alpha_ba}
\end{figure}

One of our goals in comparing multiple dynamical mass estimators is to assess the impact and importance of different modelling assumptions on the inferred properties of high-redshift galaxies.  In that regard, the main distinctions between $M_\mathrm{vir}$ and $M_\mathrm{MFL}$ are the assumptions of spherical symmetry and isotropy inherited through the application of Equations \ref{eqn.virial_mass} and \ref{eqn.virial_coeff}.  

In practice, the dynamical masses derived here are only weakly dependent on changes of the anisotropy, $\beta$, at fixed \qint{} for a range of values consistent with local early-type galaxies ($0 \leq \beta \leq 0.6$; \citealp{cappellari2007}).  It is therefore unlikely that the assumption of isotropy has a significant impact on the results presented in Figures \ref{fig.mstar_mdyn}, \ref{fig.mstar_mdyn_cumul}, and \ref{fig.mstar_fdm}, particularly given the typical uncertainties on measurements of $\sigma_\mathrm{e}$ (20--30\%). On the other hand, changes in assumed galaxy structure---for example, from spherically symmetric to oblate and axisymmetric---can systematically bias dynamical mass estimates depending on the degree of intrinsic flattening and relative importance of rotation versus pressure support. 

Crucially, there is growing observational evidence that quiescent galaxies at high redshift may indeed be rotationally flattened, violating the assumption of spherical symmetry inherent in Equation \ref{eqn.virial_coeff}. \citet{bezanson2018} showed that passive galaxies at $z\sim1$ have on average a higher proportion of rotational support (higher $V/\sigma$) than galaxies of the same mass at low redshift \citep[see also][]{van-der-wel2008b}.  These results are consistent with the observed evolution of photometrically-derived axis ratios over the same redshift range, which favour a significant portion of the quiescent galaxy population having $0.2 \leq q_\mathrm{int} \leq 0.3$ \citep{van-der-wel2011,chang2013,hill2019}.  \citet{belli2017} argued that the dynamical masses of quiescent galaxies at $z > 1.5$ are statistically consistent with a factor of $\sim$2 increase in $V/\sigma$ compared to $z = 0$ based on their correlation with observed axis ratios.  More directly, a handful of strongly-lensed passive galaxies at $z > 2$ have resolved kinematic profiles that are consistent with being rotationally-flattened disks \citep[e.g.][]{newman2015,toft2017,newman2018}.  

In the case of integrated (as opposed to resolved) absorption line kinematics, rotation is expected to manifest as a dependence of the measured velocity dispersion---and, by extension, dynamical mass---on galaxy inclination. For an oblate model observed at inclination $i$ with no azimuthal variation of the velocity ellipsoid (i.e. $\sigma_\phi/\sigma_r = 1$, with $\sigma_\phi$ and $\sigma_r$ describing velocity dispersion in the azimuthal and radial directions, respectively), the second moment of the velocity distribution $\sigma_\mathrm{obs}$ can be written as

\begin{multline}
\sigma_\mathrm{obs}^2 = \\
 \sigma^2 {(1-\beta \cos^2 i)} \left [ 1+\left (\frac{V}{\sigma}\right )_e^2 \frac{\sin^2 i}{(1-\beta \cos^2 i)^2} \right],
\label{eqn.second_moment}
\end{multline}

\noindent where $V$ and $\sigma$ are the flux-weighted mean circular velocity and velocity dispersion within the effective radius for the edge-on case ($i = 90^{\circ}$) with $(V/\sigma)^2_e \equiv V^2_\mathrm{e} / \sigma^2_\mathrm{e}$, and $\beta$ is the anisotropy parameter as defined in Section \ref{section.jam}.  Following \citet{belli2017}, substituting Equation \ref{eqn.second_moment} into Equation \ref{eqn.virial_mass} and normalising by the dynamical mass predicted for the face-on case (i.e., $i = 0^{\circ}$) gives

\begin{multline}
\frac{M_\mathrm{vir}(i)}{M_\mathrm{vir}(i=0^{\circ})} = \\ 
\frac{1-\beta \cos^2 i}{1-\beta} \left [ 1+\left (\frac{V}{\sigma}\right )_e^2 \frac{\sin^2 i}{(1-\beta \cos^2 i)^2} \right],
\label{eqn.mass_ratio}
\end{multline}

\noindent with the relationship between \qint{}, \qobs{}, and $i$ given by Equation \ref{eqn.inclination}.  In the isotropic case where $\beta = 0$, Equation \ref{eqn.mass_ratio} reduces to equation 5 of \citet{belli2017} modulo a factor $\gamma=1.51$\footnote{\citealp{belli2017} adopt the value of $\gamma$ determined by \citet{cappellari2013a} which relates the measured second moment \sigmaeff{} to the circular velocity \emph{at} $r_\mathrm{e}^\mathrm{sma}$.  Here we instead define $V/\sigma$ in terms of the flux-weighted mean \emph{within} $r_e$, so that all of $\sigma_\mathrm{obs}$, $V$, $\sigma$, and $(V/\sigma)_e$ are defined over the same aperture.}.

In Figure \ref{fig.alpha_ba} we show the dynamical-to-stellar mass ratio as a function of \qobs{} for both $M_\mathrm{vir}$ and $M_\mathrm{MFL}$ estimates.  In the virial theorem case we find evidence for a weak negative correlation between $M_\mathrm{vir}$ and \qobs{}, with a Spearman Rank coefficient $\rho = -0.23\pm0.10$ ($p=0.007$), while for the MFL models the correlation is not significant ($\rho = -0.14\pm0.10$; $p=0.068$). Individual galaxies are color coded according to their S\'{e}rsic indices as derived from the profile fits described in Section \ref{section.sersic_fits}.  In contrast to \citet{belli2017} we find no significant dependence of the dynamical-to-stellar mass ratio on S\'{e}rsic index in either case with $\rho = -0.20\pm0.10$ ($p=0.04$) and $-0.08\pm0.10$ ($p=0.44$), which appears to preclude a simple exclusion of disk-dominated systems based on their structure, and motivates a more detailed examination of the correlation between \qobs{} and dynamical-to-stellar mass ratio.

Lines in the top panel of Figure \ref{fig.alpha_ba} show predicted behavior of the dynamical-to-stellar mass ratio for a galaxy with $q_\mathrm{int} = 0.25$ observed at different inclinations as given by Equation \ref{eqn.mass_ratio}.  We set $\beta = 0.7(1-q_\mathrm{int})$  based on the results of \citet{cappellari2007} and \citet{emsellem2011} for nearby fast rotating early type galaxies.  In the case of an oblate system, $V/\sigma$ and anisotropy are related by \citep{binney1987,binney2005}

\begin{displaymath}
\beta = 1 - \frac{1 + \left(V/\sigma\right)^2}{1 - \alpha (V/\sigma)^2} \left( \frac{W_{zz}}{W_{xx}}\right),
\end{displaymath}

\noindent where $\alpha$ is a dimensionless number that quantifies the contribution of streaming motions to the line-of-sight velocity dispersion and $(W_{zz}/W_{xx})$ is a shape parameter related to the intrinsic axis ratio $q$ as

\begin{displaymath}
\left( \frac{W_{zz}}{W_{xx}}\right) = \frac{2 \left(q\sqrt{1-q^2} - q^2\arccos q\right)}{\arccos q - q\sqrt{1-q^2}}.
\end{displaymath}

\noindent We adopt a value of $\alpha=0.15$, which provides a good description for nearby galaxies \citep{cappellari2007}. Models are offset in $M_\mathrm{vir}/M_\mathrm{SPS}$ to reflect a range of dynamical-to-stellar mass ratios. The predicted trends qualitatively reproduce the observed correlation between mass ratio and \qobs{}, supporting previous statistical evidence of rotational support among a fraction of high-redshift quiescent galaxies \citep[e.g.][]{belli2017}.

In the bottom panel of Figure \ref{fig.alpha_ba} the correlation between \qobs{} and dynamical-to-stellar mass ratio for MFL models is notably weaker than in the virial theorem case, both visually and as measured by Spearman $\rho$, though galaxies with higher \qobs{} still tend towards lower dynamical-to-stellar mass ratios.  Unlike the virial theorem case, we can explicitly test the impact of intrinsic structure on our MFL mass estimates through application of a prior on \qint{} in our modeling\footnote{Functionally speaking we enforce different intrinsic axis ratios by deprojecting our MGE models at inclination $i$ given \qobs{} and \qint{} by inverting Equation \ref{eqn.inclination}.}.  The vertical lines in the bottom panel of Figure \ref{fig.alpha_ba} show the effect of assuming that galaxies are intrinsically flat, with \qint{} $=0.25$, as opposed to the default case where we adopt a uniform prior on \qint{}.  For galaxies with low \qobs{}, the effect of assuming a different intrinsic structure is minimal, but for galaxies with \qobs{}$>0.6$ ($i < 55^{\circ}$) the estimated dynamical mass can increase by as much as $\sim$65\% (0.22 dex), with a median increase of $\sim$15\% (0.06 dex).  The resulting correlation between dynamical-to-stellar mass ratio and \qobs{} is also flatter, with $\rho = 0.03\pm0.06$ ($p=0.95$).  Furthermore, assuming an intrinsically flat structure for these objects reduces the number of galaxies with dynamical masses significantly lower than their photometrically derived stellar mass $M_\mathrm{SPS}$.

In summary, the data considered here support the conclusions of previous studies suggesting that rotational support is prevalent among quiescent galaxies at high redshift \citep[e.g.][]{chang2013,newman2015,belli2017,toft2017,newman2018,hill2019}. While we expect rotational flattening to have a minimal impact on dynamical mass estimates for galaxies with \qobs{}$<0.6$, galaxies with high \qobs{} can have their dynamical masses underestimated by 0.2 dex or more depending on their intrinsic structure (i.e., if they are intrinsically spherical versus flattened systems viewed face-on). As mentioned in Section \ref{results.mdyn_mstar}, such a discrepancy between intrinsic and assumed structure can at least partially explain those galaxies in our sample that have dynamical masses formally less than their photometrically-derived stellar mass, though my not be the only factor affecting this comparison.  Finally, we note that enforcing an intrinsically flat structure for all galaxies in our sample (e.g. \qint{} $=0.25$) shifts the results presented in Section \ref{section.fdm} towards \emph{lower} central dark matter fractions, and cannot explain the apparent evolution of \fdm{} without also appealing to significant changes in the stellar initial mass function (see Section \ref{section.imf}).

\subsection{The normalization of the stellar IMF at $1.4 \leq z \leq 2.1$}
\label{section.imf}

\begin{figure}
\centering
\includegraphics[scale=1.0]{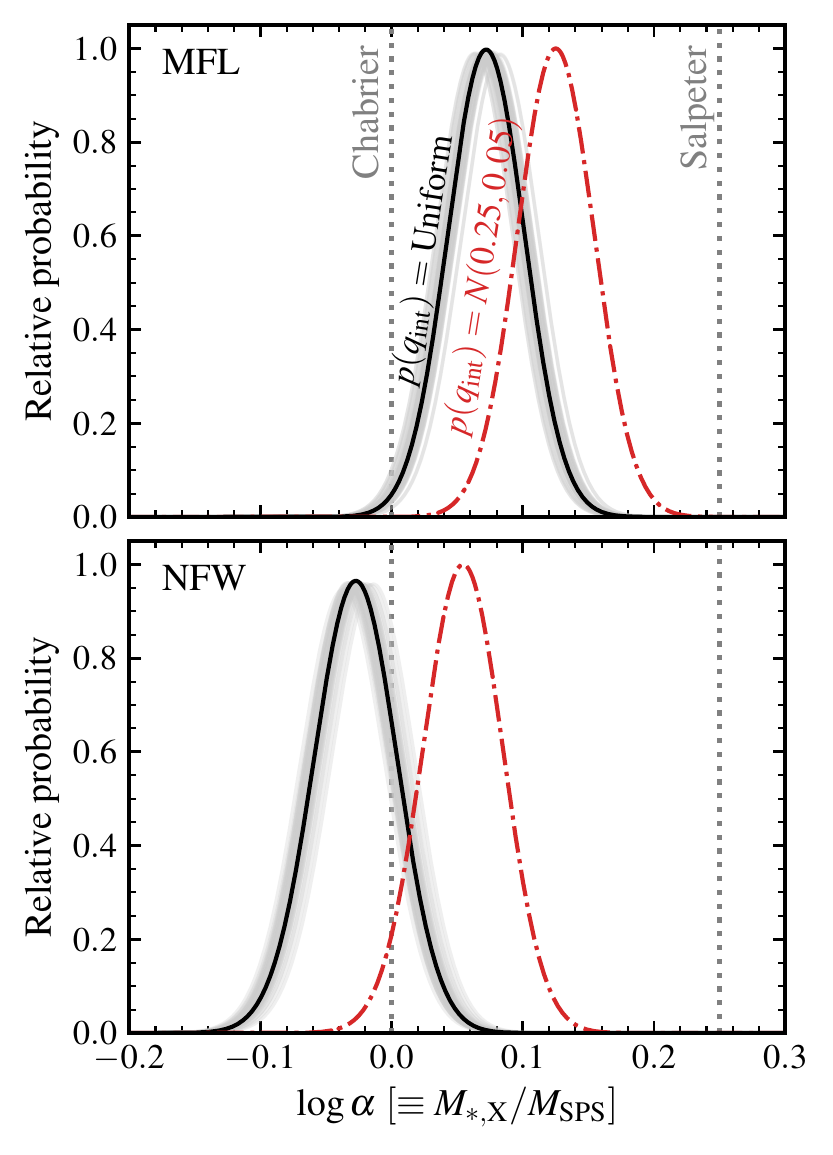}
\caption{Constraints on the IMF normalization parameter $\alpha$ derived by combining the posterior PDFs of individual galaxies in our $1.4 < z < 2.1$ sample. Top panel shows the case where mass follows the observed light profile (MFL), while the bottom panel shows results when explicitly including an NFW-like dark matter halo.  Solid curves are derived assuming a uniform prior on galaxies intrinsic axis ratio \qint{} ($\equiv b/a$), with light curves indicating variations derived from a jackknife analysis.  Dot-dashed curves show the effect of assuming that high-redshift galaxies are intrinsically flat with an axis ratio of \qint{} $= 0.25\pm0.05$, consistent with the values derived by \citet{chang2013}.  Vertical dotted lines indicate the expected values of $\alpha$ for different IMFs as indicated.}
\label{fig.dm_ob}
\end{figure}

In the case that we include an explicit dark matter component in our dynamical models, then we obtain an independent estimate of the \emph{stellar} dynamical mass, $M_\mathrm{\ast, NFW}$, that can be used to diagnose changes in the normalization of the stellar IMF.  A similar approach has been used to highlight possible IMF variation in low- and intermediate-redshift galaxies through the IMF offset parameter $\alpha\equiv M_\mathrm{\ast,dyn}/M_\mathrm{\ast,IMF}$, where $M_\mathrm{\ast,IMF}$ is the stellar mass computed for some default IMF \citep[e.g.][]{treu2010,thomas2011b,cappellari2013a,conroy2013,dutton2013,spiniello2014,smith2015}. In our case $\alpha$ is measured with respect to the \citet{chabrier2003} IMF used in our SPS models (i.e. $M_\mathrm{\ast,IMF} \equiv M_\mathrm{SPS}$).  While in principle $\alpha$ does not rely on any assumptions about \emph{how} the IMF varies, significant deviations from a Salpeter-like IMF above 1--2 $M_\odot$ are difficult to reconcile with observations of color and luminosity evolution for elliptical galaxies \citep[e.g.][]{tinsley1978,van-dokkum2008a}.  We therefore assume that any variation in the IMF occurs at stellar masses which contribute very little to the overall luminosity of the population, i.e. well below the MS turnoff, which is $<2M_\odot$ for stellar populations older than $\sim$1 Gyr \citep[the typical age for galaxies in our high-redshift sample; see, e.g.,][]{mendel2015}.

In Figure \ref{fig.dm_ob} we consider two limiting cases for the derivation of $\alpha$: one where total mass follows the light profile and \fdm{}$=0$ (top panel), and a second where we include a static NFW dark matter halo following the procedure outlined in Section \ref{section.masses} (bottom panel).  In each case we show the combined constraint obtained from stacking individual posterior probability distribution functions (PDFs) for galaxies in our high-redshift sample.  

We find that the high-redshift data prefer an overall normalization of the IMF which is lighter than reported for nearby early-type galaxies of a similar mass ($\lmass \sim 11$), which tend to favor \citet{salpeter1955} or heavier IMFs \citep[but see also \citealp{smith2015}]{conroy2012,conroy2013,cappellari2013,li2017}.  There is an offset between the MFL and NFW models such that models including an explicit dark matter halo predict $\log \alpha = -0.03\pm0.03$, consistent with a Chabrier IMF, while MFL models prefer a slightly heavier IMF normalization with $\log \alpha = 0.07\pm0.03$.  There is little evidence for the bottom-heavy IMFs that have been reported in the central regions ($\lesssim1/8 r_\mathrm{e}$) of massive nearby ETGs \citep[e.g.][]{van-dokkum2016,parikh2018}, which one might expect if \emph{all} quiescent galaxies seen at $z > 1$ are the seeds of local massive ellipticals.  We will discuss this further in Section \ref{discussion.imf}.

As highlighted by Section \ref{sect.rotation}, systematic differences in galaxy structure can influence the derivation of dynamical masses, and by extension our inferences about the IMF.  In order to estimate the magnitude of this effect we re-computed the stacked $\alpha$ PDFs, imposing a Gaussian prior on $q_\mathrm{int} = N(0.25,0.05)$ following the result of \citet{chang2013}; the results are shown as dot-dashed lines in Figure \ref{fig.dm_ob}.  Assuming an intrinsically flattened structure \emph{for all galaxies} results in a slightly heavier overall normalization of IMF, such that for the MFL (NFW) case $\log \alpha = 0.12\pm0.03$ ($0.05\pm0.03$). It therefore seems unlikely that structural evolution \emph{alone} can account for the apparent IMF differences between galaxies in our high-redshift sample and the cores of local early-type galaxies.

\subsubsection{The degeneracy between central dark matter fraction and IMF normalization}

\begin{figure}
\centering
\includegraphics[scale=1.0]{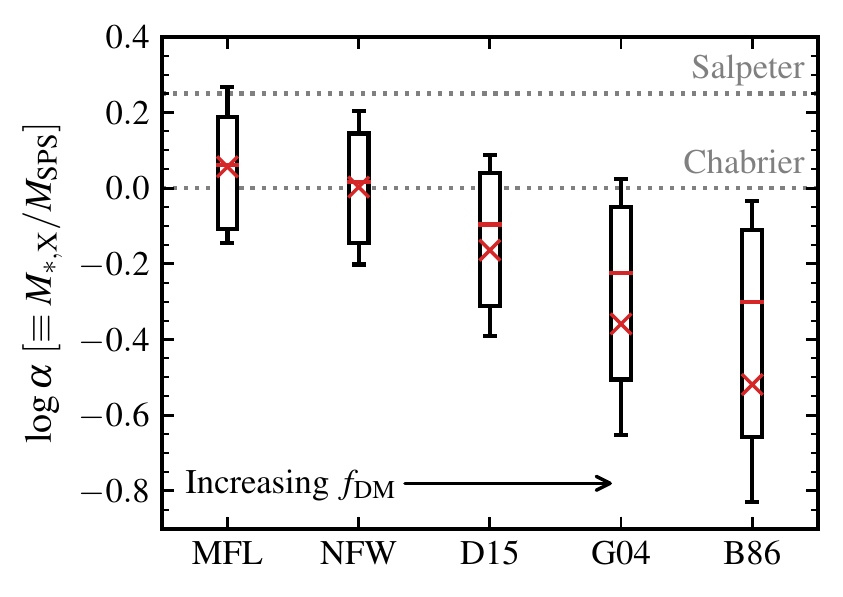}
\caption{IMF normalisation $\alpha$ ($\equiv \mathrm{M_{\ast,dyn}}/\mathrm{M_\mathrm{Chabrier}}$) for different prescriptions describe the dark matter halo response to galaxy formation.  Models are arranged from left to right in order of increasing \fdm{}: mass-follows-light, NFW, \citet{dutton2015}, \citet{gnedin2004}, and \citet{blumenthal1986}.  Red horizontal lines and crosses show the median and mean values, respectively.  Boxes indicate the interquartile range, while error bars show the 16/84$^\mathrm{th}$ percentiles of the observational data. Prescriptions that predict stronger contraction of the dark matter halo lead to higher central dark matter fractions and correspondingly lighter stellar IMFs.}
\label{fig.halo_contract}
\end{figure}

One of the key assumptions in computing $M_{\ast,\mathrm{NFW}}$ is that the dark matter halo component is well represented by an NFW profile, with no accounting for the possible influence of baryons on the dark matter profile shape.  However, if the timescale for galaxy formation is long compared to the halo dynamical time then the halo is expected to contract adiabatically as a result of baryonic collapse \citep[e.g.][]{blumenthal1986,gnedin2004}.  \citet{dutton2016} argue that the dark matter halo can contract or expand depending on the relative balance of inflows, outflows, and feedback \citep[see also][]{lovell2018}, suggesting that our assumption of a static halo may bias the derived values of $M_\mathrm{\ast,NFW}$ and, by extension, $\alpha$.  In this section we therefore explore a broader set of dynamical models that explicitly probe the effect of a variable dark matter halo response on our results.

In the case of spherical symmetry and circular dark matter particle orbits, the adiabatic invariant is given by $rM_\mathrm{tot}(r)$---where $M_\mathrm{tot}(r)$ is the total (baryonic plus dark matter) mass within radius $r$---so that $r_\mathrm{f}/r_\mathrm{i} = M_\mathrm{tot,i}(r_\mathrm{i})/M_\mathrm{tot,f}(r_\mathrm{f})$.  Therefore, given an initial mass distribution $M_\mathrm{tot,i}(r)$ and a final baryonic mass profile $M_\mathrm{bar,f}(r)$, we can derive the final dark matter profile $M_\mathrm{DM,f}(r)$.  Here we assume that the initial dark matter distribution is described by an NFW profile with mass and concentration parameter set by the scaling relations adopted in Section \ref{section.masses}, and that the baryonic mass is distributed in the same way, i.e. $M_\mathrm{bar,i}(r) = f_\mathrm{bar}M_\mathrm{tot,i}(r)$ with $f_\mathrm{bar}$ set by the stellar-to-halo mass relation.  The final baryonic profile $M_\mathrm{bar,f}(r)$ is given by the de-projected MGE luminosity density scaled to match the galaxy stellar mass.  We note that this assumes that star formation is distributed throughout the halo, and is therefore likely an upper limit on the expected contraction. 

If we assume no shell crossing of the dark matter, then $M_\mathrm{DM,i}(r_\mathrm{i}) = M_\mathrm{DM,f}(r_\mathrm{f})$, and the final mapping between $r_\mathrm{f}$ and $r_\mathrm{i}$ is given by

\begin{equation}
\Gamma = M_\mathrm{DM,i}(r_\mathrm{i}) / \left[M_\mathrm{bar,f}(r_\mathrm{f}) + (1-f_\mathrm{bar})M_\mathrm{DM,i}(r_\mathrm{i})\right]
\end{equation}

\noindent with $\Gamma = (r_f/r_i)^\nu$ following the generalized contraction formula suggested by \citet{dutton2007}.  In this framework $\nu = 1$ reproduces the standard adiabatic contraction derived by \citet{blumenthal1986}, while $\nu = 0.8$ reproduces the modified contraction scenario described by \citet{gnedin2004}.  $\nu = 0$ is equivalent to an unmodified NFW profile.  We also include a more mild model for the halo response derived from the NIHAO simulations discussed by \citet{dutton2015}, such that

\begin{equation}
r_\mathrm{f}/r_\mathrm{i} = 0.5 + 0.5(M_\mathrm{tot,i}/M_\mathrm{tot,f})^2.
\end{equation}

\noindent For each contraction model we solve for the mapping between $r_\mathrm{f}$ and $r_\mathrm{i}$, and use this modified dark matter profile as input to the JAM modelling procedure.  While we do not explicitly include any models for halo \emph{expansion} (i.e. $\nu < 0$), our default mass-follows-light models can be interpreted as the extreme case where dark matter is completely evacuated within $r_e$, setting an upper limit for the dynamical effects of an expanded halo.

In Figure \ref{fig.halo_contract} we show the distributions of $\alpha$ derived for these different models of halo response, along with the standard mass-follows-light and NFW cases described in the previous Section.  The expected trend of a decreasing stellar contribution---that is, a lighter overall IMF normalization---with increasingly strong halo contraction is clearly visible, with pure adiabatic contraction \citep[e.g.][]{blumenthal1986} predicting stellar masses which are a factor of $>$3 lighter than those obtained assuming a \citet{chabrier2003} IMF.  Even the most mild model for halo response, \citet{dutton2015}, predicts values of $\alpha$ which are $\sim$60\% lighter than Chabrier, and all of the contraction models considered here predict IMF normalisations which are lighter than observed or inferred for nearby stellar systems \citep[e.g.][]{chabrier2003,bastian2010}.  Taken together, the results in Figure \ref{fig.halo_contract} suggest that any contraction of the dark matter halo due to gas inflow should be counterbalanced by equally violent outflows during the formation process.

\section{Evolutionary trends}
\label{section.evolution}

\begin{figure*}
\centering
\includegraphics[scale=1]{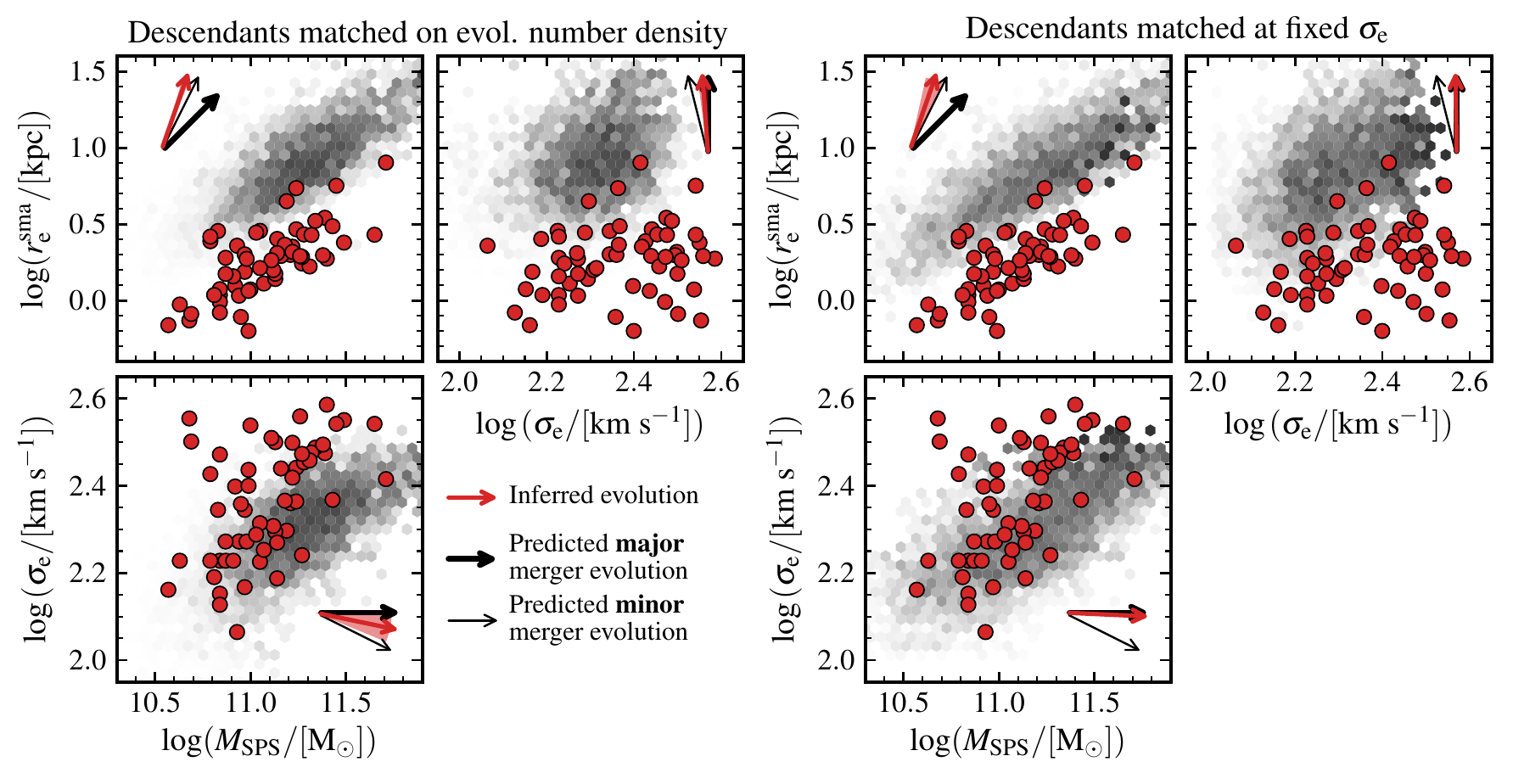}
\caption{Evolutionary trends in half-light size, stellar velocity dispersion, and stellar mass.  Circles (red) show measurements for individual quiescent galaxies at $1.4 < z < 2.1$.  Shading shows the distribution of $z\approx0$ descendants matched either on evolving number density or stellar velocity dispersion (left and right panels, respectively) as described in Section \ref{section.evolution}. In each panel, black arrows indicate the predicted evolution for each pair of parameters based on simple energetic arguments \citep[e.g.][]{bezanson2009}.  Red arrows and shading show the measured trends and their $\pm1\sigma$ uncertainties derived from jackknife resampling.  Note that by definition \sigmaeff{} does not evolve for descendants matched on \sigmaeff{}.}
\label{fig.param_evo}
\end{figure*}

The data presented in Figures \ref{fig.mstar_mdyn}, \ref{fig.mstar_mdyn_cumul}, and \ref{fig.mstar_fdm} suggest an evolution of quiescent galaxy properties from high to low redshift, with high-redshift galaxies having on average lower central dark matter fractions and/or a lighter overall IMF compared to their low-redshift counterparts.  These trends appear to persist even when considering only the oldest galaxies at low-redshift.  However, some care must be taken when making such comparisons as individual galaxies are expected to evolve over the $\sim$8 Gyrs separating our two samples. The challenge of connecting progenitor and descendant populations is therefore that it relies on having at least some a priori knowledge of \emph{how} this evolution proceeds.  Here we consider two different methods for connecting galaxies across redshift based on either their number density or central stellar velocity dispersion.

The stochastic nature of dark matter halo assembly in a $\Lambda$CDM cosmology leads to a broad range of plausible descendants for any single galaxy at high redshift. This diversity of assembly histories can bias inferred evolutionary trends when considering samples selected at either a single fixed or evolving (median) cumulative number density \citep[e.g.][]{mundy2015,wellons2017}.  As an alternative, one can identify descendants based on the full number density PDF as opposed to a single value, which we do here using \texttt{NDPredict}\footnote{Available at \url{https://github.com/sawellons/NDpredict}} as described by \citet{wellons2017}.  For each galaxy in our high-redshift sample, \texttt{NDPredict} provides an estimate of the likely descendant stellar mass distribution based on both expected evolution of the median number density \emph{and its scatter}.  This distribution is then used as a weight to select probable descendants from our low-redshift sample. We use the stellar mass functions published by \citet{muzzin2013} to translate between stellar mass and number density at any given redshift, but have verified that our results are insensitive to the particular choice of mass function.  

As an alternative to the number density matching described above, we also construct a population of descendants matched at fixed central stellar velocity dispersion. Numerical studies have shown that central stellar velocity dispersion is relatively insensitive to assembly via dissipationless mergers \citep{hopkins2009c,oser2012,nipoti2012}, which is expected to be the main growth channel for quiescent galaxies from high redshift. \citet{hopkins2009c} argue that repeated dry mergers tend to decrease stellar velocity dispersions by at most 30\%, as any increase in size leads to a corresponding increase in the dark matter content. To first order this assumption is consistent with the global trends shown in Figure \ref{fig.mstar_fdm}.  Based on these arguments we identify likely descendants as those with \sigmaeff{} within 0.05 dex of galaxies in our high-redshift sample, allowing for replacement---that is, the same low-redshift galaxies can be matched to multiple galaxies in our high-redshift sample.

\subsection{Galaxy evolution in size, stellar mass, and \sigmaeff{}}
\label{section.param_evo}

\begin{figure*}
\centering
\includegraphics[scale=1]{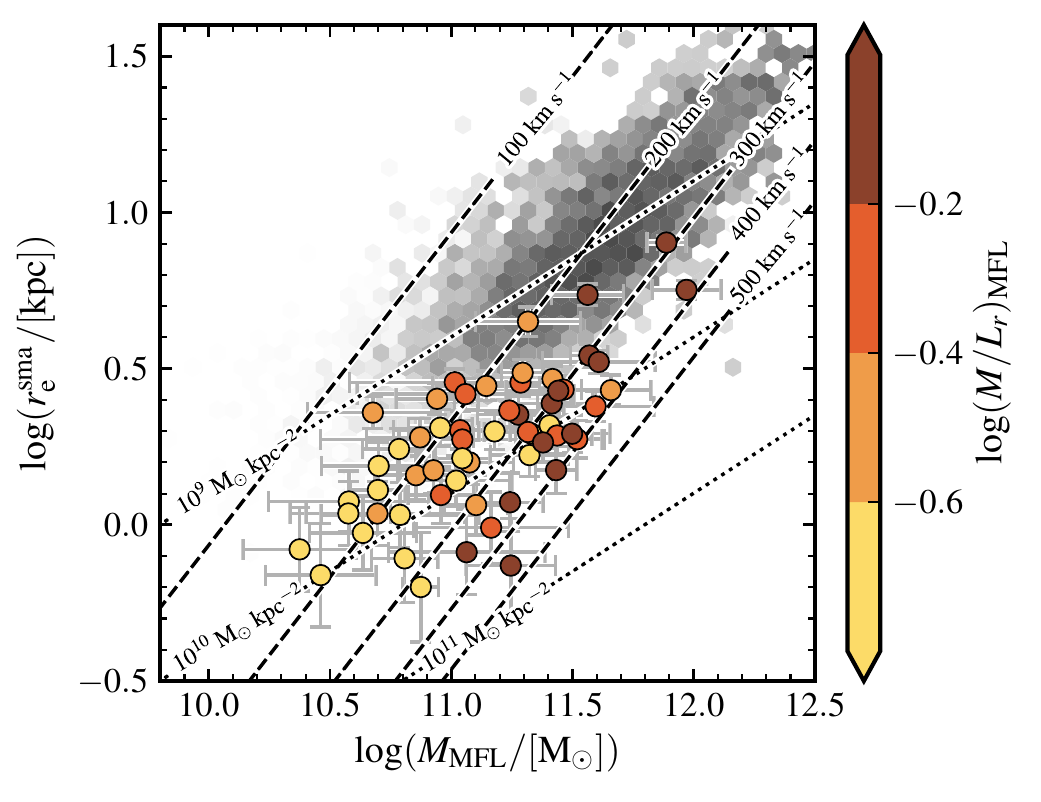}
\caption{Half-light size versus JAM-derived dynamical mass for high-redshift galaxies and their (evolving number density matched) descendants.  High-redshift galaxies, shown as circles, are colored according to their dynamical mass-to-light ratio as indicated by the scale on the right.  Shading shows the distribution of $z\approx0$ descendants derived following Section \ref{section.evolution}. Dashed lines show tracks of constant \sigmaeff{}$=$ 100, 200, 300, 400, and 500 $\mathrm{km~s^{-1}}$ assuming the scalar virial relation (Eqn. \ref{eqn.virial_mass}) with $\kappa = 5$.  Dotted lines instead show tracks of constant stellar surface density, with $\Sigma_\mathrm{e} =$ 10$^9$, 10$^{10}$, and $10^{11}$ $M_\odot~\mathrm{kpc^{-2}}$. Galaxies with high dynamical mass-to-light ratios are relatively well aligned with tracks of constant velocity dispersion, while lower M/L galaxies follow more closely tracks of constant stellar mass density.}
\label{fig.mass_plane}
\end{figure*}

Assuming high-redshift galaxies evolve to reproduce the typical properties of their matched descendants, we can use the relative difference in \rsma{}, $M_\mathrm{SPS}$, and \sigmaeff{} to study the evolutionary processes acting on galaxies over their lifetimes. In the context of $\Lambda$CDM, individual quiescent galaxies are expected to evolve from $z \sim 2$ to the present day through continued mergers after quenching.  \citet{bezanson2009} used scaling relations based on the virial theorem to argue that minor, gas-poor mergers can efficiently increase half-light sizes and decrease stellar velocity dispersions, a result which has been reiterated in a number of theoretical studies \citep[e.g.][but see also \citealp{nipoti2012}]{naab2009,hopkins2009c,oser2010,oser2012,hilz2013}.  These predictions are supported by results that show minor merging can explain the observed evolution of galaxy properties since $z \approx 2$ \citep[e.g.][]{damjanov2011,van-de-sande2013,belli2014a,belli2017}.   

\begin{deluxetable}{lcc}
\tablewidth{0pt}
\tablecolumns{3}
\tablecaption{Inferred evolution of stellar mass, velocity dispersion, and size since $1.4 < z < 2.1$.\label{table.param_evo}}
\tablehead{
\colhead{Parameter} & \colhead{Evol. number} \vspace{-0.25cm} & \colhead{Fixed \sigmaeff{}}  \\ 
 & \colhead{density} & 
} 
\startdata
$\Delta \log (M_\mathrm{SPS}/[M_\odot])$				& $\hphantom{-}0.24\pm0.01$ 	&  $0.22\pm0.03$ \\
$\Delta \log (\sigma_\mathrm{e}/[\mathrm{km~s^{-1}}])$	& $-0.05\pm0.02$ 	&  \ldots \\
$\Delta \log (r_\mathrm{e}^\mathrm{sma}/[\mathrm{kpc}])$				& $\hphantom{-}0.68\pm0.02$ 	&  $0.63\pm0.03$ 
\enddata
\end{deluxetable}

For each high-redshift galaxy we compute the average difference between its properties and those of its matched descendants.  In Table \ref{table.param_evo} we quote the median evolution of $M_\mathrm{SPS}$, \sigmaeff{}, and \rsma{} derived in this way for the full sample, with uncertainties estimated by jackknife resampling.  There is little dependance of the inferred parameters on the method used to define descendant populations, with both the number density and velocity dispersion matched samples pointing towards a dramatic increase in size relative to stellar mass (a factor of $\sim$4.5 in \rsma{} compared to only $\sim$1.5 in $M_\mathrm{SPS}$).  For samples matched on number density we can additionally infer the evolution of \sigmaeff{}, which appears to decrease by only 12\% on average from $z > 1.4$ to the present day, in good agreement the predictions of numerical simulations \citep[e.g.][]{hopkins2009c,hilz2012}.

Figure \ref{fig.param_evo} shows the pairwise distributions of $M_\mathrm{SPS}$, \sigmaeff{}, and \rsma{}, where the observed evolution can be compared to simple energetic arguments for major and minor merging as in \citet[shown as inset coordinate arrows in each panel]{bezanson2009}.  The factor of 4--5 evolution in \rsma{} discussed above is immediately apparent, as is the comparably milder evolution in $M_\mathrm{SPS}$ and \sigmaeff{}. These results are quantitatively similar regardless of how we choose to identify low-redshift descendants (i.e. evolving number density or fixed \sigmaeff{}), and are consistent with the simple predictions for evolution by predominantly minor merging, where $\Delta r_\mathrm{e} \propto \Delta M_\ast^\alpha$ with $\alpha\approx2$ (c.f. $\alpha\approx1$ for major mergers).  In contrast, the distributions of both \rsma{} vs.~\sigmaeff{} and \sigmaeff{} vs.~$M_\mathrm{SPS}$ suggest a more complicated interpretation, whereby those galaxies with $\log (M_\mathrm{SPS}/M_\odot) > 11.2$ or $\log \sigma_\mathrm{e} \gtrsim 2.4$ evolve relatively \emph{more} in $M_\mathrm{SPS}$ and/or \sigmaeff{} than those with lower masses or velocity dispersions.  The inclusion of velocity dispersion also complicates our otherwise straight-forward interpretation of $M_\mathrm{SPS}$ and \rsma{} evolution in terms of minor merging, although we note that more detailed numerical simulations find a range of behaviours depending on the properties of galaxies' host dark matter haloes \citep[e.g.][]{oser2012,nipoti2012,hilz2012,hilz2013}. We will discuss these trends further in Section \ref{discussion.imf}.

\begin{figure*}
\centering
\includegraphics[scale=1.0]{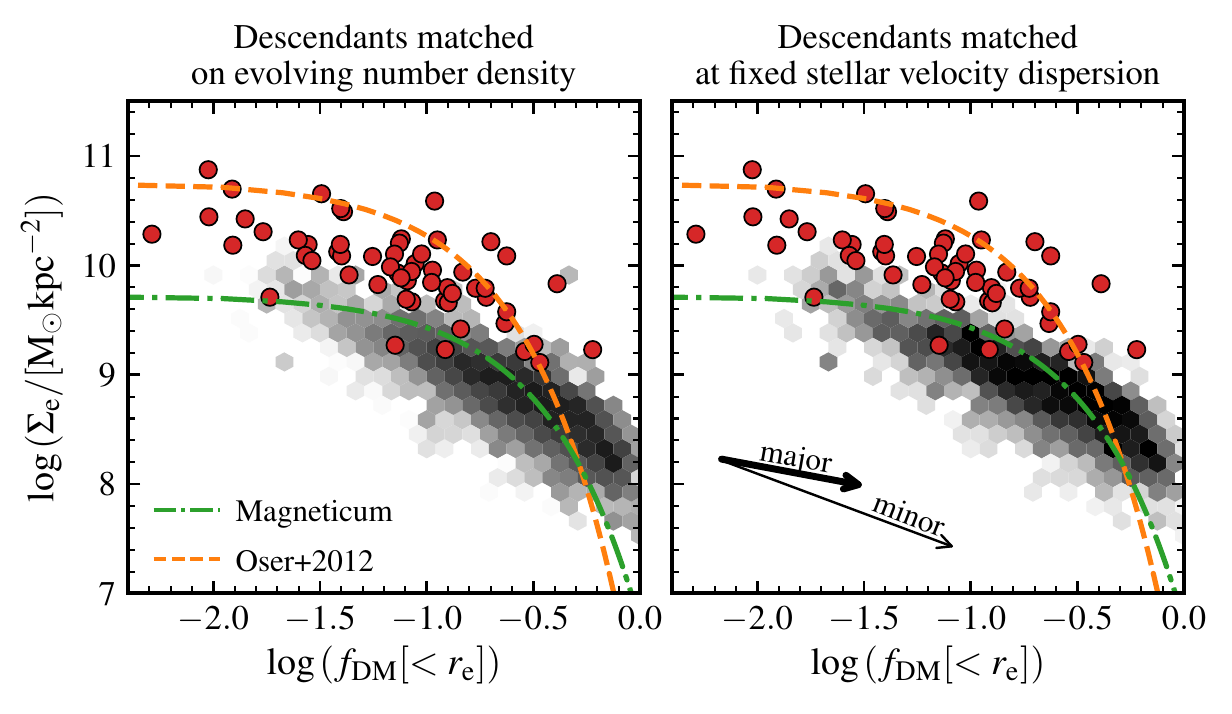}
\caption{Stellar mass surface density, $\Sigma_\mathrm{e}$, as a function of central dark matter fraction for high- and low-redshift galaxies. Filled circles indicate measurements for quiescent galaxies at $1.4 < z < 2.1$, while the background shading shows the distribution of $z\approx0$ descendants matched either by evolving number density (left panel) or central stellar velocity dispersion (right panel; see Section \ref{section.evolution} for details).  Dashed and dot-dashed curves show the trends derived by \citet{remus2017} for simulated early-type galaxies from \citet{oser2012} and Magneticum \citep[Dolag et al., in prep.]{hirschmann2014}.  In the right panel, arrows show the predicted trajectories for evolution driven by major and minor mergers described in Section \ref{section.size_evo}.  In both cases our simple model fails to match the required evolution in $\Sigma_\mathrm{e}$ and \fdm{}.}
\label{fig.sdens_evo}
\end{figure*}

In order to investigate these differences further, in Figure \ref{fig.mass_plane} we show the same galaxy samples as in Figure \ref{fig.param_evo}, however this time in terms of \rsma{} and dynamical mass, often referred to as the mass plane.  Galaxies have additionally been color-coded according to their dynamical mass-to-light ratio, $M/L$. Given the low dark matter fractions inferred for a majority of galaxies in our high-redshift sample, we expect that $M/L_\mathrm{MFL}$ is primarily a tracer of stellar population variations. The mass plane provides a useful parameter space within which to understand systematic variations of galaxy properties and their evolution, and has been used extensively to study the interrelationship between dynamics, structural properties, and stellar populations \citep[e.g.][]{cappellari2013,mcdermid2015,cappellari2016,belli2017,scott2017}.  

We find that those galaxies with the highest $M/L$ follow lines of roughly constant \sigmaeff{} (dashed lines in Figure \ref{fig.mass_plane}), while at lower $M/L$ galaxies more closely follow lines of constant stellar mass surface density, $\Sigma_\mathrm{e}$ (dotted lines in Figure \ref{fig.mass_plane}).  Under the assumption that variations in $(M/L)_\mathrm{MFL}$ are primarily driven by changes in the mean stellar age, these differences are in stark contrast to observational results at low redshift, where \sigmaeff{} is by far the best predictor of galaxy stellar populations \citep[e.g.][]{graves2009a,mcdermid2015,scott2017}.  Nevertheless, numerous studies have shown that stellar surface density is a strong predictor of galaxy quiescence at low redshift \citep[e.g.][but see also \citealp{wake2012}]{cheung2012,fang2013,woo2015}, and high densities appear to be a necessary (if not sufficient) condition for quenching at high redshift \citep[e.g.][]{barro2013,van-dokkum2015,barro2017}. It is therefore unsurprising to see such a correlation borne out in our $M/L$ measurements: at any given epoch galaxies are likely added to the quiescent population as a function of their stellar surface density.  Subsequent merger-driven assembly will then tend to move galaxies along lines of constant velocity dispersion \citep[or steeper, see e.g.][]{nipoti2012,hilz2013}.

\subsubsection{Size evolution as the primary driver of $f_\mathrm{DM}$}
\label{section.size_evo}

We showed in Section \ref{section.fdm} and Figure \ref{fig.mstar_fdm} that, at fixed stellar mass, \fdm{} increases by a factor of $\sim$2 from high redshift to the present day.  However, based on the matched progenitor and descendant samples described in Section \ref{section.evolution} we find that the growth of \fdm{} for individual galaxies is likely even larger, reaching $0.64\pm0.05$ dex ($0.52\pm0.07$ dex) on average for progenitors and descendants matched on evolving stellar mass (fixed \sigmaeff{}). 

Just as major and minor mergers are expected to have different effects on a given galaxy's evolution in size, stellar mass, and stellar velocity dispersion (e.g.~Figure \ref{fig.mass_plane}), they also have a distinct influence on the evolution of $f_\mathrm{DM}$.  \citet{hilz2012,hilz2013} show that minor mergers can dramatically increase \fdm{}, so that a factor of 2 growth in stellar mass can nearly double the central dark matter fraction.  In contrast, a single equal mass (major) merger may only increase \fdm{} by 50\%. This difference is driven by the relatively efficient growth of sizes in minor compared to major mergers; for an NFW-like halo the dark matter mass within a given (small) radius $r$ scales roughly with the virial mass of the halo as $M_\mathrm{DM}(r) \propto M_\mathrm{vir} r^2$ \citep{boylan-kolchin2005}.  In an equal mass merger where both $r_\mathrm{e}$ and the stellar mass within $r_\mathrm{e}$ double, the dark matter mass within $r_\mathrm{e}$ can increase by up to a factor of 8 (assuming that the stellar and halo mass increase by a similar amount). By comparison, a similar doubling of stellar mass through multiple minor mergers can increase $r_\mathrm{e}$ by a factor of 4, and the enclosed dark matter mass by more than a factor of $\gtrsim$30.  

These differences are particularly apparent when viewed in terms of stellar mass surface density, $\Sigma_\mathrm{e}$ ($\equiv M_\mathrm{SPS}/ 2 \pi r_\mathrm{e}^2$), and \fdm{}, which we show in Figure \ref{fig.sdens_evo}.  We find that $\Sigma_\mathrm{e}$ and \fdm{} are anti-correlated, in the sense that those galaxies with the highest stellar mass surface densities have the lowest \fdm{}; a similar anti-correlation has been demonstrated previously for both late- and early-type galaxies \citep[e.g.][]{mcgaugh2005,sonnenfeld2015}.  Figure \ref{fig.sdens_evo} additionally shows that $\Sigma_\mathrm{e}$ decreases by $\sim$1 dex on average from $z >1.5$ to $z\approx0$, which is a direct consequence of the size and stellar mass evolution discussed in Section \ref{section.param_evo} (see also Table \ref{table.param_evo} and Figure \ref{fig.param_evo}). This can be compared to the results of \citet{remus2017}, who show that simulated galaxies follow well-defined tracks in $\Sigma_\mathrm{e}$--$f_\mathrm{DM}$ regardless of redshift (shown as dashed and dot-dashed lines in Figure \ref{fig.sdens_evo}).  Nevertheless, the simulation results roughly reproduce the observational trends at any given epoch, with the primary difference between the two simulations discussed by \citet{remus2017} being their treatment of black hole feedback.

We can use simple scaling relations to better understand galaxies' expected evolution in Figure \ref{fig.sdens_evo} given various assumptions. Following Figure \ref{fig.param_evo} we adopt a simple model for size growth during mergers such that $\Delta r_\mathrm{e} \propto \Delta M_\ast^\alpha$, with $\alpha= 1$ or 2 for major and minor mergers, respectively.  We additionally assume that the enclosed dark matter mass scales with the total (virial) mass of the halo and radius as $M_\mathrm{DM}(r) \propto M_\mathrm{vir} r^2$, and that the stellar and halo mass grow at the same rate ($\Delta \log M_\mathrm{vir} = \Delta \log M_\ast$). Arrows in the right-hand panel of Figure \ref{fig.sdens_evo} show the predicted evolution for a galaxy doubling its stellar mass either through a single major merger, or successive 10:1 (minor) mergers.  As expected, the efficient size growth associated with minor mergers in our toy model drives rapid evolution in both $\Sigma_\mathrm{e}$ and \fdm{}.  However, while minor mergers can explain the observed decrease in $\Sigma_\mathrm{e}$ from high to low redshift, our toy model over-predicts the increase of \fdm{}.  \citet{tortora2018} report a similar result for galaxies at $z\approx0.7$, and suggest that allowing for variation in the stellar-to-halo mass ratio of accreted galaxies (i.e. $\Delta \log M_\mathrm{vir} \neq \Delta \log M_\ast$) can help to lessen the tension between predicted and observed evolution in $f_\mathrm{DM}$. This is particularly likely for the massive quiescent galaxies in our high-redshift sample, which are expected to trace the most massive halos at their respective redshifts \citep[e.g.][]{lin2019}.

\subsection{Evolution of the IMF at fixed \sigmaeff{}}
\label{discussion.imf}

\begin{figure*}
\centering
\includegraphics[scale=1.0]{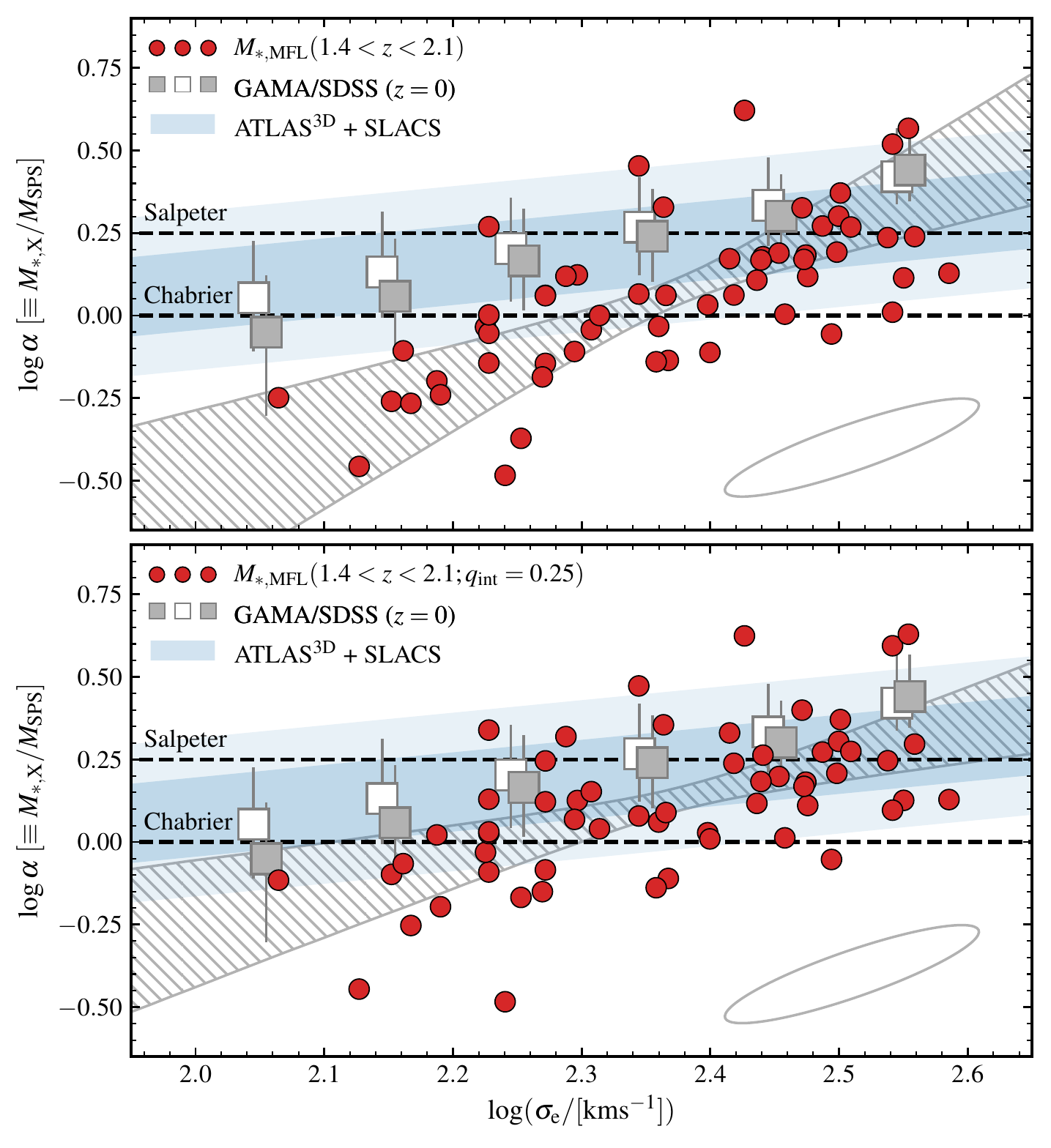}
\caption{The IMF offset parameter, $\alpha$, as a function of stellar velocity dispersion for different samples. In the top panel, filled circles show at $1.4 < z < 2.1$ assuming that mass-follows-light (MFL) and a uniform prior on the intrinsic axis ratio, \qint{}.  In the bottom panel we show the same data, but adopting a prior such that \qint{}$=0.25$. The error ellipse in the bottom right corner shows the typical uncertainty for individual galaxies, while the hatched regions illustrates the 16/84$^\mathrm{th}$ percentile confidence interval based on a linear fit to the high-redshift data. Large squares represent the median $\alpha$ derived for GAMA/SDSS data matched in either evolving number density (filled squares) or stellar velocity dispersion (open squares) as described in Section \ref{section.evolution}. Error bars indicate the 16/84$^\mathrm{th}$ percentile uncertainties on the binned values. Shaded bands show the $\alpha$-\sigmaeff{} correlation based on a joint analysis of ATLAS$^\mathrm{3D}$ and SLACS data by \citet{posacki2015}, with dark and light light regions indicating the 1 and 2$\sigma$ bounds.  There is good agreement between the matched GAMA/SDSS data and derived best-fit from \citeauthor{posacki2015}, both of which show evidence for a positive correlation between $\alpha$ and \sigmaeff{}. High-redshift galaxies show evidence for a similar correlation, albeit with a steeper slope such that objects with $\log \sigma_\mathrm{e} \gtrsim 2.45$ have $\alpha$ and \sigmaeff{} comparable to low-redshift galaxies, while at lower \sigmaeff{} further evolution is required to reproduce the low-redshift correlation.}
\label{fig.alpha_sigma}
\end{figure*}

In Section \ref{section.imf} we showed that the kinematics of massive quiescent galaxies at $z > 1.5$ are consistent with a MW-like IMF on average.  In contrast, a number of studies based on stellar absorption features, lensing, and dynamical modelling have shown that a Salpeter or even heavier IMF may be more typical in the in the inner 0.1--0.2 $R_\mathrm{e}$ of present-day early-type galaxies \citep[e.g.][but see also \citealp{smith2015}]{thomas2011b,conroy2012,cappellari2012,cappellari2013,spiniello2014,van-dokkum2016,conroy2017}, suggesting some tension between our high-redshift results and those in the nearby Universe.

Figure \ref{fig.alpha_sigma} shows a comparison of the IMF offset parameter, $\alpha$, as a function of stellar velocity dispersion, \sigmaeff{}, at different redshifts. We include in this figure constraints derived from low-redshift galaxies in the ATLAS$^\mathrm{3D}$ and SLACS samples by \citet{posacki2015}, as well as binned results for the matched GAMA/SDSS samples described in Section \ref{section.evolution}. This comparison shows that the apparent offset in mean $\alpha$ between high and low redshift depends on \sigmaeff{}, such that galaxies with the highest velocity dispersions generally have $\alpha$ values consistent with observations at low redshift, while lower velocity dispersion galaxies are offset towards lower $\alpha$---that is, towards a ``lighter'' IMF---at fixed stellar velocity dispersion. Assuming an intrinsically flat axis ratio for high-redshift galaxies (e.g. \qint $=0.25$; bottom panel of Figure \ref{fig.alpha_sigma}) reduces the apparent offset of $\alpha$ at low \sigmaeff{}, though the qualitative trend remains unchanged. We note that this offset is unlikely to be the result of (correlated) uncertainties in \sigmaeff{} and $\alpha$, shown by the error ellipse in the bottom-right corner of Figure \ref{fig.alpha_sigma}, which tend to scatter galaxies along the low-redshift $\alpha$--\sigmaeff{} correlation rather than away from it. It may be that the differential evolution seen between high- and low-\sigmaeff{} galaxies is tied to the overall buildup of the velocity dispersion function (VDF) over time.  \citet{bezanson2012} show that galaxies with $\log \sigma_\mathrm{e} \gtrsim 2.4$ form early and their number density has changed little since at least $z \approx 1.5$, while the population at lower \sigmaeff{} evolves significantly due to the addition of newly-quenched galaxies. A more detailed dissection of the $\alpha$--\sigmaeff{} relation in the context of galaxy star-formation histories will be the subject of future work.

An interesting implication of Figure \ref{fig.alpha_sigma} is that the $\alpha$-\sigmaeff{} relation found at low redshift is established early on, and any scenario invoked to explain quiescent galaxies' subsequent evolution in \rsma{}, \sigmaeff{}, and $M_\mathrm{SPS}$ (e.g. Figure \ref{fig.param_evo}) should largely preserve the underlying correlation; this is especially true at the highest stellar velocity dispersions. Using high signal-to-noise long-slit spectra of 6 nearby ETGs, \citet{van-dokkum2016} found evidence for strong radial gradients in $\alpha$ which increased from MW-like at $R > 0.4R_\mathrm{e}$ to Salpeter or heavier at $0.1R_\mathrm{e}$ \citep[see also][]{martin-navarro2015a,lyubenova2016}.  Under the assumption that gas-poor mergers primarily contribute to the buildup of an extended stellar envelope \citep[e.g.][]{hopkins2009a,van-dokkum2010,karademir2019}, then we would expect the remnants of our high-redshift sample should survive in the cores of massive nearby ellipticals. Indeed, the typical half-light size of galaxies in our sample (1--1.5 kpc) are comparable to the physical extent over which the IMF is found to vary significantly in \citet{van-dokkum2016}.

Finally, although we have tried to be comprehensive in our measurement of stellar masses, we cannot rule out that at least some part of the evolutionary trends implied by Figure \ref{fig.alpha_sigma} could be the result of redshift-dependent systematic uncertainties in our derivation of $M_\mathrm{SPS}$.  These could be, for example, due to our adopted star-formation history or stellar population synthesis models \citep[e.g.][]{pforr2012,leja2019}.  Based on the analysis of mock galaxy spectra generated from semi-analytic models, \citet{pforr2012} showed that stellar masses for passive galaxies can be recovered to better than $\sim$0.05 dex for a wide range of star-formation histories, suggesting that our results are unlikely to be due \emph{entirely} to the details of our SFH modelling.  Furthermore, while \citealp{pforr2012} show that mismatches in metallicity can lead to systematic offsets of up 0.2--0.3 dex, direct observations at $z>1.5$ support our adoption of a solar metallicity template library \citep[e.g.][]{onodera2015,kriek2016}.

\section{Conclusions}
\label{section.conclusions}

We present an analysis of 58 massive quiescent galaxies at $1.4 < z < 2.1$ with measured stellar velocity dispersions and deep near-infrared HST imaging.  We use these data to study the evolution of dynamical masses and the dynamical-to-stellar mass ratio, which are sensitive to the central dark matter fraction and normalization of the stellar IMF.  We find that:

\begin{itemize}
\item[i)]{The median dynamical-to-stellar mass ratio of quiescent galaxies is lower by $\sim$0.2 dex at $1.4 < z < 2.1$ compared to low redshift.  In Figures \ref{fig.mstar_mdyn} and \ref{fig.mstar_mdyn_cumul} we showed that this offset appears to be independent of the method used to derive dynamical masses (e.g.~Jeans models vs.~virial mass estimates).  The observed evolution is consistent with an decrease in the fraction (by mass) of dark matter within the galaxy effective radius, \fdm{}, which is lower by a factor of roughly two in our high-redshift sample compared to nearby galaxies in the SDSS/GAMA surveys (\fdm{}$=6.6\pm1.0$\% at $z\approx 1.8$ vs. $16.3\pm0.3$\% at $z = 0$) at fixed stellar mass. These measurements appear consistent with recent results based on the rotation curves of high-redshift star-forming galaxies.  Based on the matching of progenitor and descendant populations, we argue in Section \ref{section.size_evo} that the evolution of \emph{individual} galaxies is likely even larger, with \fdm{} increasing by a factor of 4--5 from high redshift to the present day.}

\item[ii)]{Under the assumption that central dark matter fractions are intrinsically low in high-redshift galaxies, the dynamical-to-stellar mass ratio can be used as a probe of the stellar IMF.  For mass-follows-light models, we find that high-redshift data are consistent with a Kroupa-like IMF on average, while models including an explicit NFW dark matter halo are consistent with a Chabrier IMF (see Figure \ref{fig.dm_ob}).  We find a correlation between stellar velocity dispersion and the IMF offset parameter, $\alpha$, at high redshift that is consistent with low-redshift data, suggesting that any subsequent evolution should act to preserve this underlying correlation  (see Figure \ref{fig.alpha_sigma}).  We argue that minor mergers are the most likely drivers of galaxies' growth in \rsma{} and $M_\mathrm{SPS}$, as they primarily add material at large radii while preserving the stellar populations in the inner regions.}

\item[iii)]{Simple models for the contraction of dark matter haloes in response to baryonic collapse predict high central dark matter fractions.  In the most extreme case of pure adiabatic contraction, such models require IMF normalizations a factor of $\sim$3 \emph{lighter} than Chabrier to explain the observed kinematics (see Figure \ref{fig.halo_contract}).  Significant contraction of the dark matter halo is difficult to accommodate given current observational constraints \emph{unless} other baryonic process (e.g. outflows or AGN feedback) act to reduce the central dark matter content after collapse.}

\item[iv)]{A comparison of kinematics and structural properties between high-redshift galaxies and their likely descendants at low-redshift supports minor merging as the dominant evolutionary pathway after quenching; however those galaxies with the highest stellar masses and/or stellar velocity dispersions appear to evolve relatively more than lower mass/dispersion objects.  This separation is apparent for descendants matched both on evolving number density as well as at fixed \sigmaeff{} (see Figure \ref{fig.param_evo}). } 

\item[v)]{In the two dimensional parameter space of size and dynamical mass---the so-called ``mass plane'' (Figure \ref{fig.mass_plane})---high redshift galaxies are both smaller and have lower dynamical masses than their low-redshift descendants.  Separating the galaxy population in terms of total mass-to-light ratio ($M/L$), which we interpret here as a proxy for stellar population age, galaxies with the highest $M/L$ follow lines of roughly constant \sigmaeff{}, while those with lower $M/L$ follow more closely lines of constant stellar surface density.  We interpret these differences as being driven by two separate phases of passive galaxy formation, whereby galaxies first quench as a strong function of their stellar mass surface density and their subsequent evolution on the mass plane is driven by minor merging.}

\end{itemize}

Taken together, our results point towards an evolutionary scenario for massive quiescent galaxies that sees their formation occurring rapidly at $z > 2$.  Subsequent quenching of star formation appears to preserve the disc-like structural and kinematic signatures associated with massive star-forming galaxies at those redshifts \citep[e.g.][]{wuyts2011,wisnioski2015,wisnioski2019} as well as their dark matter properties \citep[e.g.][]{wuyts2016,lang2017,genzel2017,genzel2020}.  Subsequent evolution through continued merging is then required to transform both their structural and kinematics properties to reproduce the massive, predominantly slowly-rotating galaxies that constitute their likely descendants at $z=0$ \citep[e.g.][]{veale2017}.  Based on galaxies' evolution in size, stellar mass, stellar velocity dispersion, and central dark matter fraction it appears that the most likely channel for this evolution is through the accretion of relatively lower-mass galaxies (i.e. minor mergers). While these results are based on the best currently available kinematic and photometric data, future spectroscopic observations with the \emph{James Webb Space Telescope} and 30+ meter ground-based observatories will enable systematic surveys of high-redshift stellar kinematics.

\acknowledgements

We thank the referee for constructive feedback that helped to improve the overall quality of this manuscript, as well as Sandesh Kulkarni and Kaushi Bandara for their help with KMOS observations for the VIRIAL survey.  JTM acknowledges the support of the Australian Research Council Centre of Excellence for All Sky Astrophysics in 3 Dimensions (ASTRO 3D), through project number CE170100013.  DJW and MF acknowledge the support of the Deutsche Forschungsgemeinschaft via Project IDs 3871/1-1 and 3871/1-2.  MF has received funding from the European Research Council (ERC) under the European Union's Horizon 2020 research and innovation programme (grant agreement No 757535).

This work is based on observations taken by the CANDELS Multi-Cycle Treasury Program (GO 12060 and 12099) and 3D-HST Treasury Program (GO 12177 and 12328) with the NASA/ESA HST, which is operated by the Association of Universities for Research in Astronomy, Inc., under NASA contract NAS5-26555.  It also includes data obtained from the Hubble Legacy Archive, which is a collaboration between the Space Telescope Science Institute (STScI/NASA), the Space Telescope European Coordinating Facility (ST-ECF/ESA) and the Canadian Astronomy Data Centre (CADC/NRC/CSA).

GAMA is a joint European-Australasian project based around a spectroscopic campaign using the Anglo-Australian Telescope. The GAMA input catalogue is based on data taken from the Sloan Digital Sky Survey and the UKIRT Infrared Deep Sky Survey. Complementary imaging of the GAMA regions is being obtained by a number of independent survey programmes including GALEX MIS, VST KiDS, VISTA VIKING, WISE, Herschel-ATLAS, GMRT and ASKAP providing UV to radio coverage. GAMA is funded by the STFC (UK), the ARC (Australia), the AAO, and the participating institutions.  The VISTA VIKING data used here were taken using ESO Telescopes at the La Silla Paranal Observatory under programme ID 179.A-2004. The GAMA website is http://www.gama-survey.org/ .

Funding for the SDSS and SDSS-II has been provided by the Alfred P. Sloan Foundation, the Participating Institutions, the National Science Foundation, the U.S. Department of Energy, the National Aeronautics and Space Administration, the Japanese Monbukagakusho, the Max Planck Society, and the Higher Education Funding Council for England. The SDSS Web Site is http://www.sdss.org/.

The SDSS is managed by the Astrophysical Research Consortium for the Participating Institutions. The Participating Institutions are the American Museum of Natural History, Astrophysical Institute Potsdam, University of Basel, University of Cambridge, Case Western Reserve University, University of Chicago, Drexel University, Fermilab, the Institute for Advanced Study, the Japan Participation Group, Johns Hopkins University, the Joint Institute for Nuclear Astrophysics, the Kavli Institute for Particle Astrophysics and Cosmology, the Korean Scientist Group, the Chinese Academy of Sciences (LAMOST), Los Alamos National Laboratory, the Max-Planck-Institute for Astronomy (MPIA), the Max-Planck-Institute for Astrophysics (MPA), New Mexico State University, Ohio State University, University of Pittsburgh, University of Portsmouth, Princeton University, the United States Naval Observatory, and the University of Washington.

\bibliographystyle{aasjournal}
\bibliography{biblio}

\clearpage

\appendix
\twocolumngrid

\section{Additional tests of velocity dispersion measurements}
\label{appendix.modeling}

\begin{figure}
\centering
\includegraphics[scale=1.0]{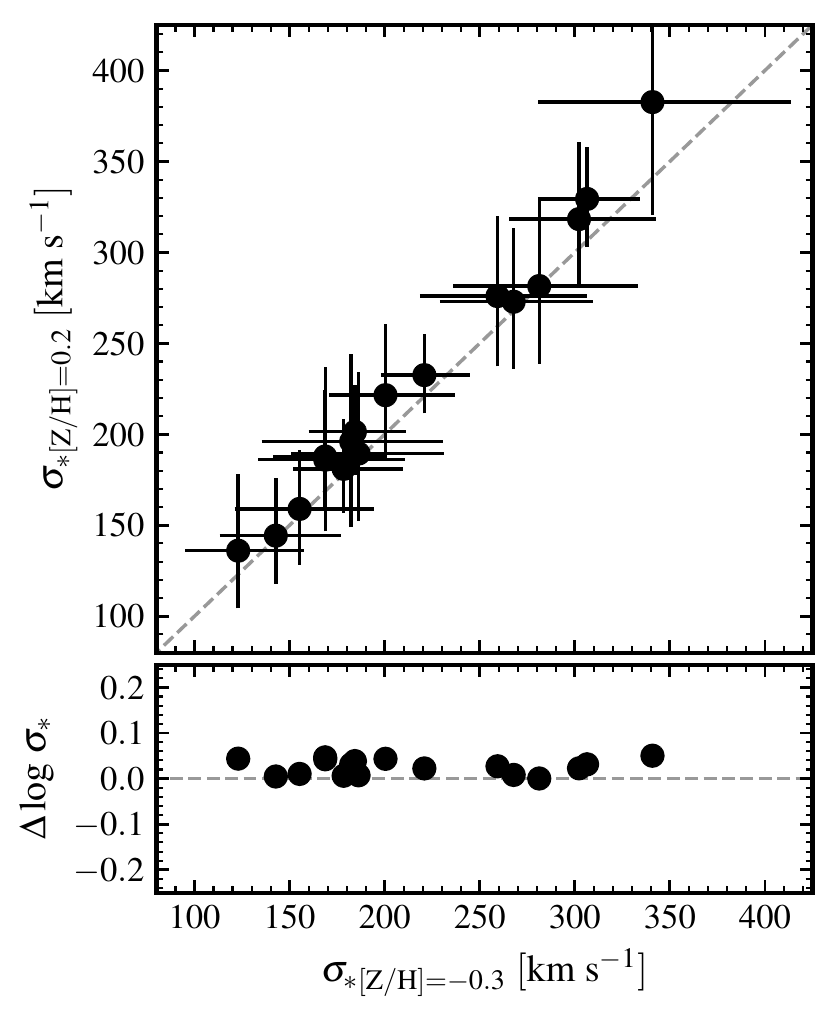}
\caption{The effects of changing template metallicity on the derived stellar velocity dispersion.  In this case we consider the effects of changing the assumed metallicity by a factor of $\sim$3, from [Z/H]=-0.3 to 0.2 dex.  The overall impact is small, with the mean dispersion increasing by less that 2\% when going from low to high metallicity.}
\label{fig.met_comp}
\end{figure}

\begin{figure}
\centering
\includegraphics[scale=1.0]{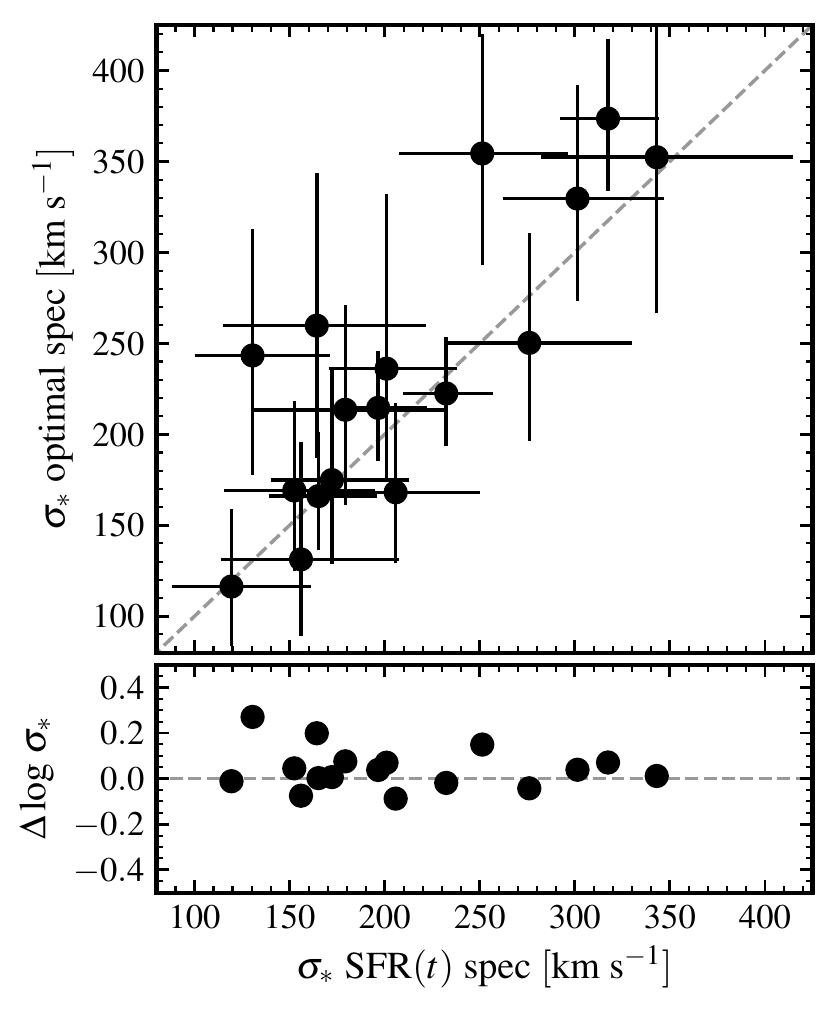}
\caption{The impact of using ``optimal'' vs. parametric template models on the derived stellar velocity dispersion.  The optimal templates are derived from a linear combination of single stellar population models, while the parametric models assume an explicit description of SFR as a function of cosmic time as given by Eqn. \ref{eqn.ssfr} }
\label{fig.templ_comp}
\end{figure}

In this Appendix we explore possible systematics in our derived velocity dispersions stemming from assumptions made in our spectrophotometric modelling approach.  There are two effects in particular that we are concerned with: the impact of assuming a fixed solar metallicity for our SED template library, and the effect of our adopted (parametric) star-formation history.

\subsection{Metallicity effects}

While the majority of low-redshift massive early-type galaxies are consistent with solar metallicity or higher \citep[e.g.][]{gallazzi2006,thomas2010}, the picture at high redshift is still uncertain, with different results suggesting variations in total metallicity of up to 0.8 dex \citep[e.g.][]{onodera2015,lonoce2015,kriek2016,morishita2018}.  Given our adoption of a fixed solar metallicity for our template grid, it is therefore worthwhile investigating the possible impact of this assumption on our derived velocity dispersions.  In Figure \ref{fig.met_comp} we show a comparison of stellar velocity dispersions obtained with templates a factor of $\sim$2 higher or lower in metallicity.  While there is a clear systematic shift in the resulting values of $\sigma_\ast$,the offsets are of order a few per cent, significantly smaller than the typical 20\% uncertainties on our measurements of $\sigma_\ast$, and are therefore unlikely to bias our results.

\subsection{Star-formation history effects}

While the adoption of a parametric star-formation history to describe the observed photometry is common practice, the use of such models in performing kinematic measurements is less common, and bears further investigation.  In brief, we modified the MCMC fitting code described in Section \ref{section.obs} to construct the best-fit template from a linear combination of SSP templates, as opposed to the parametric star-formation histories adopted previously.  This approach mimics the internal fitting procedure adopted in well-known fitting codes such as  pPXF \citep{cappellari2002}, while at the same time providing improved handling of low signal-to-noise data thanks to the MCMC sampling of the LOSVD.  We show the results of this re-fitting in Figure \ref{fig.templ_comp}.  

There is significant scatter between the two velocity dispersion estimates, but in nearly all cases the derived $\sigma_\ast$ values are consistent within uncertainties.  The values derived in our default modelling are larger by $\sim$0.06 dex on average, however this does not significantly affect our results.

\section{Comparisons with previous structural parameter measurements}
\label{appendix.photometry}

In this Section we present a comparison of the galaxy structural parameters derived here with those available in the literature (where available).

\subsection{High redshift}
\label{appendix.highz}

\begin{figure*}
\centering
\includegraphics[scale=1]{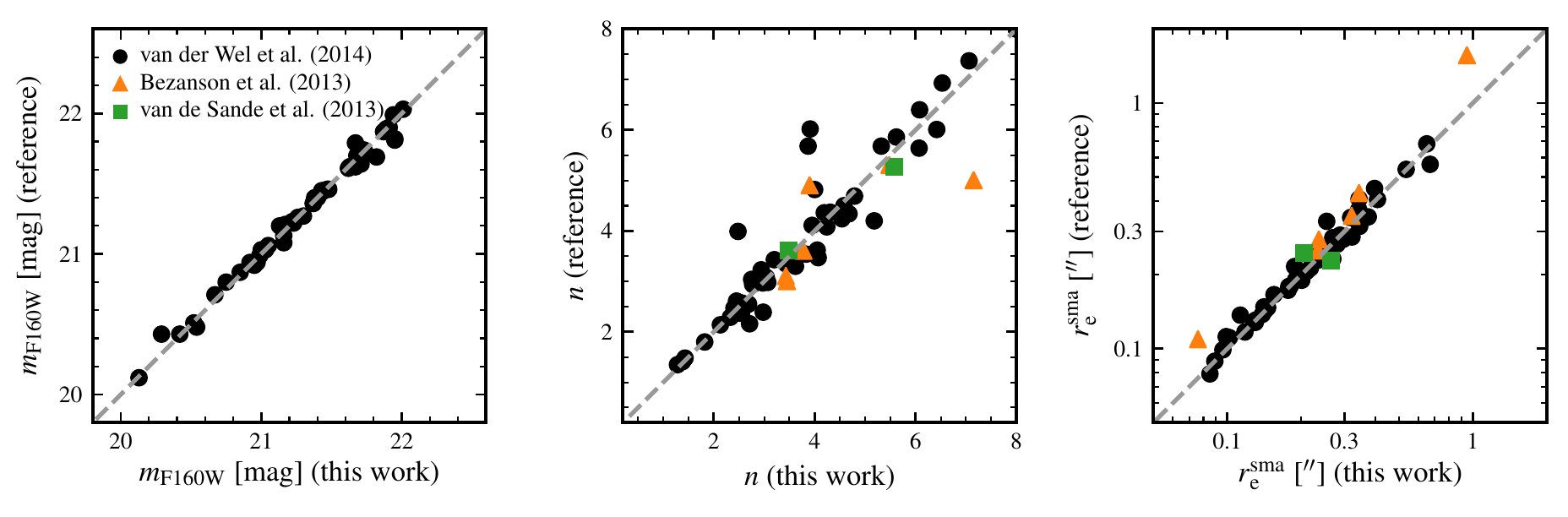}
\caption{Comparison of total magnitude $m_r$, S\'ersic index $n$, and semi-major axis size $r_\mathrm{e}^\mathrm{sma}$ for galaxies in our high-redshift sample from different sources.}
\label{fig.highz_comp}
\end{figure*}

The comparison of apparent magnitude, S{\'e}rsic index, and size for galaxies in our high-redshift sample is shown in Figure \ref{fig.highz_comp}.  For most galaxies these quantities were taken from \citet{van-der-wel2014}, which are based on the same CANDELS/3D-HST WFC3/F160W imaging.  As discussed in Section \ref{section.literature}, several galaxies from \citet{bezanson2013} and \citet{van-de-sande2013} fall outside the CANDELS/3D-HST footprint, and there our measurements are based on pipeline-processed data retrieved from the HLA.  In most cases the agreement between different measurements is excellent.

\subsection{Low redshift}
\label{appendix.lowz}

\begin{figure*}
\centering
\includegraphics[scale=1]{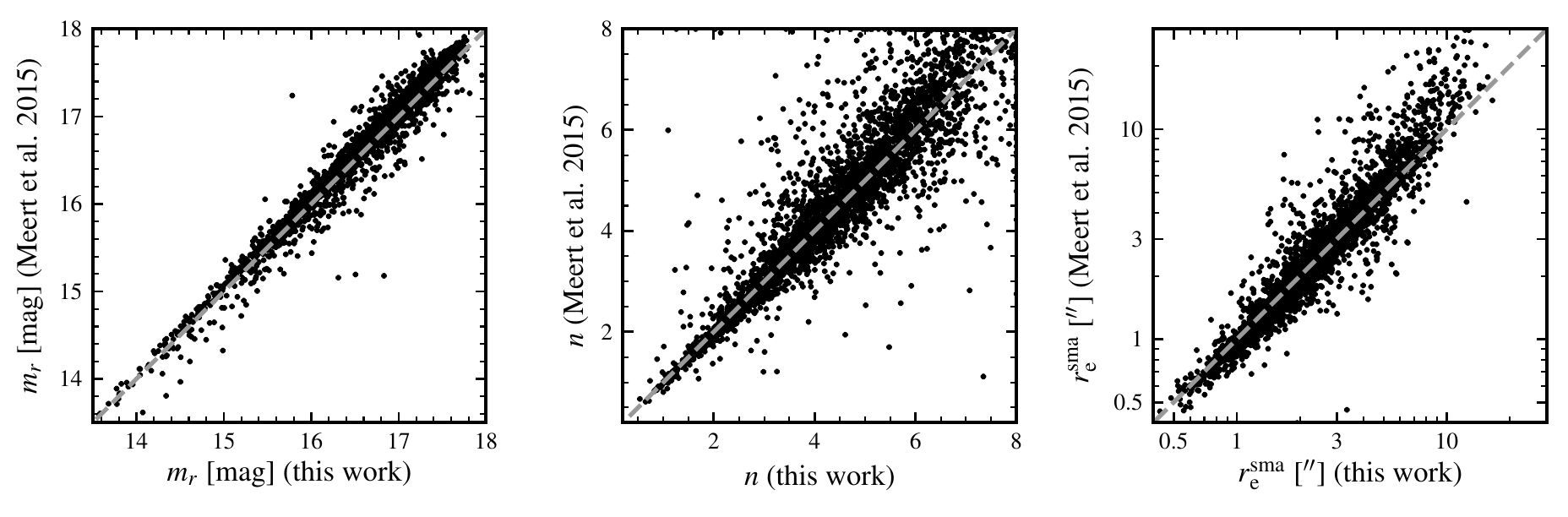}
\caption{Comparison of total magnitude $m_r$, S\'ersic index $n$, and semi-major axis size $r_\mathrm{e}^\mathrm{sma}$ for low-redshift galaxies derived in this work vs. \citet{meert2015}.}
\label{fig.meert_comp}
\end{figure*}

\begin{figure*}
\centering
\includegraphics[scale=1]{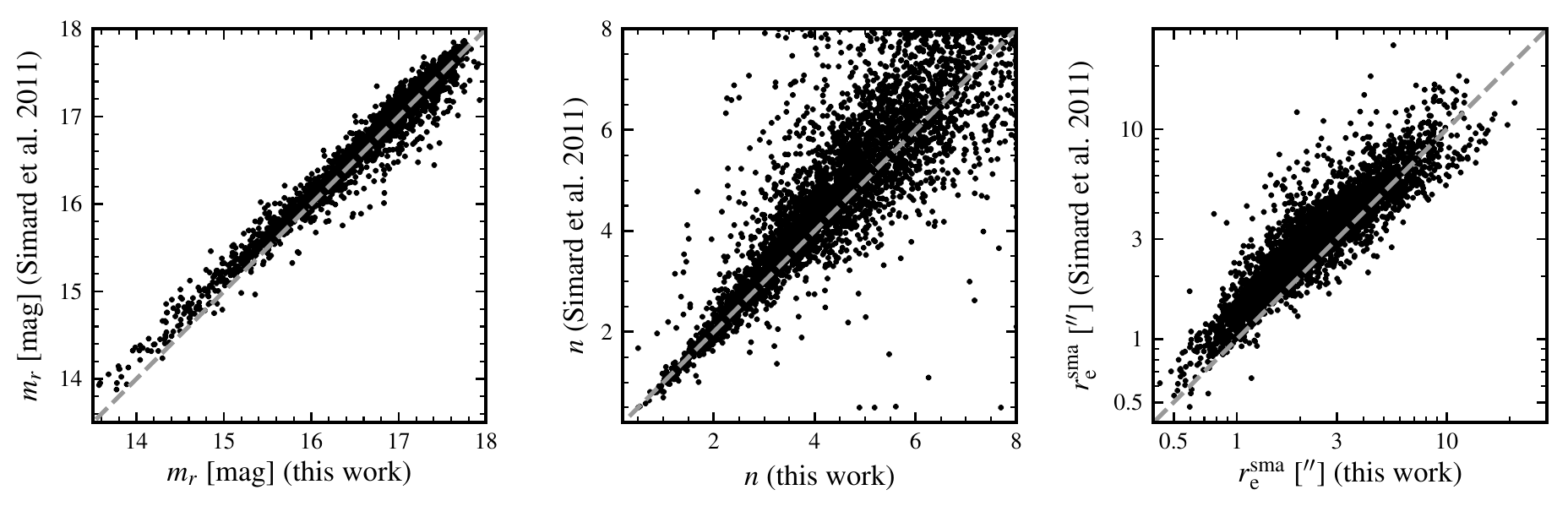}
\caption{Comparison of total magnitude $m_r$, S\'ersic index $n$, and semi-major axis size $r_\mathrm{e}^\mathrm{sma}$ for low-redshift galaxies derived in this work vs. \citet{simard2011}.}
\label{fig.simard_comp}
\end{figure*}

Figures \ref{fig.meert_comp} and \ref{fig.simard_comp} show a comparison of apparent $r$-band magnitude, S{\'e}rsic index, and size for galaxies in our low-redshift SDSS/GAMA sample with measurements from \citet{meert2015} and \citet{simard2011}, respectively.  While there is generally good agreement between measurements in the different structural catalogs, there is significant scatter driven by differences in approach used to mask/model neighboring objects, the size of fitted images, and the method for measuring the sky background.  In most cases any systematic biases are relatively small, $\lesssim20$\%, and do not effect the conclusions of this work.

\section{Photometric fits for high-redshift galaxies}
\label{appendix.phot_figs}

In Figure \ref{fig.mp0} we show the Sersic and MGE fits derived for each galaxy in the high redshift sample.

\begin{figure*}
\centering
\includegraphics[scale=1,angle=90]{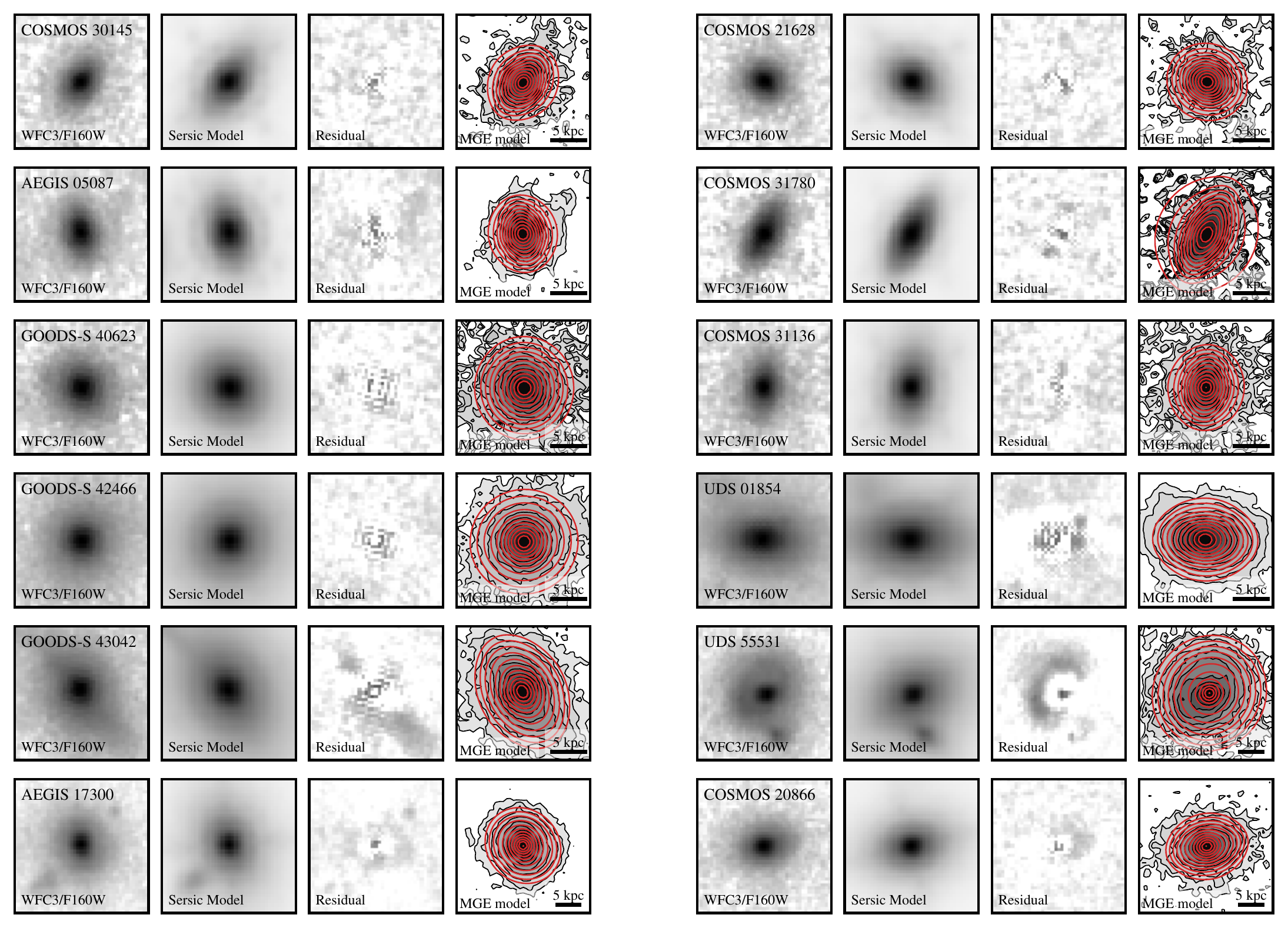}
\caption{Individual galaxy images and fits, following the format of Figure \ref{fig.sb_example}.  From left to right, panels show the observed HST WFC3/F160W image, the best-fit S\'ersic model derived using \texttt{galfit}, residual maps, and an overlay of MGE contours on the observed galaxy image.  Images are plotted in surface brightness units, and contours are evenly spaced in steps of 0.5 mag arcsec$^{-2}$.}
\label{fig.mp0}
\end{figure*}

\begin{figure*}
\centering
\figurenum{\ref{fig.mp0}}
\includegraphics[scale=1,angle=90]{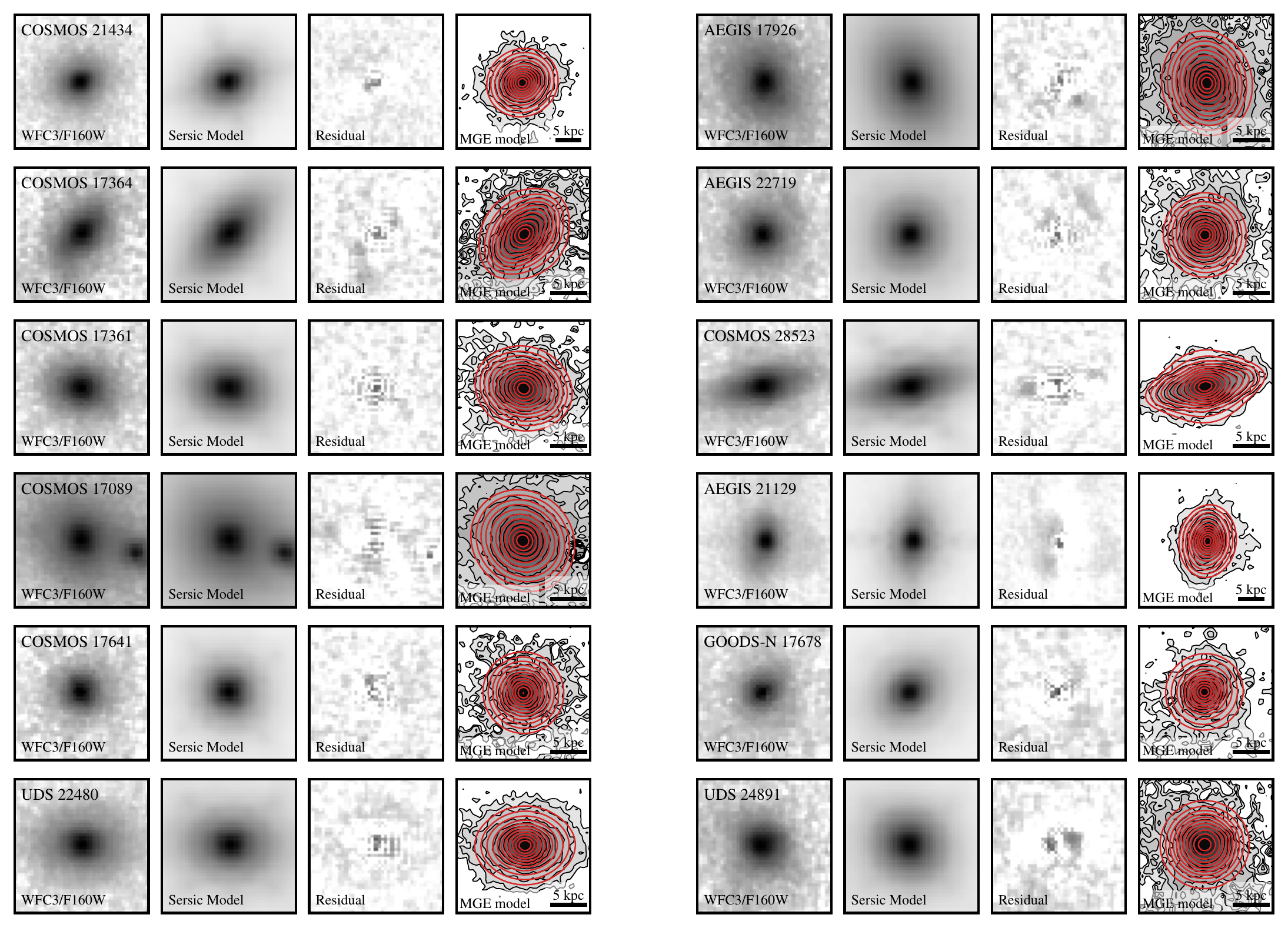}
\caption{(Continued)}
\label{fig.mp1}
\end{figure*}

\begin{figure*}
\centering
\figurenum{\ref{fig.mp0}}
\includegraphics[scale=1,angle=90]{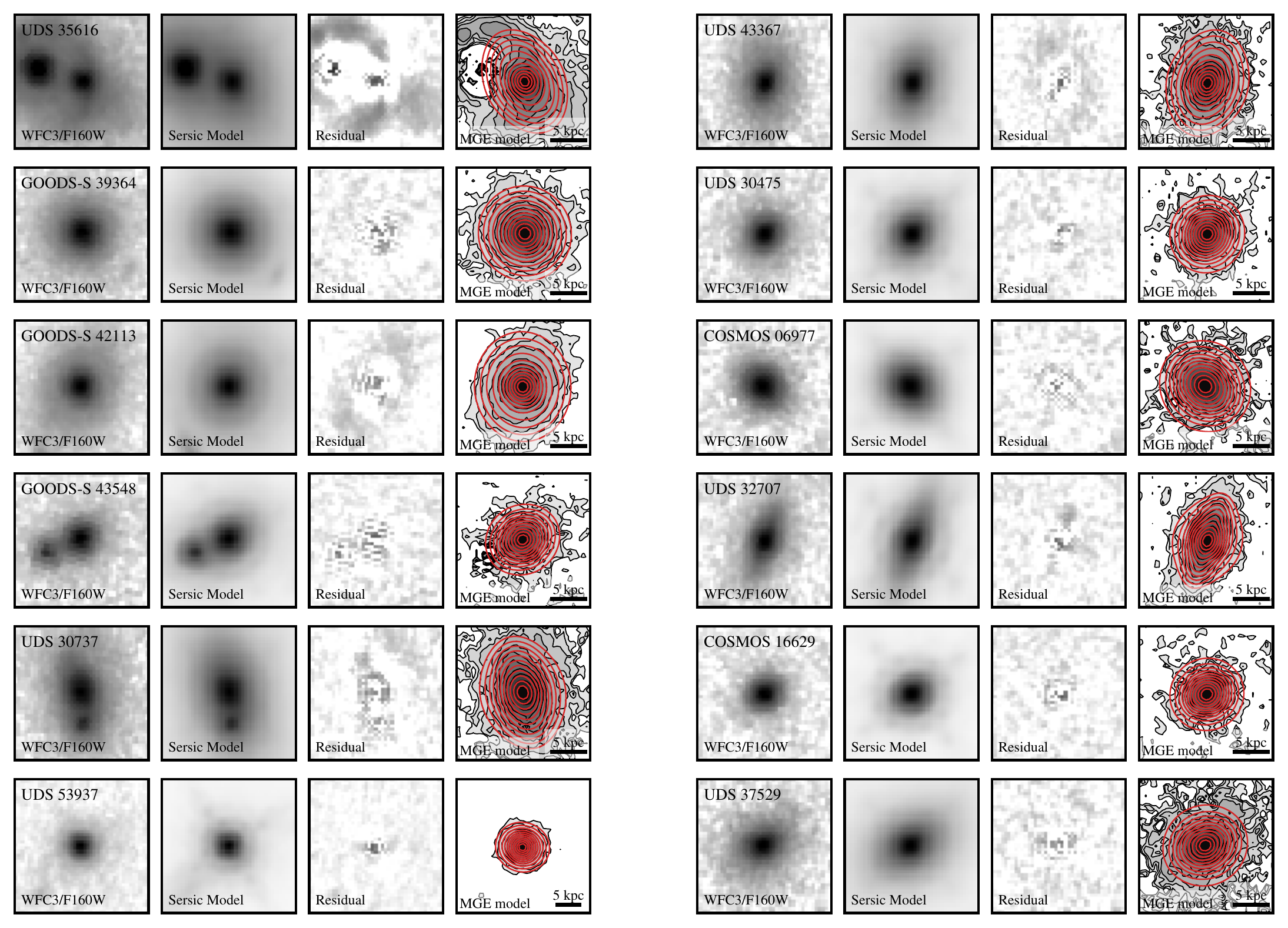}
\caption{(Continued)}
\label{fig.mp2}
\end{figure*}

\begin{figure*}
\centering
\figurenum{\ref{fig.mp0}}
\includegraphics[scale=1,angle=90]{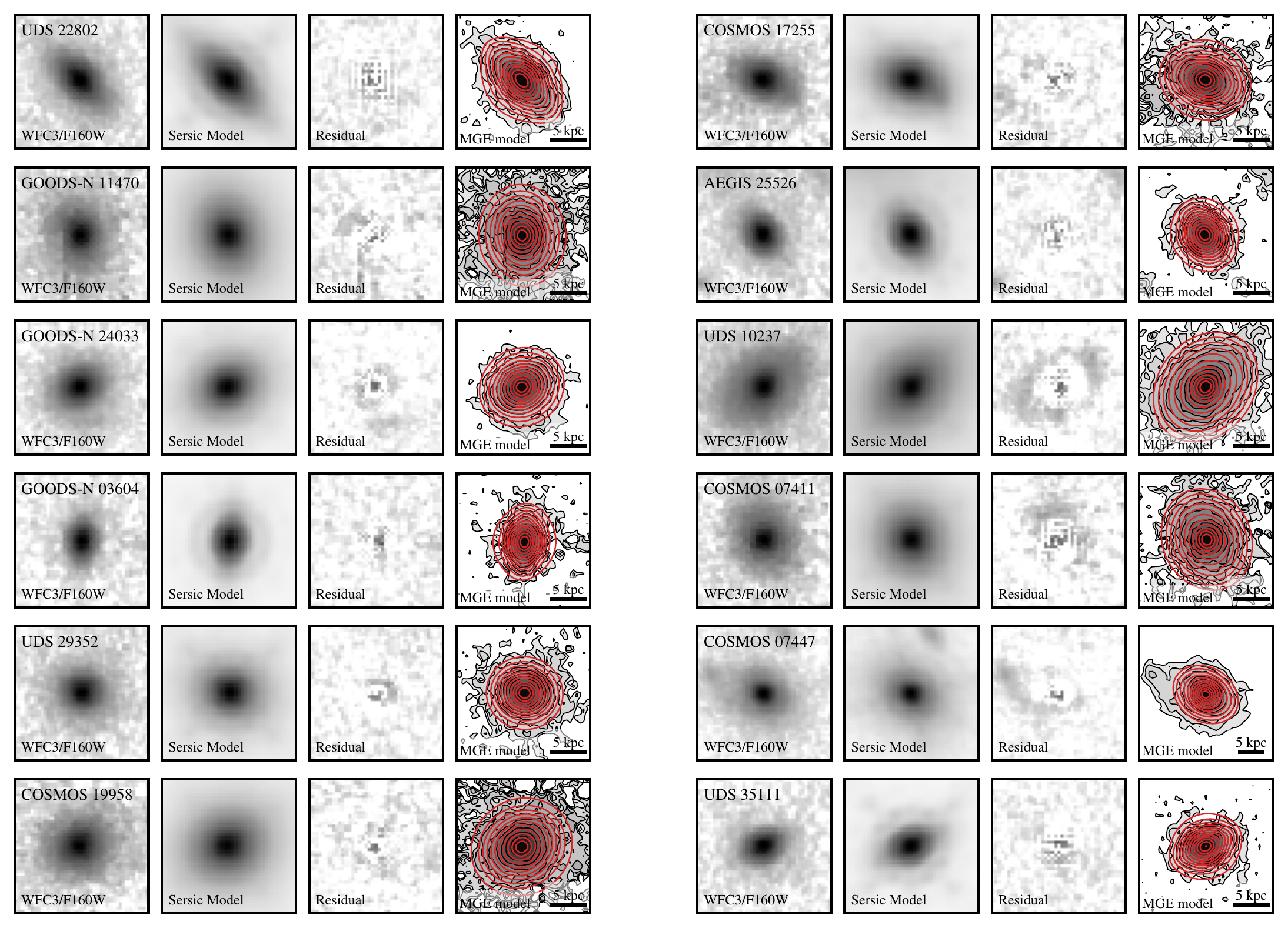}
\caption{(Continued)}
\label{fig.mp3}
\end{figure*}

\begin{figure*}
\centering
\figurenum{\ref{fig.mp0}}
\includegraphics[scale=1,angle=90]{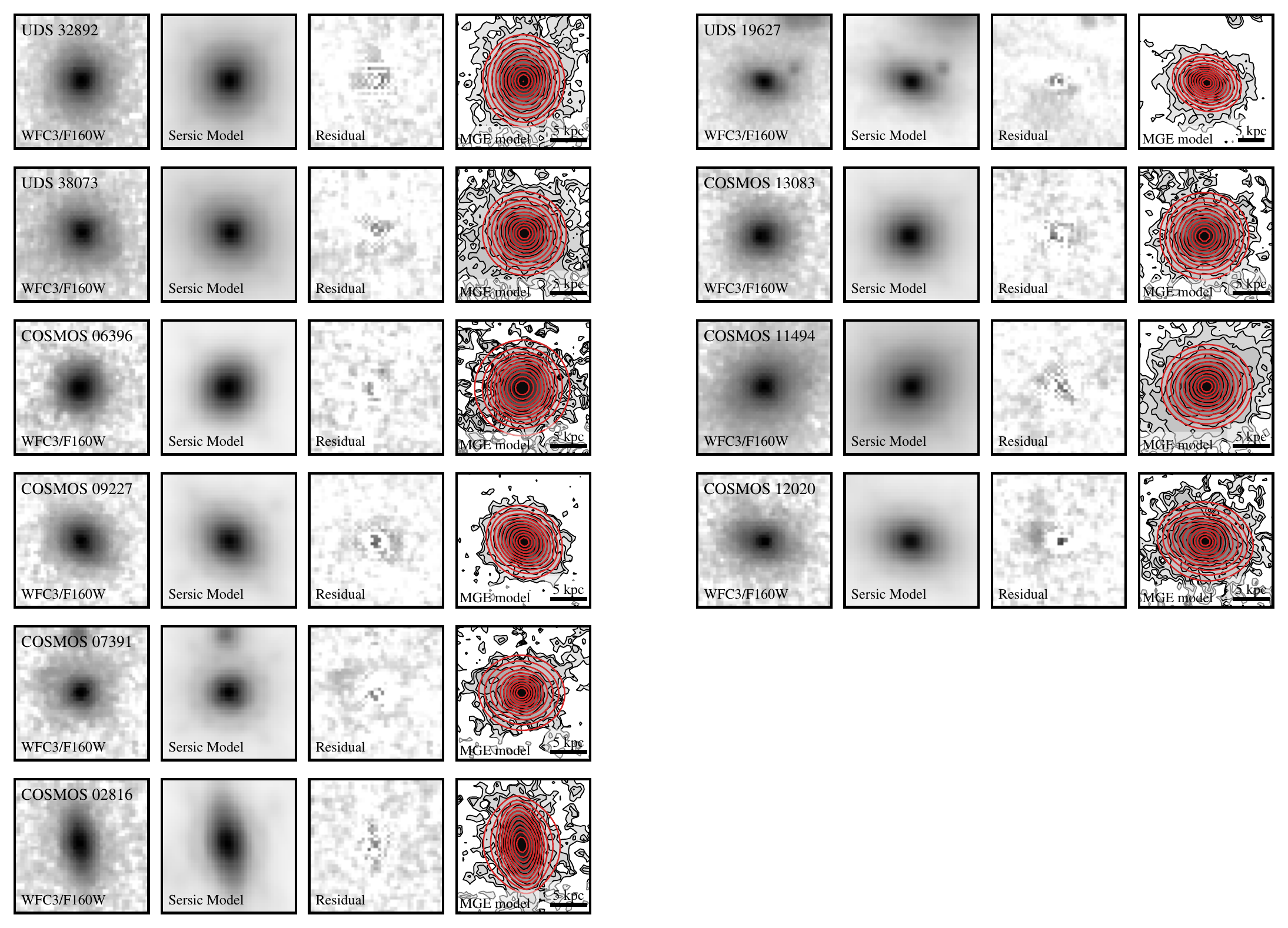}
\caption{(Continued)}
\label{fig.mp4}
\end{figure*}

\end{document}